\documentclass[twocolumn]{aastex6}

\usepackage{lineno}

    {\pagebreak[4]\global\pdfpageattr\expandafter{\the\pdfpageattr/Rotate 90}}%
    {\pagebreak[4]\global\pdfpageattr\expandafter{\the\pdfpageattr/Rotate 0}}%

\let\pwiflocal=\iffalse \let\pwifjournal=\iffalse
\usepackage{amsmath,amssymb}
\usepackage{epsfig}    
\usepackage{graphicx}    
\usepackage{lineno}
\usepackage{natbib}
\usepackage{bigints}
\usepackage[outdir=./]{epstopdf}
\usepackage[flushleft]{threeparttable}

\usepackage[T1]{fontenc}
\pwifjournal\else
  \usepackage{microtype}
\fi

\pwifjournal\else
  \makeatletter
  \renewcommand\plotone[1]{%
    \centering \leavevmode \setlength{\plot@width}{0.95\linewidth}
    \includegraphics[width={\eps@scaling\plot@width}]{#1}%
  }%
  \makeatother
\fi

\makeatletter

\newcommand\@simpfx{http://simbad.u-strasbg.fr/simbad/sim-id?Ident=}

\newcommand\MakeObj[4][\@empty]{
  \pwifjournal%
    \expandafter\newcommand\csname pkgwobj@c@#2\endcsname[1]{\protect\object[#4]{##1}}%
  \else%
    \expandafter\newcommand\csname pkgwobj@c@#2\endcsname[1]{\href{\@simpfx #3}{##1}}%
  \fi%
  \expandafter\newcommand\csname pkgwobj@f#2\endcsname{#4}%
  \ifx\@empty#1%
    \expandafter\newcommand\csname pkgwobj@s#2\endcsname{#4}%
  \else%
    \expandafter\newcommand\csname pkgwobj@s#2\endcsname{#1}%
  \fi}%

\newcommand\MakeTrunc[2]{
  \expandafter\newcommand\csname pkgwobj@t#1\endcsname{#2}}%

\newcommand{\obj}[1]{%
  \expandafter\ifx\csname pkgwobj@c@#1\endcsname\relax%
    \textbf{[unknown object!]}%
  \else%
    \csname pkgwobj@c@#1\endcsname{\csname pkgwobj@s#1\endcsname}%
  \fi}
\newcommand{\objf}[1]{%
  \expandafter\ifx\csname pkgwobj@c@#1\endcsname\relax%
    \textbf{[unknown object!]}%
  \else%
    \csname pkgwobj@c@#1\endcsname{\csname pkgwobj@f#1\endcsname}%
  \fi}
\newcommand{\objt}[1]{%
  \expandafter\ifx\csname pkgwobj@c@#1\endcsname\relax%
    \textbf{[unknown object!]}%
  \else%
    \csname pkgwobj@c@#1\endcsname{\csname pkgwobj@t#1\endcsname}%
  \fi}

\makeatother

\pwifjournal\else
  \usepackage{etoolbox}
  \makeatletter
  \patchcmd{\NAT@citex}
    {\@citea\NAT@hyper@{%
       \NAT@nmfmt{\NAT@nm}%
       \hyper@natlinkbreak{\NAT@aysep\NAT@spacechar}{\@citeb\@extra@b@citeb}%
       \NAT@date}}
    {\@citea\NAT@nmfmt{\NAT@nm}%
     \NAT@aysep\NAT@spacechar\NAT@hyper@{\NAT@date}}{}{}
  \patchcmd{\NAT@citex}
    {\@citea\NAT@hyper@{%
       \NAT@nmfmt{\NAT@nm}%
       \hyper@natlinkbreak{\NAT@spacechar\NAT@@open\if*#1*\else#1\NAT@spacechar\fi}%
         {\@citeb\@extra@b@citeb}%
       \NAT@date}}
    {\@citea\NAT@nmfmt{\NAT@nm}%
     \NAT@spacechar\NAT@@open\if*#1*\else#1\NAT@spacechar\fi\NAT@hyper@{\NAT@date}}
    {}{}
  \makeatother
\fi

\providecommand{\adsurl}[1]{\href{#1}{ADS}}

\newcommand{\kepler}{{\it Kepler}}

\newcommand\teff{\ensuremath{T_\text{eff}}}

\slugcomment{Accepted to AAS Journals}

\shorttitle{Parallaxes of Cool Planet Hosts}

\shortauthors{Mann et al.}

\begin{document}

\title{The Gold Standard: Accurate Stellar and Planetary Parameters for Eight \kepler\ M dwarf Systems Enabled by Parallaxes}

\author{Andrew W. Mann,\altaffilmark{1,2} Trent Dupuy,\altaffilmark{1} Philip S. Muirhead,\altaffilmark{3} Marshall C. Johnson,\altaffilmark{1,4} Michael C. Liu,\altaffilmark{5} Megan Ansdell,\altaffilmark{5} Paul A. Dalba,\altaffilmark{3} Jonathan J. Swift,\altaffilmark{6} Sam Hadden\altaffilmark{7}} 

\affil{\vspace{0pt}\\ 
$^{1}$Department of Astronomy, The University of Texas at Austin, Austin, TX 78712, USA\\
$^{2}$Hubble Fellow\\
$^{3}$Department of Astronomy, Boston University, 725 Commonwealth Ave., Boston, MA 02215, USA\\
$^{4}$Department of Astronomy, The Ohio State University, 140 West 18th Ave., Columbus, OH 43210 USA\\
$^{5}${Institute for Astronomy, University of Hawaii, 2680 Woodlawn Drive, Honolulu, HI 96822, USA}\\
$^{6}$The Thacher School, 5025 Thacher Rd., Ojai, CA 93023, USA\\
$^{7}${Department of Physics \& Astronomy, Northwestern University, Evanston, IL 60208, USA} \\
}

\begin{abstract}
We report parallaxes and proper motions from the Hawaii Infrared Parallax Program for eight nearby M~dwarf stars with transiting exoplanets discovered by \kepler. We combine our directly measured distances with mass-luminosity and radius--luminosity relationships to significantly improve constraints on the host stars' properties. Our astrometry enables the identification of wide stellar companions to the planet hosts. Within our limited sample, all the multi-transiting planet hosts (three of three) appear to be single stars, while nearly all (four of five) of the systems with a single detected planet have wide stellar companions. By applying strict priors on average stellar density from our updated radius and mass in our transit fitting analysis, we measure the eccentricity probability distributions for each transiting planet. Planets in single-star systems tend to have smaller eccentricities than those in binaries, although this difference is not significant in our small sample. In the case of Kepler-42bcd, where the eccentricities are known to be $\simeq$0, we demonstrate that such systems can serve as powerful tests of M~dwarf evolutionary models by working in $L_\star-\rho_\star$ space. The transit-fit density for Kepler-42bcd is inconsistent with model predictions at 2.1$\sigma$ (22\%), but matches more empirical estimates at 0.2$\sigma$ (2\%), consistent with earlier results showing model radii of M dwarfs are underinflated. Gaia will provide high-precision parallaxes for the entire Kepler M~dwarf sample, and {\it TESS} will identify more planets transiting nearby, late-type stars, enabling significant improvements in our understanding of the eccentricity distribution of small planets and the parameters of late-type dwarfs.
\end{abstract}


\maketitle

\section{Introduction}\label{sec:intro}

Accurate and precise fundamental properties of stars, such as mass, radius, and luminosity, are critical for determining accurate and precise fundamental properties of orbiting planets. A stellar radius is needed to derive both the planetary radius and semi-major axis from transit observables, stellar luminosity is critical to constrain planet insolation, and stellar mass is required to determine the planetary mass from radial velocity observations. In the case of precision radial velocity or transit measurements, or stars unlike the Sun, planet properties are often limited by our understanding of their host star \citep[e.g.,][]{Bastien2014,2014ApJS..211....2H,2014MNRAS.443.1810B}.

There are a wide range of methods to measure stellar properties. Photometry alone provides a reasonable estimate of stellar effective temperature, and metallicity for Sun-like stars \citep[e.g.,][]{2011A&A...530A.138C}. This can be combined with assumptions about a star's evolutionary state based on, for example, the reduced proper motion, to provide an accurate but imprecise estimate for the stellar mass and radius \citep{Lepine:2011vn,Huber2016}. Spectroscopy provides further information: more accurate determination of effective temperature and metallicity, as well as means to estimate stellar surface gravity, all of which further reduce errors on the stellar radius, luminosity, and mass \citep[e.g.,][]{2015ApJ...805..126B}. For warm and/or evolved stars with precision light curves, asteroseismology can probe a star's internal structure through seismic oscillations, yielding some of the most precise stellar radii and masses available \citep[e.g.,][]{2013ApJ...767..127H,2015MNRAS.452.2127S}.

In the absence of direct mass and radius measurements, and for stars where asteroseismology is impractical or impossible, parallaxes provide the more accurate means to determine stellar parameters. When combined with multi-band photometry, parallaxes provide a measurement of a star's intrinsic luminosity, a parameter directly tied to stellar mass and age.  For most stars, luminosity from parallax and temperature and metallicity from spectroscopy fully constrain the location of a star on an evolutionary track on a Hertzsprung-Russell diagram, providing the fundamental stellar properties. While this does not work for all stars owing to complications of age \citep[e.g.,][]{2011ApJ...739L..49L} and activity/spots \citep[e.g.,][]{Gully-Santiago2017}, parallaxes are especially effective for measuring stellar properties of very low-mass stars (late K and M~dwarfs). Once settled on the main sequence, M~dwarf stars exhibit little change in stellar temperature and luminosity over tens of billions of years \citep[][]{Laughlin1997}. 

The slow evolution of (main-sequence) M~dwarfs means that age has a negligible influence on the star's position on a color-magnitude diagram, creating a tight relation between the chemical abundance, luminosity, and radius of M~dwarfs. Empirical studies of M~dwarf parameters have taken advantage of this to derive relations between absolute magnitude and fundamental parameters (e.g., mass, radius). \citet{Henry:1993fk}, \citet{Delfosse2000}, and \citet{Benedict:2016aa} measured dynamical masses of visual M~dwarf binaries, which they used to constrain the relation between absolute $V$ or $K$ magnitude ($M_{V,K}$) and stellar mass. Because of the relative tightness of the relation with $M_K$, \citet{Delfosse2000} argued that metallicity increased the scatter at shorter wavelengths. \citet{Mann2015b} used nearby stars with precise parallaxes, bolometric fluxes, and effective temperatures calibrated to reproduce stars with interferometric radii \citep{Boyajian2012} to develop empirical relations between $M_K$ and stellar radius. \citet{Mann2015b} also calculated a $M_K$-mass relation by interpolating their parameters onto a stellar evolution model \citep{Dotter2008}. 

With accurate stellar parameters and a high-precision light curve, it is possible to constrain the eccentricity of a transiting planet. \citet{Seager:2003lr} derived the relation between transit observables and stellar density ($\rho_\star$), although the result is degenerate with the planet's eccentricity ($e$). If $e$ is known independently, the transit-derived $\rho_\star$ can be used to test stellar evolutionary models \citep[e.g.,][]{2016arXiv161204379B}. If $\rho_\star$ is known independently, this method can be reversed to derive the eccentricity of the orbiting planet. Early efforts to apply this technique on \kepler\ planet hosts \citep[e.g.,][]{Moorhead:2011lr,2012MNRAS.425..757K} have been complicated by imprecise or systematically inaccurate stellar parameters common to \kepler\ target stars \citep[e.g.,][]{Gaidos2013,Bastien2014, 2014PASP..126...34P}. Two approaches have been more successful; (1) identifying high eccentricity planets, which is possible even with relatively large errors in $\rho_\star$ due to their extreme transit durations \citep[e.g.,][]{Dawson:2012fk}, and (2) subsamples of transiting planet hosts with extremely precise stellar densities, such as those with stellar parameters from asteroseismology \citep[e.g.,][]{Van-Eylen2015}. An intermediate case, where bulk stellar parameters are known with moderate precision, are also useful for statistical eccentricity constraints over large samples of planets \citep{2016PNAS..11311431X}, although this is less useful on individual planets. 

\kepler\ M~dwarf planet hosts are appealing for eccentricity studies because of the combination of a large target sample (including multi-planet systems), continuous photometric monitoring for $\simeq$4.5 years, and exceptional photometric precision for most objects. \kepler\ target selection is public and well established \citep{Batalha:2010fk}, and planet-detection completeness can be precisely measured \citep[e.g.,][]{Dressing2015,2016ApJ...828...99C}, making \kepler\ targets the ideal dataset for statistical studies of exoplanets. The small size of M dwarfs makes them attractive for studies of small planets, as the transit depths are larger when compared to Sun-like stars. However, their stellar parameters are more poorly constrained. \kepler{} M~dwarf radii are typically determined by measuring their effective temperature and [Fe/H] from spectra or photometry, then interpolating the radius onto a model-based or empirical relation between \teff, [Fe/H], and $R_\star$ \citep[e.g.,][]{Dressing2013,Muirhead2014,Newton2015A,Gaidos2016a}, which provides radii precise only to 10-25\%. While {\it K2} targets are statistically closer, few have parallaxes, and like \kepler\ targets, require follow-up to constrain the stellar parameters \citep[e.g.,][]{Dressing2017,Martinez:2017aa}. The stellar parameter problem could be eliminated with parallax measurements of \kepler\ M~dwarfs. 

The Gaia spacecraft \citep{2001A&A...369..339P} is expected to provide parallaxes for all \kepler{} M~dwarfs accurate to $\lesssim$100\,$\mu$as, yielding distances accurate to better than 2\% \citep{2012Ap&SS.341...31D}. These data will enable studies of the eccentricity distribution of small exoplanets, currently only possible around higher-mass stars with asteroseismic densities \citep[e.g.,][]{Van-Eylen2015}, and offer a wealth of systems to test stellar models. Until this time we are limited by the small number of systems for which we can measure parallaxes from the ground, but even such a small sample is sufficient to test and refine our methods in preparation for the impending Gaia release. 

In this paper we present parallaxes from the Hawaii infrared parallax program for eight M~dwarfs with planets detected by \kepler\ (Section~\ref{sec:target}, \ref{sec:obs}). In Section~\ref{sec:params} we combine these distances with newly developed relations between luminosity, mass, and radius to provide more precise stellar (and exoplanet) parameters for each object. We describe our MCMC fit to the transit light curves of each system and the resulting planetary parameters in Section~\ref{sec:transit}. Utilizing our improved stellar parameters, updated transit fits, and literature adaptive optics information, in Section~\ref{sec:fpp} we verify the planetary nature of two of the three planets in our sample currently lacking confirmation.  We provide details on individual systems in Section~\ref{sec:systems}. In Section~\ref{sec:42} we take advantage of external constraints on the eccentricities of Kepler-42bcd to test models of low-mass stars and the empirical mass/radius-luminosity relations used in this paper. We conclude in Section~\ref{sec:discussion} with a brief summary and discussion of how Gaia parallaxes will enable more precise calculations for most of the planet sample from \kepler{} and eventually the Transiting Exoplanet Survey Satellite mission \citep[{\it TESS},][]{Ricker2014}.

\section{Target Selection}\label{sec:target}

The sample was chosen from spectroscopic distances measured as part of the Characterizing the Cool KOIs survey \citep{Muirhead2012a, Muirhead2014}. Distances were estimated for late-type planet-candidate hosts by interpolating spectroscopically determined effective temperature (\teff) and metallicity ([M/H]) onto custom low-mass-star isochrones similar to the Dartmouth Stellar Evolution Program \citep[DSEP; ][]{Dotter2008, Feiden2013, Feiden2014a}.  The interpolation provided a prediction for the absolute $K_S$-band ($M_{K_S}$) magnitudes of the stars, which we compared to apparent $K_S$ magnitudes measured by the Two Micron All Sky Survey \citep[2MASS; ][]{Cutri2003, Skrutskie2006} to estimate distances to the stars. However, because of of complexity in the cool atmospheres of M~dwarfs \citep{1976A&A....48..443M}, \teff, and [M/H] are difficult to constrain precisely. Combined with a steep relation between \teff-[Fe/H] and $M_{K_S}$ for cool stars, resulting distances are subject to large uncertainties. 

The 20 closest stars listed in \citet{Muirhead2014} are the following: KOI-314 \citep[Kepler-138, ][]{Rowe2014}, KOI-2453, KOI-2842 \citep[Kepler-446, ][]{Muirhead2015}, KOI-1702 B (Kepler-1651 B), KOI-961 \citep[Kepler-42, ][]{Muirhead2012}, KOI-249 B, KOI-3497 \citep[Kepler-1512, ][]{2016ApJ...822...86M}, KOI-1725 A (Kepler-1651 A), KOI-249 A \citep[Kepler-504, ][]{2016ApJ...822...86M}, KOI-463 \citep[Kepler-560, ][]{2016ApJ...822...86M}, KOI-3119, KOI-2662 \citep[Kepler-1308, ][]{2016ApJ...822...86M}, KOI-571 \citep[Kepler-186, ][]{Quintana2014}, KOI-1702 (Kepler-1650), KOI-2542, KOI-2705 \citep[Kepler-1319, ][]{2016ApJ...822...86M}, KOI-3749, KOI-4290 \citep[Kepler-1582, ][]{2016ApJ...822...86M}, KOI-251 \citep[Kepler-125][]{Rowe2014} and KOI-1422 \citep[Kepler-296, ][]{Torres2015}. Revisions to stellar properties of KOI-2704 \citep[Kepler-445, ][]{Muirhead2015} led us to include this in the nearby list as well. 

Of these 21 nearby targets, we observed nine based on how accurately their trigonometric parallaxes could be measured (e.g., NIR brightness, distance, number of nearby stars). This prioritizes the coolest systems, as they are still relatively bright in the near-infrared (where the astrometry is measured) but are statistically closer due to cuts imposed by \kepler\ at the bluer $K_P$-band. The final selected systems are Kepler-42, 138, 445, 560, 1319, 1650, 1651, and KOI-, 2453, and 2542. Both components for likely binaries (e.g., Kepler-1651AB) are included, owing to the field of view of our detector (see Section~\ref{sec:obs}). 

KOI-2542 was later removed from the sample due to the presence of a 0\farcs75 companion, which is near the seeing limit of our observations, and hence complicates the astrometric measurement. While four other targets are confirmed to be in wide binaries, the next tightest system is 1\farcs88 (Kepler-1319) and hence easily resolvable. All targets also have adaptive optics imaging and non-redundant aperture masking from \citet{Kraus2016a}, which rules out any companions bright enough to significantly impact the astrometry. 

\section{Observations and Reduction}\label{sec:obs}
Astrometric observations of each target were taken using the facility near-IR camera WIRCam \citep{2004SPIE.5492..978P} on the Canada-France-Hawaii Telescope from 2012 to 2015 as part of the Hawaii Infrared Parallax Program. Our observing strategy and analysis methods are described in detail in \citet{Dupuy2012}, and recent updates to the pipeline used here are described in \citet{2017arXiv170305775D}. For all targets here we used WIRCam's narrow $K$-band filter centered at 2.122\,\micron\ and typically obtained 18--20 dithered images at each observation epoch that had typical seeing of 0$\farcs$5--0$\farcs$6. We computed the mean and standard error on the mean as the astrometric measurement and corresponding uncertainty for each star at a given epoch. After solving for linear transformations between epochs while masking the target and reference stars with high proper motion, we used 2MASS to determine the absolute astrometric terms (e.g., pixel scale, orientation). We used this final absolute astrometry to determine initial parallaxes and proper motions of our targets using Markov Chain Monte Carlo analysis. Because stars used to build the astrometric reference grid in the WIRcam images have non-zero parallaxes and proper motions, the resulting parallaxes and proper motions are only relative. We convert relative to absolute astrometry using corrections estimated from the Besan\c{c}on model of the Galaxy \citep{2003A&A...409..523R}. 

In addition to the target of interest, we attempt to solve for proper motions and parallaxes of all stars in the WIRCam field of view. We identify four stars where the derived parallax and proper motions agree with that of the target within 1.5$\sigma$ (Kepler-560, 1319, 1651, and KOI-2453). Two of these systems, Kepler-560AB and KOI-2453AB were previously confirmed as wide binaries in \citet{2016MNRAS.455.4212D}; the other two were missed because they land inside the \citet{2016MNRAS.455.4212D} 6\arcsec\ minimum search radius. 

\begin{figure}
 \includegraphics[width=84mm]{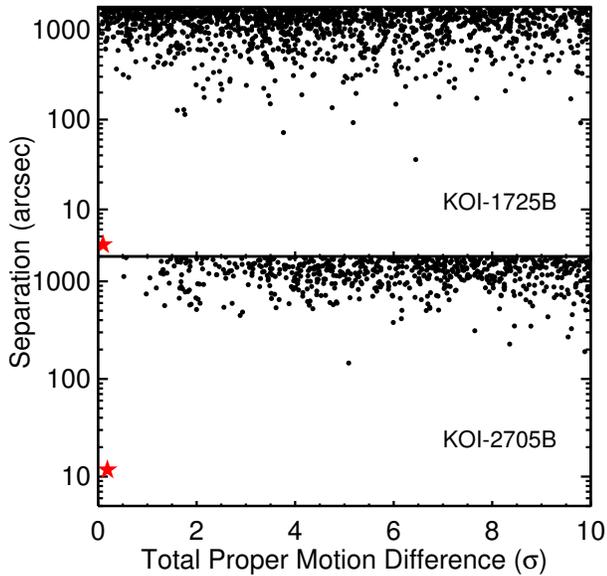}
 \caption{ Separation vs. total proper motion difference (in standard deviations) for stars near KOI-1725 (Kepler-1651B, top) and KOI-2705 (Kepler-1319B, bottom). The red star indicates the target we identify as a true companion, while black points are unassociated stars. For relatively high-proper-motion targets like these, the probability that an unassociated star has comparable proper motion and is nearby on the sky is extremely small, which is evident by the large region of unfilled parameter space. } 
  \label{fig:pm}
 \end{figure}

\citet{2016MNRAS.455.4212D} derive a relation for the probability that two stars have consistent proper motions by chance, as a function of their proper motion, separation, and density of stars in that region of the sky. \citet{2016MNRAS.455.4212D} empirically calibrate their relation by shifting the positions of stars by several degrees \citep{Lepine:2007qy}, after which any nearby targets with consistent proper motions will be chance alignment. Following this method, we find that both Kepler-560AB and Kepler-1319AB have $<$0.01\% probability of begin unassociated pairs (see Figure~\ref{fig:pm}). The wide binary to Kepler-560 (KOI-463) is brighter than the planet host; to avoid confusion, we denote this as KOI-463A (and the planet host as KOI-463B). The three other binaries we denote as Kepler-1319B (KOI-2705B), Kepler-1651B (KOI-1725B), and KOI-2453B. 

\begin{figure}
 \includegraphics[width=84mm]{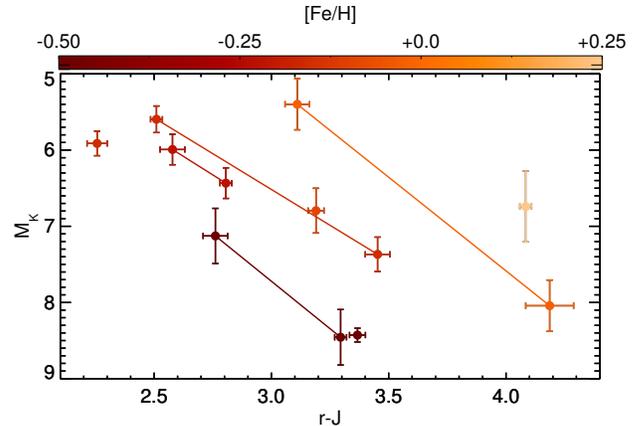}
 \caption{Color-magnitude diagram of the targets and their primary/companion stars. Points are colored by their [Fe/H] values. Lines connect the components of binary systems. } 
  \label{fig:HR}
 \end{figure}

Final parallaxes and proper motions for the targets and identified companions are given in Table~\ref{tab:plx}, with binary parameters (separation, position angle, $\Delta K$) in Table~\ref{tab:binary}. Plots of the astrometric fits are shown in Figure~\ref{fig:plx}. For our analysis of KOI-2453AB, Kepler-1319AB, and 1651AB, we use the mean of the primary and companion parallax as the system parallax. Since parallax measurements of targets near each other are highly correlated, we adopt the minimum error of the primary and companion as the error on the system parallax. For Kepler-560AB, the components are widely separated enough (40\arcsec) that the parallaxes are largely uncorrelated, so we adopt the weighted mean for system parallax and error. We show a simple color-magnitude diagram of all targets (including identified binaries) in Figure~\ref{fig:HR}.

\begin{deluxetable*}{l l l l l l l l}
\tablecaption{Summary of Binary Parameters  \label{tab:binary}}
\tablewidth{0pt}
\tablehead{
\colhead{KOI} &  \multicolumn{2}{c}{Separation} & \colhead{PA }& \colhead{$\Delta$K} \\
 & \colhead{(\arcsec)} & \colhead{(au)} & \colhead{($^\circ$)} & \colhead{(mag)} 
}
\startdata
KOI-463AB  &     $40.436\pm0.007$  &$4084\pm371$&    $101.02\pm0.03$ &     $-0.44\pm0.06$  \\
KOI-1725AB & \phn$4.055\pm0.003$ &\phn$281\pm 21$& \phn$98.50\pm0.03$ & \phs$1.46\pm0.01$ \\
KOI-2453AB &     $11.726\pm0.003$ &$1465\pm238$ & $204.67\pm0.02$ &    \phs$1.61\pm0.01$ \\
KOI-2705AB &  \phn$1.880\pm0.007$  &\phn$205\pm31$&     $304.07\pm0.09$  & \phs$2.45\pm0.03$ \\
\enddata
\end{deluxetable*}

\section{Stellar Parameters}\label{sec:params}

\subsection{Revised Parameters from Parallaxes}
For each planet host and stellar companion we derived updated $R_\star$, $L_\star$, $M_\star$, and \teff\ values using our parallaxes. Because evolutionary and atmospheric models poorly reproduce radii of low-mass eclipsing binaries and observed spectra of M dwarfs \citep[e.g.,][]{Mann2013c,2014MNRAS.437.2831Z}, we prefer using more empirically driven methods where possible. Although few techniques are completely free of model assumptions, we consider a method `empirical' if potential errors introduced by model assumptions are likely small compared to measurement or other uncertainties. For example, radii and temperatures derived from long-baseline optical interferometry \citep[LBOI; e.g.,][]{Berger2006,Kane2017} depend only on model limb-darkening values (corrections of 2-5\%), and atmospheric models in the ultraviolet and infrared to calculate the total bolometric flux (\teff\ corrections of $<2\%$). While these corrections are comparable to or larger than typical measurement uncertainties (2-10\%), the {\it errors} on these corrections are $\ll$100\%, and thus introduce model-dependent errors well below typical measurement uncertainties. 

Iron abundances ([Fe/H]) for each target were taken from \citet{Mann2013b} and \citet{Muirhead2014}. Both \citet{Mann2013b} and \citet{Muirhead2014} utilize relations between the equivalent width of spectroscopic features and the metallicity of the M~dwarf, calibrated using nearby M~dwarf companions to FGK dwarfs \citep[e.g.][]{2010ApJ...720L.113R,2012A&A...538A..25N,Terrien:2012lr} under the assumption that binary components have identical metallicities. Because the different calibrations use overlapping samples and similar methods, the [Fe/H] from each source generally agree within errors. For stars with determinations in both we use the weighted mean of the two. Lower limits on the error are taken to be 0.09~dex (Gaussian) due to underlying calibration errors common to both. While the calibration of M dwarf abundance measurements still relies on atmospheric models to determine the abundances of the FGK primary, the model-fitting procedure utilized for these FGK dwarfs reproduces independent asteroseismic and interferometric measurements of surface gravity \citep{2015ApJ...805..126B}.

To determine the radius ($R_\star$) of each star, we first calculated the $M_{K_S}$ from our measured parallax (Section~\ref{sec:obs}) and $K_S$ magnitudes from 2MASS. For Kepler-1319AB the components were unresolved in 2MASS, so we used the $\Delta K$ value from \citet{Kraus2016a} to calculate component magnitudes. For Kepler-1651, the $K_S$ magnitude is also partially contaminated by its nearby companion, so we instead use the 2MASS $J$-band measurement, which is negligibly affected. Radii for each of the stars were then computed using the relation between $M_{K_S}$ (or $M_J$ for Kepler-1651), [Fe/H], and $R_\star$ from \citet{Mann2015b}. The \citet{Mann2015b} $R_\star$ relation is derived from a local sample of M~dwarfs with precise ($<5\%$) parallaxes, bolometric fluxes from flux-calibrated spectra, \teff\ values from optical spectra, and $R_\star$ calculated from the Stefan-Boltzmann relation. While the \teff\ determinations depend on comparing spectra to atmospheric models, the comparison down-weights regions of the optical spectrum poorly reproduced by models, and the final values are calibrated to reproduce radii and \teff{} from direct radii measurements using long-baseline optical interferometry \citep{Boyajian2012,2014MNRAS.438.2413V}. 

We determined the masses of each object using the $M_{K_S}$-$M_\star$ relations from \citet{Benedict:2016aa} and \citet{Mann2015b}. The relation from \citet{Mann2015b} is more precise (residuals of $\lesssim2\%$), but is derived by interpolating [Fe/H], \teff, and bolometric luminosities onto a model grid from an updated version of the DSEP models \citep{Dotter2008, Feiden2013,Feiden2014a}. These masses reproduce the empirical mass-radius relation derived from low-mass eclipsing binaries \citep{Feiden2012a,Mann2015b}. However, they are not as model-independent as those from \citet{Benedict:2016aa}, which are derived from astrometric monitoring of low-mass visual binaries. Hence we also include masses derived using the \citet{Benedict:2016aa} relation. Values from both methods are consistent within 1$\sigma$ for all stars. 

We combined our distances with bolometric corrections to derive $L_\star$, which we utilize to calculate \teff. To this end, we drew SDSS $r'$ magnitudes for each star from the Carlsberg Meridian Catalog \citep[CMC15,][]{2014AN....335..367M}, the \kepler-INT survey \citep{Greiss2012}, the eighth data release of APASS \citep{Henden:2012fk}, or \citet{Muirhead2012} for Kepler-42. All $J$ magnitudes come from 2MASS. For Kepler-1319AB, we used $\Delta i'$ measurements from \citet[][Robo-AO; ]{2016AJ....152...18B} and synthetic magnitudes from \citet{Mann2015b} to estimate $\Delta r$ and calculate component magnitudes. We used the $r-J$ colors to calculate a $J$-band bolometric correction following \citet{Mann2015b}, providing us with a bolometric luminosity ($L_\star$) for each target.  From $L_\star$ and $R_\star$ we derive a revised \teff\ using the Stefan-Boltzmann relation.

Errors on stellar parameters were determined by a an MC analysis. We generated a sample of 10$^4$ parallaxes, $K_S$ magnitudes, and [Fe/H] values according to the prescribed errors. We fed these posteriors through each formula, which provided a range of values for each parameter. We account for errors in the given relations (e.g., bolometric corrections) by adding the literature uncertainties in quadrature. We report the median, 15.9\% and 84.1\% values of the cumulative distributions (equivalent to $\pm1\sigma$ for normal distributions) for each parameter in Table~\ref{tab:stellar}.

\begin{deluxetable*}{l l l c c c c c c c c }
\tablecaption{Updated Stellar Parameters \label{tab:stellar}}
\tablewidth{0pt}
\tablehead{
\colhead{KOI} & \colhead{KIC} & \colhead{Kepler} & \colhead{$M_{\star,Mann}^a$} & \colhead{$M_{\star,Ben}^b$} &\colhead{$R_\star$} & \colhead{$\rho_\star$} & \colhead{$L_\star$} & \colhead{\teff} & \colhead{Distance} & \colhead{[Fe/H]\tablenotemark{c}} \\
\colhead{} & \colhead{} & \colhead{\#} & \colhead{($M_\odot$)} & \colhead{($M_\odot$)} & \colhead{($R_\odot$)} & \colhead{($\rho_\odot$)} & \colhead{($\log\left(L/L_\odot \right)$)} & \colhead{(K)} & \colhead{(pc)} 
}
\startdata
\multicolumn{11}{c}{Likely Planet Hosts}\\
 314$^d$ &    7603200 & 138 & $0.468^{+0.030}_{-0.028}$ & $0.511^{+0.035}_{-0.033}$ & $0.454^{+0.025}_{-0.023}$ & $5.00^{+0.52}_{-0.50}$ & $-1.451^{+0.066}_{-0.062}$ & $3726.^{+44.}_{-40.}$ & $ 52.4\pm  3.8$ & $-0.16$ \\
 463B &    8845205 & 560 & $0.381^{+0.034}_{-0.030}$ & $0.412^{+0.043}_{-0.039}$ & $0.378^{+0.028}_{-0.025}$ & $7.03^{+0.92}_{-0.87}$ & $-1.717^{+0.082}_{-0.075}$ & $3502.^{+44.}_{-39.}$ & $101.0\pm  9.2$ & $-0.23$ \\
 961$^d$ &    8561063 & 42 & $0.144^{+0.007}_{-0.006}$ & $0.141^{+0.008}_{-0.008}$ & $0.175^{+0.006}_{-0.006}$ & $26.80^{+1.77}_{-1.72}$ & $-2.511^{+0.037}_{-0.035}$ & $3258.^{+21.}_{-21.}$ & $ 40.5\pm  1.6$ & $-0.50$ \\
1702 &    7304449 & 1650 & $0.326^{+0.047}_{-0.039}$ & $0.344^{+0.059}_{-0.048}$ & $0.334^{+0.039}_{-0.032}$ & $8.73^{+1.67}_{-1.57}$ & $-1.871^{+0.123}_{-0.107}$ & $3410.^{+56.}_{-46.}$ & $120.5\pm 16.0$ & $-0.11$ \\
1725A &   10905746 & 1651 & $0.522^{+0.033}_{-0.031}$ & $0.564^{+0.034}_{-0.034}$ & $0.503^{+0.030}_{-0.027}$ & $4.10^{+0.47}_{-0.44}$ & $-1.368^{+0.069}_{-0.064}$ & $3713.^{+57.}_{-53.}$ & $ 69.4\pm  5.3$ & $-0.16$ \\
2453A &    8631751 &  & $0.279^{+0.055}_{-0.042}$ & $0.287^{+0.068}_{-0.050}$ & $0.291^{+0.044}_{-0.034}$ & $11.34^{+2.70}_{-2.48}$ & $-1.897^{+0.153}_{-0.130}$ & $3603.^{+67.}_{-53.}$ & $125.0\pm 20.3$ & $-0.48$ \\
2704$^d$ &    9730163 & 445 & $0.334^{+0.080}_{-0.059}$ & $0.354^{+0.097}_{-0.072}$ & $0.347^{+0.068}_{-0.049}$ & $8.01^{+2.46}_{-2.22}$ & $-1.939^{+0.202}_{-0.165}$ & $3219.^{+89.}_{-63.}$ & $149.3\pm 31.2$ & $+ 0.27$ \\
2705A$^e$ &   11453592 & 1319 & $0.557^{+0.063}_{-0.055}$ & $0.592^{+0.042}_{-0.049}$ & $0.539^{+0.062}_{-0.050}$ & $3.56^{+0.75}_{-0.71}$ & $-1.327^{+0.142}_{-0.121}$ & $3673.^{+103.}_{-81.}$ & $116.3\pm 17.6$ & $-0.03$ \\
\hline
\multicolumn{11}{c}{Companion/Primary to Planet Hosts}\\
 463A &    8845251 &  & $0.454^{+0.036}_{-0.033}$ & $0.496^{+0.042}_{-0.039}$ & $0.440^{+0.031}_{-0.027}$ & $5.32^{+0.69}_{-0.66}$ & $-1.519^{+0.083}_{-0.075}$ & $3637.^{+52.}_{-46.}$ & $101.0\pm  9.2$ & $-0.23$ \\
1725B &   10905748 &  & $0.247^{+0.030}_{-0.027}$ & $0.249^{+0.036}_{-0.032}$ & $0.269^{+0.024}_{-0.022}$ & $12.73^{+2.00}_{-1.76}$ & $-1.982^{+0.069}_{-0.065}$ & $3569.^{+108.}_{-104.}$ & $ 69.4\pm  5.3$ & $-0.16$ \\
2453B &    8631743 &  & $0.143^{+0.030}_{-0.019}$ & $0.140^{+0.029}_{-0.017}$ & $0.174^{+0.028}_{-0.020}$ & $27.16^{+7.03}_{-6.19}$ & $-2.553^{+0.153}_{-0.130}$ & $3193.^{+56.}_{-55.}$ & $125.0\pm 20.3$ & $-0.48$ \\
2705B$^e$ &   11453591 &  & $0.175^{+0.036}_{-0.025}$ & $0.170^{+0.038}_{-0.025}$ & $0.208^{+0.031}_{-0.023}$ & $19.42^{+4.41}_{-4.08}$ & $-2.425^{+0.143}_{-0.122}$ & $3141.^{+47.}_{-43.}$ & $116.3\pm 17.6$ & $-0.03$ \\
\enddata
\tablenotetext{a}{Mass derived from the \citet{Mann2015b} semi-empirical mass-luminosity relation.}
\tablenotetext{b}{Mass derived from the \citet{Benedict:2016aa} empirical mass-luminosity relation.}
\tablenotetext{c}{[Fe/H] is limited by the underlying calibration \citep{Rojas-Ayala:2012uq,Mann2013a}, so we adopt 0.09~dex Gaussian errors for all targets.}
\tablenotetext{d}{Multi-transiting planet host.}
\tablenotetext{e}{Either component could be the host, but the primary is favored (see Section~\ref{sec:systems}).}
\end{deluxetable*}

\subsection{Comparison to spectroscopic parameters}
We compare our updated parameters to those derived spectroscopically for M~dwarf KOIs by \citet{Mann2013b, Mann2013c,Muirhead2014} and \citet{Newton2015A}. Each of these studies followed similar methods with a few key differences, which we briefly summarize here. All studies utilized low-resolution ($700\le R \le2000$) spectra: \citet{Mann2013b, Mann2013c} used a mix of optical for \teff\ and [Fe/H] and NIR for [Fe/H], while \citet{Muirhead2014} and \citet{Newton2015A} used exclusively NIR spectra for both \teff\ and [Fe/H].  All studies computed [Fe/H] of \kepler\ M~dwarf planet hosts using empirical metallicity calibrations from nearby binaries \citep{Rojas-Ayala:2012uq,Mann2013a,Newton:2014}. \citet{Mann2013c} and \citet{Newton2015A} both anchored their \teff\ and $R_\star$ values to those from nearby stars with interferometric temperatures and radii from \citet{Boyajian2012}, while \citet{Muirhead2014} used atmospheric models \citep{2012RSPTA.370.2765A} following \citet{Rojas-Ayala:2012uq} to constrain \teff, and evolutionary models \citep{Dotter2008} for $R_\star$. \citet{Mann2013c} relies on first computing \teff\ and deriving $R_\star$ using an empirical \teff-$R_\star$ relation, while \citet{Newton:2014} uses an empirical relation between NIR atomic indices and $R_\star$ (eliminating the need for \teff\ or [Fe/H]). \citet{Mann2013c} $R_\star$ values do not properly account for [Fe/H], so we use the updated numbers from \citet{Mann2015b}. 

\begin{figure}
 \includegraphics[width=84mm]{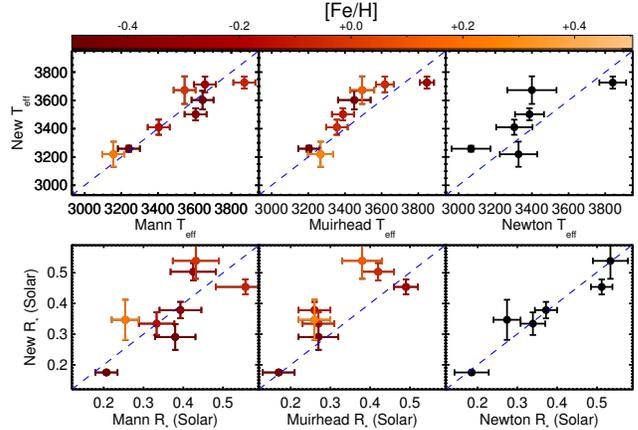}
 \caption{Comparison between \teff\ and $R_\star$ derived using the parallax (Section~\ref{sec:params}) and those derived using NIR or optical spectroscopy from \citet{Mann2013b, Mann2013c} (left), \citet{Muirhead2014} (middle), and \citet{Newton2015A} (right). Points are color-coded by [Fe/H] with the values from the literature source. \citet{Newton2015A} does not provide independent [Fe/H] measurements so the right panels are left uncolored. } 
  \label{fig:parameters}
 \end{figure}

We show the resulting comparison in Figure~\ref{fig:parameters}. \citet{Muirhead2014} measure systematically smaller $R_\star$ and \teff\ values, with reduced $\chi^2$ ($\chi^2_{\nu}$) values of 2 and 2.3, respectively. This is a consequence of known differences between models and empirical estimates of \teff\ and $R_\star$ \citep[e.g.,][]{Boyajian2012, 2014MNRAS.437.2831Z}. \citet{Mann2013c} \teff\ values are the most consistent ($\chi_\nu^2=1.0$), with similarly consistent radii ($\chi^2_{\nu}$=1.2). \citet{Newton:2014} \teff\ values show a large scatter ($\chi^2_{\nu}$=2.3) when compared to our own, but achieve highly consistent $R_\star$ values ($\chi^2_{\nu}$=0.6), likely an advantage of skipping the intermediate [Fe/H] and \teff\ step. Stronger agreement between our values and \citet{Newton:2014} and \citet{Mann2013c} over \citet{Muirhead2014} is somewhat expected, as the former two both anchor their results to the same set of stars with radii from interferometry. 

\citet{Dressing2013}, \citet{2014ApJS..211....2H}, and \citet{Gaidos2016a} also provide revised stellar parameters for the \kepler\ M~dwarfs. All three of these papers are based on photometry and proper motions, and thus are generally less precise than the spectroscopic methods we compare to above, but cover a larger number of stars. More importantly, each of these papers calibrate against or adopt the stellar parameters from \citet{Mann2013b, Mann2013c}, \citet{Muirhead2014} and \citet{Newton2015A}, so a comparison would not be independent. A detailed analysis of these assignments would be instructive to test how well the field sample is characterized without spectroscopic follow-up, but will require Gaia parallaxes to provide a large enough sample. 

\section{Transit Fits}\label{sec:transit} 
\subsection{Methodology}
We downloaded Q1-17 light curves (DR25) for the eight KOIs of interest from the Barbara A. Mikulski Archive for Space Telescopes (MAST). We used the Pre-search Data Conditioning \citep[PDCSAP,][]{Jenkins:2010qy,Stumpe2012,2012PASP..124..963K} light curves, excluding Kepler-1319, for which we used the Simple Aperture Photometry (SAP) data (see below). We download short-cadence data in place of long-cadence where available. For one system (Kepler-445) the KIC incorrectly contains an additional, equal-brightness star within the \kepler{} aperture, which affects the measured transit depth and parameters \citep{2017AJ....153...59D}. However, this issue was corrected in the most recent data release. 

We removed stellar and instrumental variations from the light curves using a running median with a width of 10 hours, masking out the transits and outlier points, and with a cutoff at each gap in the data (between quarters). Outliers were identified as points $>6\sigma$ outside the running median. Less than 1\% of points were removed this way, and visual inspection suggests most follow the expected shape for a flare (a sharp increase followed by a slow drop-off in flux). Once outliers are identified and masked, the running median is re-run to identify and remove longer-term stellar variability. To identify transit locations we pulled values for orbital period ($P$), transit midpoint ($T_0$), and transit duration from \citet{Swift2015}. Once identified, we fit a third-order polynomial to the running median, covering three on hours on either side of the transit, to fit out stellar variations during transit. We tested higher-order polynomial fits and longer/shorter out-of-transit windows, and found that this performed best in terms of smallest scatter in the bottom of transit points for the majority of systems. A more robust method would be to include parameters that describe the out-of-transit fit as part of the MCMC analysis, as done in \citep{Gazak:2012vn}, although tests with different fits to the out-of-transit data suggest this effect is negligible, and hence can be ignored. After fitting out stellar variability and masking outliers we fed the final flattened light curve to our transit fitting software. 

Transiting timing variations (TTVs), if left uncorrected, spread out the observed ingress/egress of a transit light curve when phase-folded assuming a linear ephemeris. Depending on the size and shape of the TTV signal, this will typically result in an erroneously small transit depth and large duration. \citet{Swift2015} searched for TTVs in all the \kepler\ M dwarf planet candidates, and for the sample considered here, identified significant TTVs only in Kepler-138bcd. This is consistent with earlier analyses of the same systems \citep[e.g.,][]{2013ApJ...772...74W,2013ApJS..208...16M,2014ApJ...784...28K}. Because transit timing constraints from \kepler\ photometry are typically quite precise ($\lesssim$minutes) it is unlikely derived parameters of other systems will be impacted by TTVs below the detection threshold. 

The Kepler-138 planets have TTV amplitudes ranging from 5m for planet c, to 50m for planet b \citep{2015Natur.522..321J}. To correct for this, we use the transit midpoints derived from the $N$-body simulations of \citet{2016arXiv161103516H} both when removing stellar variability and fitting the transit. This does not account for uncertainties in the transit times. A more robust method would be to treat individual transit times as free parameters, or include $N$-body simulations inside the transit-fitting MCMC. However, these options are computationally expensive, and unlikely to significantly improve the results given the relatively small errors on individual transit times.

A more detailed description of our light curve fitting method is given in \citet{Mann2016a} and \citet{Mann:2017aa}, which we summarize here. We fit each system using a Monte Carlo Markov Chain (MCMC) with the \textit{emcee} python module \citep{Foreman-Mackey2013} and the \textit{batman} tool \citep{Kreidberg2015}, based on the \citet{MandelAgol2002} transit model. To account for morphological effects of integration times \citep{Kipping:2010lr}, models for long-cadence data were binned and resampled to match the observations. Free parameters for each fit were planet-to-star radius ratio ($R_P/R_\star$), impact parameter ($b$), $P$, $T_0$, two parameters that describe the eccentricity and argument of periastron ($\sqrt{e}\sin(\omega)$ and $\sqrt{e}\cos(\omega)$), linear and quadratic limb-darkening (\citet[$q1$, $q2$][]{Kipping2013}), and stellar density ($\rho_\star$). Planets orbiting a common star were fit together with common limb-darkening and $\rho_\star$ parameters. MCMC chains were run with 100 walkers for systems with a single transiting planet, and 200 walkers for those with multiple transiting planets; each chain contains 100,000 steps including a burn-in phase of 10,000 steps. Examination of the posteriors suggests that this was more than sufficient for each chain to converge. 

We applied a prior drawn from the model-derived limb-darkening coefficients using the LDTK toolkit \citep{2015MNRAS.453.3821P} using the stellar parameters and errors derived in Section~\ref{sec:params}. Based on the variation in limb-darkening parameters drawn from different model grids, we add an additional error of 0.05 (in quadrature) on all limb-darkening coefficients. We report the adopted limb-darkening priors in Table~\ref{tab:limb}. We used the triangular sampling method of \citet{Kipping2013} in order to uniformly sample limb-darkening over the physically allowed parameter space. We uniformly sampled $|b|<1+R_P/R_\star$ (allows grazing transits), $P>0$ (no upper bound), $T_0$ over one orbital period, $\sqrt{e}\sin(\omega)$ and $\sqrt{e}\cos(\omega)$ from -1 to 1, and $R_P/R_\star$ from 0 to 0.5. We employed a Gaussian prior on $\rho_\star$ using our $R_*$ and $M_*$ values derived in Section~\ref{sec:params}, and $\rho_\star$ was not allowed to explore below zero, but had no upper bound.

\begin{table}
\caption{Adopted Limb-Darkening Priors  \label{tab:limb}}
\begin{center}
\begin{tabular}{l c c }
KOI &  $\mu_1$ & $\mu_2$ \\
\hline
 314 & 0.35$\pm$0.05  & 036$\pm$0.05 \\
 463 & 0.32$\pm$0.06 & 0.38$\pm$0.06 \\
 961 & 0.33$\pm$0.05  & 0.41$\pm$0.05 \\
1702 & 0.36$\pm$0.07 &  0.35$\pm$0.08  \\
1725 &  0.32$\pm$0.06 & 0.37$\pm$0.07 \\
2453 & 0.31$\pm$0.06 & 0.34$\pm$0.07   \\
2704 &  0.43$\pm$0.07 & 0.31$\pm$0.06  \\
2705A &  0.37$\pm$0.07 &  0.35$\pm$0.08  \\
2705B &  0.40$\pm$0.07 & 0.32$\pm$0.07   \\
\hline
\end{tabular}
\end{center}
\end{table}

If a companion star lands within the \kepler\ aperture, the additional light will dilute the transit signal, altering both the transit depth and shape \citep[e.g.,][]{Johnson:2012fk,2015ApJ...805...16C}. Component stars of Kepler-560AB, and 2453AB are sufficiently separated ($\simeq$40\arcsec\ and 11\arcsec), that the \kepler\ apertures do not include the companions. The aperture for Kepler-1651 does include its 4\arcsec\ companion. However, both targets resolved in the \kepler\ input catalog \citep{Brown2011} and \kepler\ pixel-level data, so the contaminating flux has been removed as part of generating the PDCSAP light curve; we extract a consistent light curve when attempting to correct for contaminating flux ourselves. The PDCSAP flux also accounts for changes in the contamination between quarters as varying levels of light from the companion falls outside the aperture. So we do not attempt any further corrections. 

For Kepler-1319 the companion is tight (1\farcs8) enough to land within the \kepler\ aperture, and a comparison of SAP (which contains no contamination correction) and PDCSAP light curves suggests the companion flux has not been completely removed in the PDCSAP data.  We first re-extracted the light curve for this star using the SAP flux, and applied relevant corrections to reproduce PDCSAP flux (e.g., co-trending basis vectors) using the \textit{PyKe} software \citep{Still2012}, but without removing any contaminating flux. We then fit the transit as above, but with an additional free parameter, $\Delta K_P$, which describes flux contamination from the non-host star, following the method from \citet{2011ApJ...730...79J}. We apply a Gaussian prior on $\Delta K_p$ using the estimate from \citet{Gaidos2016b}, derived from the AO data taken in \citet{Kraus2016a}. 

The small separation of Kepler-1319AB compared to the \kepler\ pixel size ($\simeq4\arcsec$) also makes it unclear around which star the planet orbits. To handle this, we fit the transit light curve twice, each time assuming a different host star (adjusting limb-darkening, $\rho_\star$, and $\Delta K_p$ priors). We briefly discuss the parentage of this planet further in Section~\ref{sec:systems}.

\subsection{Transit Parameters}\label{sec:transitparams}
We report the resulting $T_0$, $P$, $R_P/R_\star$, $|b|$, $\sqrt{e}\sin(\omega)$, $\sqrt{e}\cos(\omega)$, and $\rho_\star$ from of our transit fits in Table~\ref{tab:planets}. For each parameter we report the median value with the 84.1 and 15.9 percentile values (corresponding to $\pm1\sigma$ for Gaussian distributions). Distributions of $\sqrt{e}\sin(\omega)$, $\sqrt{e}\cos(\omega)$ are highly correlated with each other, so the simple median and 1$\sigma$ reporting can be highly misleading, and also less physically useful than $e$. We calculate the $e$ posterior by adding the square of  $\sqrt{e}\sin(\omega)$, $\sqrt{e}\cos(\omega)$ for each step in the chain. However, the median of $e$ is positively biased due to the cutoff at zero. Thus we instead report the mode (and 84.1 and 15.9 percentile) for $e$. We combine our $R_P/R_\star$ posteriors with $R_\star$ posteriors from Section~\ref{sec:params} to calculate $R_P$, which we report in Table~\ref{tab:planets}. To highlight correlations between parameters, we show distributions of $R_P/R_\star$, $|b|$, $e$, and $\rho_\star$ and mutual correlations for each fit in Figure~\ref{fig:transitfit}. 

Resulting $R_P$ values are generally extremely precise. Of the 13 planets, 7 have radii constrained to better than 10\%, 3--5\% or better, and all 13 to better than 20\%. Previous determinations were rarely better than 10\% because of large errors on $R_\star$. For Kepler-42bcd, which has both the most precise parallax ($\simeq4\%$) and some of the smallest planets (0.67-0.76$R_\earth$), the radii are constrained to $\simeq0.03R_\earth$, or $\simeq$200\,km. At this level, errors that are not considered (e.g., imperfect removal of stellar variability, incomplete removal of faint background stars in the field of view) may become important and should be considered. 

Our values are also generally more precise and accurate than previous transit fits on the same targets \citep[e.g.,][]{Swift2015,2015arXiv150400707R} due to additional constraints on limb-darkening, $\rho_\star$, simultaneous fit of short- and long-cadence data, and overall improvements in \kepler\ light curve reduction. Other than for Kepler-445 and Kepler-1319, our transit fit parameters are consistent with those from \citet{Swift2015} (e.g., $R_P/R_\star$ values within 2$\sigma$), who also focus on M~dwarf KOIs. The two discrepant systems are due to differences in the input light curves. The light curve available to \citet{Swift2015} for Kepler-445 had flux incorrectly subtracted out from a star in the Kepler Input Catalog that does not exist \citep[see][for further details]{2017AJ....153...59D}. Thus, our resulting $R_P/R_\star$ values are smaller by $>3\sigma$ for all three planets. \citet{Swift2015} did not know Kepler-1319AB is a binary, and hence could not account for contaminating flux; thus our resulting $R_P/R_\star$ is significantly larger. However, our parameters for Kepler-1319AB b agree well with those from \citet{Gaidos2016b}, who did account for the companion flux.

\subsection{Planetary Eccentricities}
\begin{figure}
 \includegraphics[width=84mm]{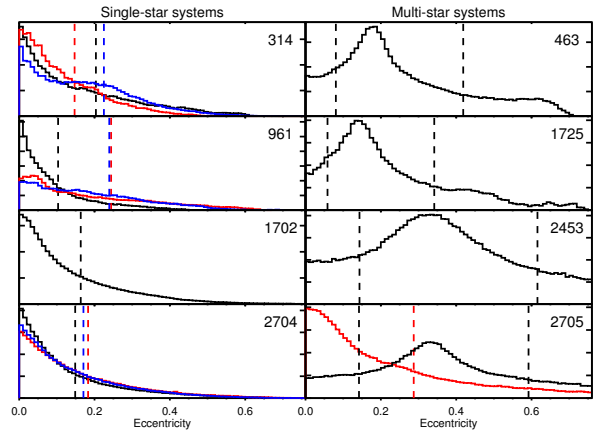}
 \caption{Eccentricity posteriors for planets in single-star systems (left column) and binaries (right column). Dashed lines mark the 68.2th percentile of points around the statistical mode. For all of the single-star systems this includes the lowest $e$ bin, so a single line (68.2\% upper limits) is indicated instead. Colors are applied to the histograms in multi-planet systems to distinguish the distributions; for Kepler-1319 (KOI-2705) the black distribution is if the planet orbits the primary, while the red is assuming it orbits the companion.} 
  \label{fig:eccen}
 \end{figure}

The eccentricity posteriors appear to fall into two qualitative groups; 10 planets in 4 systems all have posteriors peaked at $e\simeq$0, while the other 4 planets in four systems peak at larger values (0.15$<e<$0.4). While, $e=0$ cannot be ruled out at 3$\sigma$ for any planet, broad tails are not unexpected given measurement uncertainties and degeneracy with $\omega$ \citep[see][for some discussion of this]{Van-Eylen2015,Mann:2017aa}. Interestingly, the low-eccentricity planets are all in single-star systems, while the latter group all reside in binary systems, a trend which was also noted by \citet{2017arXiv170105664M}. We highlight the eccentricity differences in Figure~\ref{fig:eccen}. 

To test the significance of any difference between single- and multi-transiting systems, we compared our eccentricity posteriors to two model eccentricity distributions; (1) a single Rayleigh distribution with $\sigma$=0.05 \citep[as found by][]{Van-Eylen2015}, and (2) a double-Rayleigh distribution with $\sigma_{1,2}$=0.05, 0.25. We then compared each model to the results of our transit fits, treating the eccentricity posteriors as likelihoods. For (2), the $\sigma_1$ eccentricity distribution is only applied to the planets in single-star systems, and $\sigma_2$ to those in multi-star systems. For this test, we consider the primary of Kepler-1319 to be the parent star. We found a Bayes factor (the ratio of the likelihoods for the two models) of 11, strongly favoring the two-distribution model. However, if we instead allow $\sigma$ to float in (1), we find that the two-distribution model is only weakly favored (Bayes factor = 2.7) over a single-distribution model (with best-fit $\sigma=0.083$). Allowing both $\sigma_1$ and $\sigma_2$ to float only marginally changes the comparison (Bayes factor = 3.3). This is highly suggestive, but not definitive at this time, especially considering data-quality differences between the two samples, which we explore below. 

\begin{figure}
 \includegraphics[width=84mm]{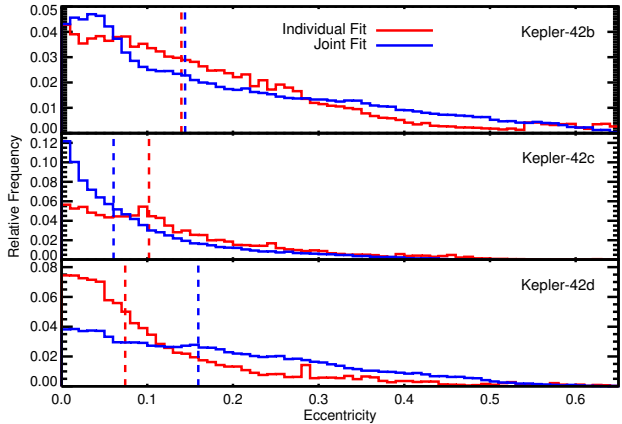}
 \caption{Comparison of eccentricity posteriors for Kepler-42bcd when fitting each planet individually (blue) or fitting simultaneously, forcing common limb-darkening parameters and stellar density (red). Bin sizes and locations are identical. The median of each distribution is shown with a vertical dashed line with the appropriate color. }
  \label{fig:e_comp}
 \end{figure}

 The apparent difference in eccentricity distributions for single-star vs. multi-star systems could also be due to differences in data-quality rather than astrophysical ones. Three out of four of the single-star systems are also multi-planet systems, which provides additional constraints on the stellar density compared to systems with a single transiting planet. To test this we re-run the transit fit for each of the Kepler-42 planets individually, without forcing a common stellar density or limb-darkening. We compare the resulting eccentricity posteriors to those from the common-star fit in Figure~\ref{fig:e_comp}. The differences are marginal, and do not preferentially favor larger or smaller eccentricities. All differences are significantly smaller than the difference between single-star and multi-star system eccentricity posteriors. This may be because, for most systems considered here, eccentricity constraints are limited more by the width of the input density prior and degeneracy with $\omega$ than by information derived directly from the light curve.

 Another complicating factor is varying quantities of short-cadence photometry between single- and multi-planet systems. As detailed in \citet{Van-Eylen2015}, changing $e$ and $b$ can have similar effects on the shape of the transit ingress/egress, especially for small, close-in planets like those studied here. It is therefore difficult to resolve this degeneracy with long-cadence data, as the effective integration times for long-cadence ($\simeq$30m) data are longer than the typical ingress/egress time (1-10m). For reference, we report the number of days of short- and long-cadence data for each system in Table~\ref{tab:data}. 

\begin{table}
\caption{Summary of Light Curve Data  \label{tab:data}}
\begin{center}
\begin{tabular}{l r r }
KOI & Long-Cadence & Short-Cadence \\
& \# of days & \# of days \\
\hline
 314 &  373 &  908\\
 463 & 1124 &  175\\
 961 &  591 &  685\\
1702 & 1272 &    0\\
1725 &  317 &  243\\
2453 & 1303 &    0\\
2704 &  561 &   24\\
2705 &  763 &    0\\
\hline
\end{tabular}
\end{center}
\end{table}

\begin{figure}
 \includegraphics[width=84mm]{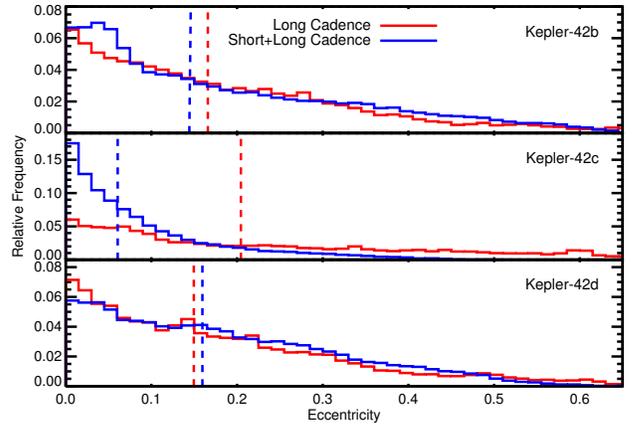}
 \caption{ Comparison of eccentricity posteriors for Kepler-42bcd using just long-cadence data (red) or all (long and short-cadence) data (blue). Bin sizes and locations are identical. The median of each distribution is shown with a vertical dashed line with the appropriate color. }
  \label{fig:e_comp2}
 \end{figure}

 The correlation between $b$ and $e$ can be seen our transit-fit posteriors (Figure~\ref{fig:transitfit}). As expected, the two systems with the most short-cadence data (Kepler-42 and Kepler-138) show little or no correlation between these parameters. However, this cannot fully explain the differences between multi-star and single-star systems. Kepler-1650 has no short-cadence data, and Kepler-445 has only 24 days of photometry at short-cadence, yet both light curves yield relatively tight eccentricity posteriors favoring $e\simeq$0. Similarly, Kepler-560 and 1651 both have significant amounts of short-cadence data, yet show broader $e$ posteriors favoring higher values. We further tested this by fitting Kepler-42bcd transit light curves using only long-cadence data (including \kepler\ binned short-cadence data). We compare the resulting eccentricity distribution in Figure~\ref{fig:e_comp2}. As with the comparison of fitting multi-planet systems individually or jointly, using only long-cadence data does broaden the posteriors, particularly for Kepler-42c. However, the overall effect is small compared to the difference between the eccentricity posteriors of single-star and multi-star systems. We likely benefit from the short period of the exoplanets considered; over many transits \kepler\ photometry will sample different fractions of the ingress/egress, providing some constraints on the transit shape even if individual long-cadence points provide little to none. 

Our fits do not take into account the fact that eccentric planets are more likely to transit, which drives the observed eccentricity distribution of transiting planets to larger values than the true distribution \citep{2014MNRAS.444.2263K,2014MNRAS.440.2164K}. We test this effect by re-running \kepler-445 and \kepler-560 (one single-star system and one in a binary), including approximate expressions for the priors on $\omega$ and $e$ from \citet{2014MNRAS.444.2263K}. The resulting posteriors on $e$ did not change significantly (statistical mode changed by $<0.02$), but the net result was a marginally larger discrepancy between the $e$ values for planets in the single compared to the binary system. This suggests that the effect is negligible for our analysis, given the small number of systems. However, these changes could become significant for analyses on larger samples, and hence should be taken into account for future studies.

\section{False-positive Analysis}\label{sec:fpp}
Five of the eight systems investigated here are already confirmed planets, while the other three are still considered planet candidates (Kepler-1651, 1652, and KOI-2453). However, the more precise stellar and transit parameters, and the addition of recently published adaptive optics data significantly improves the false-positive probability (FPP) calculations. 

We determine the FPP for these three systems using the \textit{VESPA} software \citep{2012ApJ...761....6M, 2015ascl.soft03011M}. To briefly summarize, \textit{VESPA} calculates the FPP by comparing the likelihood of astrophysical false positives (background eclipsing binaries, bound eclipsing binaries, and hierarchical eclipsing systems) to that of a planet, using constraints from the transit shape and depth, properties of the star, and external constraints from high-resolution imaging (where available). We included stellar, transit-fit parameters, and flattened light curves as described in Sections~\ref{sec:params} and \ref{sec:transit}. For the imaging, we added in results from \citet{Kraus2016a}, which include contrast curves from adaptive optics imaging and non-redundant masking for all three systems considered here.

The resulting FPP probabilities are $8.7\times10^{-6}$, $4.8\times10^{-6}$, and $3.0\times10^{-2}$ for Kepler-1650, 1651, and KOI-2453, respectively. This confirms the planetary nature for Kepler-1650 and 1651. The light curve of KOI-2453 has a short duration, which is only marginally resolved in the long-cadence data. Hence it is possible to fit the curve with a V-shaped eclipse indicative of an eclipsing binary. The planet hypothesis is still  favored (97\% to 3\%), as it provides a much better fit to the light curve, but this does not meet the threshold for validation, and the system remains a planet candidate. 

 Nearby companions to KOI-2453 and Kepler-1651 do not affect these calculations. KOI-2453B falls outside the \kepler\ aperture and cannot affect the transit signal. Kepler-1651B falls within the aperture; however, the separation is large enough to rule out the companion using the difference image \citep[flux out-of-transit $-$ flux in-transit][]{2010ApJ...713L.103B}. The centroid of a difference image is a measure of the location of the transit. For Kepler-1651AB the difference image centroid matches the primary location to within 0\farcs5 for all quarters (comparable to position and proper motion errors), but agreement with the companion is no better than 3\farcs5 for the entire \kepler\ mission\footnote{\href{http://exoplanetarchive.ipac.caltech.edu//data/KeplerData/010/010905/010905746/dv/kplr010905746-20160209194854_dvr.pdf}{Kepler-1651 Data Validation report.}}. This confirms that the transit signal comes from the primary for both systems, and that companions can be ignored for FPP calculations.

\section{Discussion of individual systems}\label{sec:systems}
\textbf{Kepler-138 (KOI-314)} is a bright ($K_P$=12.9) star with three small planets, which has made it the subject of significant follow-up and analysis \citep[e.g.,][]{Pineda:2013fk,2014ApJ...784...28K}. The planets show significant transit timing variations \citep{Swift2015}, making it possible to constrain the planet masses \citep[e.g.,][]{2015Natur.522..321J,2016arXiv161103516H}. Our derived $\rho_\star$ is $\simeq30\%$ smaller than the value in \citet{2015Natur.522..321J}, although the difference is not significant given the large errors (23\%) determined by \citet{2015Natur.522..321J}. Our revised $\rho_\star$ provides a significantly tighter constraint on the eccentricity of these planets, suggesting values close to zero, which, when combined with our improved stellar parameters should provide more precise constraints on the TTV-based masses of the three planets. 

\textbf{Kepler-560 (KOI-463)} is a wide companion (40\arcsec, 4000\,au) to KIC 8845251 with a single transiting planet. It is the second-largest planet in the sample (third if Kepler-1319 orbits the fainter component) at 1.93$\,R_\earth$. The transit duration suggests a non-zero $e$ at $\simeq2\sigma$, allowing e$\simeq$0 only if $|b|\gtrsim0.6$. This degeneracy between $b$ and $e$ can be mitigated significantly with higher-cadence data, but Kepler-560 has only two quarters of short-cadence observations with \kepler. The star is too faint ($V\simeq16$) and the transit too shallow ($\simeq$3\,mmag) for ground-based follow-up at sufficiently high cadence.

\textbf{Kepler-42 (KOI-961)} is a cool (\teff=3269$\pm19$\,K) star with three sub-Earth size planets, discussed extensively in \citet{Muirhead2012}. At just 40\,pc, this is both the closest star in our sample, and one of the closest planets in the \kepler\ sample as a whole. The parallax-based stellar parameters for this system agree well with those from \citet{Muirhead2012}, who used Gliese 699 (Barnard's Star) as a proxy to derive $R_\star$ and hence $R_P$ for the three planets. The agreement between the two results is a strong validation of this method of stellar characterization by proxy, often used for cool dwarfs that lack trigonometric parallaxes \citep[e.g.,][]{Ballard:2013,Muirhead2015}. We discuss this system in more detail in Section~\ref{sec:42}, where we outline how Kepler-42 can be used to measure precise stellar parameters of late-type stars. 

\textbf{Kepler-1651 (KOI-1702)} is the only single-transiting system in our sample with an eccentricity posterior indicating $e\simeq0$ (statistical mode = 0). Like the other low $e$ systems, it has no stellar companion. Kepler-1650 does not undergo significant TTVs indicative of additional, non-transiting planets \citep{Swift2015}, although this only rules out a narrow range of parameter space (e.g., planets near orbital resonances). This could be a dynamically cool multi-planet system where the other planets happen not to transit. More distant planets could easily be missed even in systems with $\simeq0$ mutual inclination depending on our line of sight and the orbital architecture. 

\textbf{Kepler-1651 (KOI-1725)} has a single transiting planet and a companion star at modest separation (4\arcsec, 280\,au). Like Kepler-560, the transit duration is significantly discrepant from the expected value for $e=0$ and $b=0$. Zero eccentricity is not ruled out, but is only possible if $b>0.55$. At lower impact parameters, $e\simeq0.2$ is favored. Also like Kepler-560, the planet is one of the largest in the sample (1.84$R_\earth$). 

\textbf{KOI-2453} falls in the same category as Kepler-560 and 1651, with a wide companion (11\farcs7, 1500\,au), and a transit duration indicative of $e>0$. The target has no short-cadence data, and the transit duration is short ($\sim$40 minutes) compared to other systems in our sample. The resulting constraint on $e$ is poorer than other targets in the sample, with $e$ values as large as 0.8 allowed. Poor sampling is also the reason we cannot statistically confirm this is a planet; a poorly resolved transit looks similar to the eclipse of a tightly orbiting companion.

\textbf{Kepler-445 (KOI-2704)} is one of a set of `compact multiples' (a densely packed multi-planet system) as discussed in \citep{Muirhead2015}, the others being Kepler-42 and Kepler-446. The transit analysis presented in \citet{Muirhead2015} was based on a \kepler\ light curve with incorrect corrections for stellar contamination \citep[see][]{2017AJ....153...59D}. As a result, our planet radii are significantly smaller. However, Kepler-445bcd are still some of the largest planets in the sample considered here, possibly because the star is one of the most metal-rich ([Fe/H]$\simeq0.3$). As with Kepler-42, these planets may be useful as a test of late-type stellar parameters. The lack of TTVs, despite much larger (and likely more massive) planets than Kepler-42bcd, suggests low $e$ for all three systems, although external arguments (e.g., $N$-body simulations) could provide more quantitative constraints. 

\textbf{Kepler-1319 (KOI-2705AB)} has a single known transiting planet as well as a tight (1\farcs88, 205\,au) stellar companion. Because \kepler\ pixels are $\simeq$4\arcsec\ wide and each aperture is several pixels, it is difficult to confirm around which star the planet orbits. \citet{Gaidos2016b} argue that the planet likely orbits the companion, based on a comparison between the transit-fit and spectroscopic densities. However, \citet{Gaidos2016b} caution that this conclusion assumes $e\simeq0$, and that the planet could orbit the primary with $e\gtrsim$0.2, consistent with our own analysis. Centroid offsets between the in- and out-of-transit light curves can constrain the position to $<$1\arcsec\ \citep{Borucki2010, Bryson2013}. This is complicated by Kepler-1319AB's large proper motion ($\simeq140$mas\,year$^{-1}$), although the primary is marginally favored. If the planet orbits the companion, it corresponds to a planet radius of $2.9^{+0.5}_{-0.4}R_\earth$. Planets with $R_P>2R_\earth$ are rare around stars with $M_\star<0.3$ (while Kepler-1319B is $0.17M_\star$), due to a trend of decreasing planet radius with stellar mass \citep{Gaidos2016b,Dressing2015}. The only known exceptions are GJ 1214b \citep{Charbonneau:2009rt} and K2-25b \citep{Mann2016a}. So while the primary star is more likely (and assumed for our analysis), the parentage of this planet remains unproven. 

\section{When eccentricity is known: testing stellar models with Kepler-42}\label{sec:42}
\citet{Muirhead2012} argue that $e\simeq0$ for all three planets orbiting Kepler-42, as the age of the star is much greater than the tidal circularization timescale for each of the planets. If $e=0$ for all three planets, Kepler-42 is a useful test of  evolutionary models, as there are only a few empirically well-characterized systems as cool (\teff=3269\,K) and metal-poor ([Fe/H]=-0.5) as Kepler-42. Because it is a multi-transiting system, the transit-derived $\rho_\star$ derived by locking $e$ to zero could rival the precision provided by similarly cool eclipsing binaries \citep[e.g.][]{Feiden2012a} and long-baseline interferometry \citep[e.g.][]{Boyajian2012, Kane2017}. Combined with our parallax, this system could provide constraints on the mass-luminosity relation \citep{Benedict:2016aa} or radius-luminosity relation \citep{Mann2015b} for mid-M~dwarfs.

Our transit-fit is consistent with $e=0$ for all three planets, but higher $e$ values are not ruled out. Further, the tidal circularization argument presented in \citet{Muirhead2012} does not account for eccentricity maintained by planet-planet interactions (forced eccentricity). As an additional test, we ran a series of $N$-body simulations of the system using the \textit{Mercury 6} software \citep{1999MNRAS.304..793C} with the included Bulirsch--Stoer integrator. The goals were twofold: (1) to test the expected TTV signal assuming a non-zero $e$ for any planets, and (2) to see if the forced eccentricity values are significantly non-zero. We assumed a stellar mass from our stellar parameters derived in Section~\ref{sec:params}, and orbital parameters from our transit fits (Section~\ref{sec:transit}) with the exception of $e$ and $\omega$. The masses of the three planets were fixed at values derived using our radius values and the mass-radius relation from \citet{2014ApJ...783L...6W}. For each simulation we assign $e$ values to the individual planets from 0 to 0.2 in increments of 0.05. For each unique set of $e$ values we run 10 different simulations with randomly assigned $\omega$ values for all planets. Simulations are run for 500 orbits, except for the cases where $e=0$ initially for all planets, which we run for 5000 orbits. 

\begin{figure}
 \includegraphics[width=84mm]{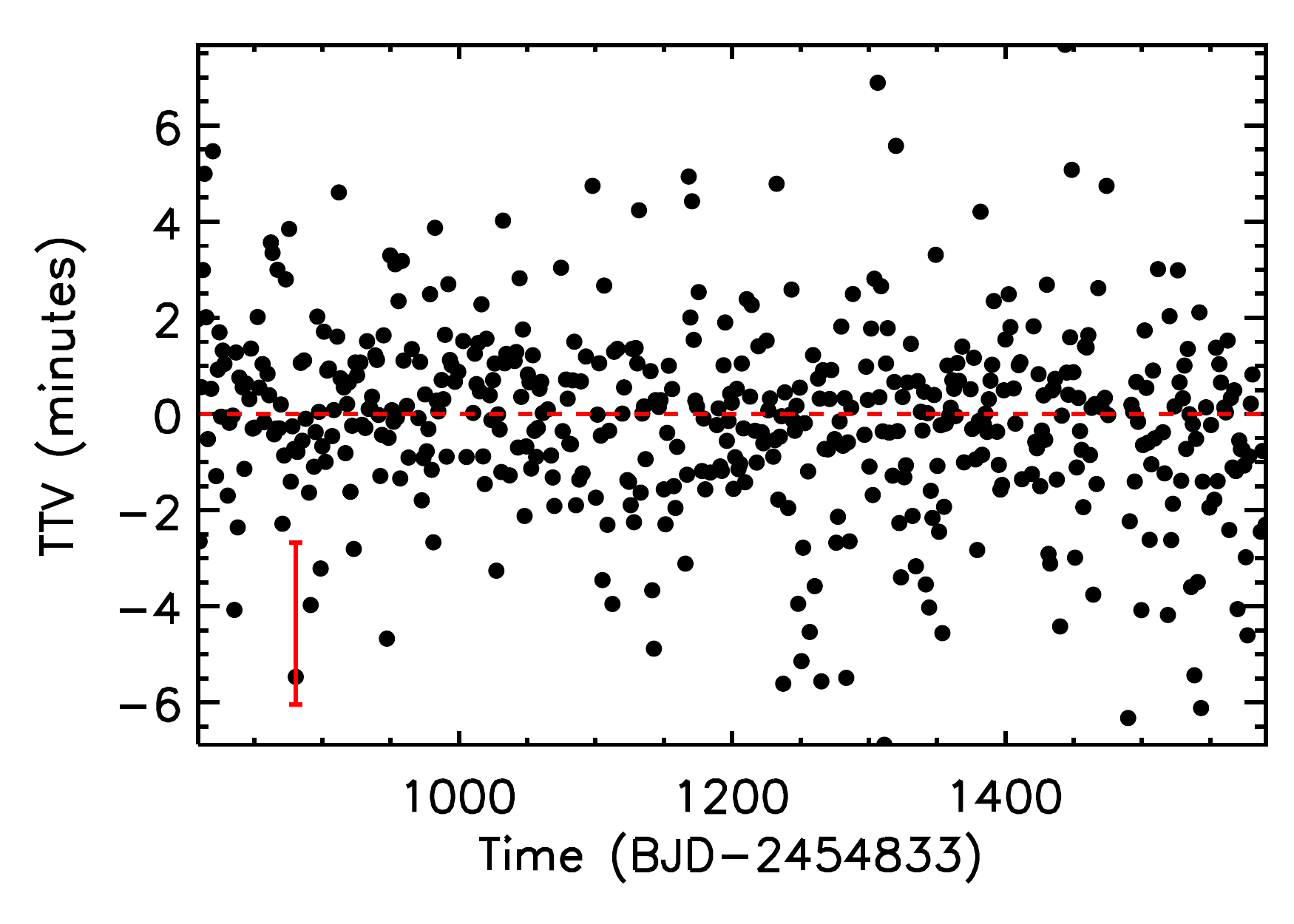}
 \caption{ Transit timing variations for Kepler-42b measured from short-cadence photometry. Partial transits and those with flares or other data artifacts were manually removed. A typical error on the transit time is shown in red. A linear ephemeris is shown as a red dashed line for reference. Measured times are taken from \citet{Swift2015}, and calculated times from our updated fits. }
  \label{fig:ttv}
 \end{figure}

For our long run starting at $e=0$, low-eccentricity values are maintained throughout the simulation, suggesting forced eccentricity values of $<0.01$ for all planets, assuming tidal dampening removed any initial eccentricity. The shorter runs with non-zero eccentricities also suggest low eccentricities for all planets. Following \citet{Mann:2010fj}, we used the simulation output to measure any deviation from the expected transit time. All simulations with $e\ge0.1$ produce TTV amplitudes $>3$~minutes, which is ruled out observationally by \citet{Swift2015}. We show an example TTV measurement for Kepler-42b in Figure~\ref{fig:ttv}. Combinations of $e=0.05$ can create TTVs below detection in the \kepler\ data, and hence cannot be ruled out from the lack of TTVs alone. We conclude that the $e$ of all planets must be $<0.1$, and $\lesssim0.01$ if our assumptions about tidal dampening are valid. 

Motivated by the tidal dissipation arguments from \citet{Muirhead2012} as well as our $N$-body simulations and the lack of TTVs, we refit the transits of Kepler-42, this time fixing $\sqrt{e}\sin(\omega)$ and $\sqrt{e}\cos(\omega)$ to zero, and with a uniform prior on $\rho_\star$. Otherwise this fit is unchanged from what is described in Section~\ref{sec:transit}. We compare the output $\rho_\star$ value from this fit to estimates from the  $M_\star-M_{K_S}$ relation of \citet{Benedict:2016aa}, the $R_\star-M_{K_S}$ relation from \citet{Mann2015b} (see Section~\ref{sec:params}), and the value predicted by the DSEP models. For the DSEP value, we consider models with [$\alpha$/Fe] = 0 and age $>$ 3 Gyr \citep{Muirhead2012}. We then interpolate (bilinear) our [Fe/H] and $L_\star$ posteriors (Section~\ref{sec:params}) for Kepler-42 onto the model grid, yielding a corresponding posterior on $\rho_\star$. Exploring model ages down to 1 Gyr and [$\alpha$/Fe] within 0.2 dex of solar do not significantly change the results, nor does using higher-order interpolation.  

\begin{figure}
\includegraphics[width=0.45\textwidth]{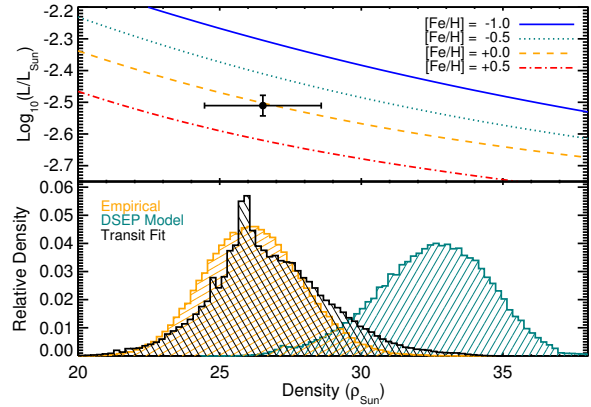}
\caption{Top: DSEP model tracks of $L_\star$ vs. $\rho_\star$ for different metallicities, with the parallax-based $L_\star$, transit-based $\rho_\star$, and corresponding errors for Kepler-42 shown as a black point. While the more empirical estimate matches the [Fe/H]=0 isochrone, Kepler-42 is metal-poor ([Fe/H]=$-$0.5). Models are restricted to $>$3\,Gyr and [$\alpha$/Fe]=0. Bottom: posteriors for $\rho_\star$ from the transit-fit (black), empirical calibration from nearby stars (orange), and DSEP model (teal, matching [Fe/H]=-0.5 on top panel). }
 \label{fig:961_model}
\end{figure}

We show the DSEP $\rho_\star-L_\star$ sequence for different metallicities alongside the $L_\star$ and transit-fit $\rho_\star$ for Kepler-42 and our empirically calibrated values in Figure~\ref{fig:961_model}. The transit-fit $\rho_\star$ agrees remarkably well with the more empirical methods results despite no external constraint on $\rho_\star$. The DSEP values, however, yield a $\rho_\star$ $\simeq$22\% higher than that derived from the transit. Examination of the model interpolation indicates that this discrepancy is entirely due to a model radius $\simeq$6\% smaller than empirical measurements for this luminosity (the predicted mass is consistent to 0.5$\sigma$), or a model \teff\ too hot by $\simeq$2\%. This is consistent with earlier studies comparing M~dwarf evolutionary models to empirical determinations of mass, luminosity, and radius \citep[e.g.,][]{Boyajian2012,2014MNRAS.437.2831Z}. However, errors in both the transit-fit and model $\rho_\star$ are $\simeq$10\%, so the resulting difference is only 2.1$\sigma$.

\section{Summary \& Discussion}\label{sec:discussion} 
As part of the Hawaii Infrared Parallax Program, we measured trigonometric parallaxes and proper motions for eight M~dwarfs with transiting planets discovered by the \kepler\ mission. We used this astrometry to identify companion (primary for Kepler-560) stars to four of the planet hosts. We also used our parallaxes to derive updated stellar parameters for all eight targets and the four associated stars. We compared our parallax-based stellar parameters to those previously derived using moderate-resolution spectroscopy. Our stellar parameters are generally more precise and accurate than spectroscopic values, in part because of lower sensitivity to [Fe/H] when the distance is known. We found the best agreement between our parallax-based values with $R_\star$ from \citet{Newton2015A} and \teff\ from \citet{Mann2013c}, although reasonable agreement is found for all three studies to which we compared \citep{Mann2013c,Muirhead2014,Newton2015A} given known errors and systematics in the methods used.

We refit the transits of each of the eight planetary systems using the full mission of \kepler\ data (short- and long-cadence). While many of these systems have been investigated previously \citep[e.g.,][]{Swift2015,Muirhead2015}, this is the first to directly investigate the eccentricity of these planets using the transit light curve and our derived $\rho_\star$. Because stellar parameters are generally poorly constrained for M~dwarfs, this has only been done previously on a few extremely well-studied, nearby M~dwarfs \citep[e.g.,][]{2013A&A...551A..48A,2014MNRAS.443.1810B}. Our revised transit fits are generally consistent with, but more precise than, previous determinations. The resulting $e$ constraints vary based on the quality of the light curve; in particular, multiple quarters of short-cadence data significantly improved the constraints. The best constrained systems (Kepler-138, 42) have $e$ values known to $\pm$0.1, which is rare to achieve without radial velocity constraints. 

Five of the eight systems are already considered confirmed planets. We ran the other three through the \textit{VESPA} software with the goal of confirming their planetary nature. We find FPP below 1\% for two of the three systems, effectively confirming their planetary nature. However, the analysis of KOI-2453 yielded a larger FPP (3\%) and hence remains a planet candidate. 

For Kepler-42bcd, we argue that $e$ $\simeq0$ based on the lack of TTVs, our own simulations, and tidal dissipation arguments presented in \citet{Muirhead2012}. This is consistent with our transit fits, which suggest low $e$ values for all three planets. Thus we refit the transits of Kepler-42bcd with $e$ fixed at zero, but replace our $\rho_\star$ prior with a uniform one. The resulting $\rho_\star$ accurately reproduces our assigned stellar parameters derived from empirical relations of nearby stars, but is 2.1$\sigma$ less dense than predicted by the DSEP models. We attribute this discrepancy to a larger radius (hence smaller density for a fixed mass) than predicted by the models, consistent with earlier findings. 

The Kepler-42 result is a useful case for testing stellar models because it is below the fully convective boundary \citep{Chabrier:1997}, metal-poor ([Fe/H]$\simeq-0.5$), and has a light curve of sufficient baseline and precision to place constraints on the spot coverage. This combination is a rare for stars with precise empirical constraints on their radii and/or masses. Disagreement between model and empirical radii of M~dwarfs has often been attributed to spots and magnetic activity, in particular to explain inflation seen in low-mass eclipsing binaries \citep[e.g.,][]{2001ApJ...559..353M, Kraus2011,Spada2013}. However, the presence of similar inflation in single, inactive stars, raises questions about this interpretation \citep[e.g.,][]{Spada2013,Mann2015b}. While still only a 2$\sigma$ discrepancy, the variability in Kepler-42's light curve is $<1\%$, suggesting that activity is not entirely responsible for the disagreement with models. Assuming a Gaia-precision parallax (with the value unchanged) this result will become $>3\sigma$, mainly because of a tighter model posterior, but also because of improvements in the limb-darkening prior to the transit fit. 

One suggestive trend in the resulting transit fits is that the four single stars (Kepler-138, 42, 445, and 1650) all have $e$ posteriors favoring $e\simeq0$ see Figure~\ref{fig:transitfit}), with significant tails at higher $e$ values. The stars with companions are reversed, with $e$ posteriors favoring non-zero values, but each exhibit a tail that makes $e=0$ impossible to rule out (see Figure~\ref{fig:transitfit}). The eccentricity distribution is best fit by two Rayleigh distributions with $\sigma=0.05$ for the single-star systems and $\sigma=0.25$ for the multi-star systems. This model is strongly favored (Bayes factor = 11) over a single distribution with $\sigma=0.05$, as found for small planets by \citet{Van-Eylen2015}. However, it is only weakly favored (Bayes factor = 2.7-3.3) over a single distribution with $\sigma=0.08$. Further complicating the matter, there are data-quality differences between the two samples (e.g., amount of short-cadence data, number of planets) that could slightly bias the results, and this depends on the assumption that Kepler-1319b orbits the primary star. So we consider the relation between planetary eccentricity and stellar multiplicity highly suggestive, but we require a larger sample to disentangle between these two scenarios conclusively.

It is possible that our ``single-star'' systems have undetected companions. However, for the distances of our targets, adaptive optics imaging available for all systems from \citet{Kraus2016a} rules out all but the faintest companions (brown dwarfs) beyond $\simeq$10\,au, and most companions down to $\sim3$\,au. Equal-mass companions are also ruled out at tight separations by the agreement between our distance-based parameters and spectroscopic parameters; an equal-mass companion would make the star appear brighter, but have a minor effect on the spectroscopy. Even low-mass tight companions are unlikely, as these are known to suppress the formation or survival of planetary systems \citep[e.g.,][]{2007ApJ...660..807Q,2012ApJ...753...91K,2014ApJ...791..111W,Kraus2016a}. We conclude that is is unlikely that missed companions are diluting the comparison.

If real, this trend would suggest that single-star systems are dynamically cooler than those in multi-star systems. Consistent with this, three of the four single-star systems are multi-planet systems, while the four stars with detected companions have only a single transiting planet detected. This could be due to a correlation between the number of planets in a system and their eccentricity \citep{2015PNAS..112...20L}, or a correlation between eccentricity and mutual inclination \citep[e.g.,][]{2016ApJ...822...54D, 2017arXiv170105664M}, such that additional planets in multi-star systems are present, but do not transit. 

A connection between $e$, binarity, and mutual inclination is also predicted by other theoretical and observational considerations. High-resolution imaging and radial velocity studies of the \kepler\ field have shown that tight ($<$100\,au) binaries suppress planet formation and/or survival \citep[e.g.,][]{2015ApJ...806..248W,Kraus2016a}. Even relatively wide binaries can have their orbits change due to interactions with passing stars, facilitating angular momentum exchange with the planetary system and driving planets to higher eccentricities and mutual inclinations or ejecting them from the system completely \citep[e.g.,][]{2006ApJ...644..543H, 2012ApJ...753...91K,Kaib:2013}. Binary companions also have a significant impact on the formation of evolution of protoplanetary disks \citep[e.g.,][]{2007ApJ...666..436H,2010ApJ...710..462A} and hence the final planetary system. Moderate- to wide-separation ($>$100\,au) binary companions could partially explain the overabundance of \kepler\ systems with only a single detected planet \citep[the \kepler\ dichotomy; ][]{Ballard2016,2016ApJ...832...34M}. Thus the connections between stellar multiplicity and planetary properties seen here, while not yet statistically significant, could be critical to understanding the evolution and diversity of planetary systems, and warrant study with a broader sample. 

Thanks to upcoming results from the Gaia and {\it TESS} missions, we will soon be able to expand the sample of planetary systems considered as well as improve the precision on individual systems. Gaia parallaxes will provide parallaxes with $\lesssim100\mu$as precision for all M~dwarfs observed by \kepler\ \citep{2012Ap&SS.341...31D}, including $\gtrsim$100 systems with known transiting planets. Using Kepler-42 as an example, if we assume that our parallax value is unchanged, but with Gaia-like errors ($100\mu$as), the difference between the model and empirical parameters will grow to 3.8$\sigma$. 

External constraints on $\rho_\star$ from the parallax will also significantly improve enabling tighter constraints on eccentricity from the transit fits. With Gaia parallaxes, stellar parameters will be limited primarily by the underlying $M_\star$-$M_{K_S}$ or $R_\star$-$M_{K_S}$ relations from which they are derived. Assuming no improvements in calibration, these generally give $M_\star$ and $R_*$ to 2-5\%, or $\rho_\star$ to 6-15\%. Assuming that the systems studied here are representative, the massive increase in sample size provided by Gaia will be more than sufficient to determine the overall eccentricity of small planets around M~dwarfs to better than 0.1. 

Similarly, the {\it TESS} mission will identify transiting planets statistically closer and brighter than those found by \kepler. {\it TESS} will also observe $>$200,00 stars at one-minute cadence\footnote{https://tess.gsfc.nasa.gov/overview.html}, including a large number of M~dwarfs. Such short-cadence data will be critical to accurately determine impact parameter and transit duration. Combined with Gaia proper motions to identify any faint stellar companions, this dataset will be ideal to study the interplay between eccentricity, planet multiplicity, and the presence of stellar companions. 

\acknowledgements
We thank the anonymous referee for their extremely thoughtful comments, which helped a great deal to make this a better manuscript. 

Based on observations obtained with WIRCam, a joint project of CFHT, Taiwan, Korea, Canada, France, and the Canada-France-Hawaii Telescope (CFHT) which is operated by the National Research Council (NRC) of Canada, the Institute National des Sciences de l'Univers of the Centre National de la Recherche Scientifique of France, and the University of Hawaii.  The observations at the Canada-France-Hawaii Telescope were performed with care and respect from the summit of Maunakea which is a significant cultural and historic site.  

Support for Program number HST-HF2-51364.001-A was provided by NASA through a grant from the Space Telescope Science Institute, which is operated by the Association of Universities for Research in Astronomy, Incorporated, under NASA contract NAS5-26555.

Some of the data presented in this paper were obtained from the Mikulski Archive for Space Telescopes (MAST). STScI is operated by the Association of Universities for Research in Astronomy, Inc., under NASA contract NAS5-26555. Support for MAST for non-HST data is provided by the NASA Office of Space Science via grant NNX09AF08G and by other grants and contracts.  This paper includes data collected by the Kepler mission. Funding for the Kepler mission is provided by the NASA Science Mission directorate. The authors acknowledge the Texas Advanced Computing Center (TACC) at The University of Texas at Austin for providing HPC resources that have contributed to the research results reported within this paper. URL: \href{http://www.tacc.utexas.edu}{http://www.tacc.utexas.edu}

\facility{CFHT (WIRCam)} \facility{Kepler}

\bibliography{$HOME/Dropbox/fullbiblio.bib}

\begin{thebibliography}{133}
\providecommand\natexlab[1]{#1}
\providecommand\JournalTitle[1]{#1}

\bibitem[{{Allard} {et~al.}(2012){Allard}, {Homeier}, \&
  {Freytag}}]{2012RSPTA.370.2765A}
{Allard}, F., {Homeier}, D., \& {Freytag}, B. 2012,
  \href{http://dx.doi.org/10.1098/rsta.2011.0269}{\JournalTitle{Royal Society
  of London Philosophical Transactions Series A}, 370, 2765}

\bibitem[{{Andrews} {et~al.}(2010){Andrews}, {Czekala}, {Wilner}, {Espaillat},
  {Dullemond}, \& {Hughes}}]{2010ApJ...710..462A}
{Andrews}, S.~M., {Czekala}, I., {Wilner}, D.~J., {et~al.} 2010,
  \href{http://dx.doi.org/10.1088/0004-637X/710/1/462}{\JournalTitle{\apj},
  710, 462}

\bibitem[{{Anglada-Escud{\'e}} {et~al.}(2013){Anglada-Escud{\'e}},
  {Rojas-Ayala}, {Boss}, {Weinberger}, \& {Lloyd}}]{2013A&A...551A..48A}
{Anglada-Escud{\'e}}, G., {Rojas-Ayala}, B., {Boss}, A.~P., {Weinberger},
  A.~J., \& {Lloyd}, J.~P. 2013,
  \href{http://dx.doi.org/10.1051/0004-6361/201219250}{\JournalTitle{\aap},
  551, A48}

\bibitem[{{Ballard} \& {Johnson}(2016)}]{Ballard2016}
{Ballard}, S., \& {Johnson}, J.~A. 2016,
  \href{http://dx.doi.org/10.3847/0004-637X/816/2/66}{\JournalTitle{\apj}, 816,
  66}

\bibitem[{{Ballard} {et~al.}(2013){Ballard}, {Charbonneau}, {Fressin},
  {Torres}, {Irwin}, {Desert}, {Newton}, {Mann}, {Ciardi}, {Crepp}, {Henze},
  {Bryson}, {Howell}, {Horch}, {Everett}, \& {Shporer}}]{Ballard:2013}
{Ballard}, S., {Charbonneau}, D., {Fressin}, F., {et~al.} 2013,
  \href{http://dx.doi.org/10.1088/0004-637X/773/2/98}{\JournalTitle{\apj}, 773,
  98}

\bibitem[{{Baranec} {et~al.}(2016){Baranec}, {Ziegler}, {Law}, {Morton},
  {Riddle}, {Atkinson}, {Schonhut}, \& {Crepp}}]{2016AJ....152...18B}
{Baranec}, C., {Ziegler}, C., {Law}, N.~M., {et~al.} 2016,
  \href{http://dx.doi.org/10.3847/0004-6256/152/1/18}{\JournalTitle{\aj}, 152,
  18}

\bibitem[{{Bastien} {et~al.}(2014){Bastien}, {Stassun}, \&
  {Pepper}}]{Bastien2014}
{Bastien}, F.~A., {Stassun}, K.~G., \& {Pepper}, J. 2014,
  \href{http://dx.doi.org/10.1088/2041-8205/788/1/L9}{\JournalTitle{\apjl},
  788, L9}

\bibitem[{{Batalha} {et~al.}(2010{\natexlab{a}}){Batalha}, {Rowe}, {Gilliland},
  {Jenkins}, {Caldwell}, {Borucki}, {Koch}, {Lissauer}, {Dunham}, {Gautier},
  {Howell}, {Latham}, {Marcy}, \& {Prsa}}]{2010ApJ...713L.103B}
{Batalha}, N.~M., {Rowe}, J.~F., {Gilliland}, R.~L., {et~al.}
  2010{\natexlab{a}},
  \href{http://dx.doi.org/10.1088/2041-8205/713/2/L103}{\JournalTitle{\apjl},
  713, L103}

\bibitem[{{Batalha} {et~al.}(2010{\natexlab{b}}){Batalha}, {Borucki}, {Koch},
  {Bryson}, {Haas}, {Brown}, {Caldwell}, {Hall}, {Gilliland}, {Latham},
  {Meibom}, \& {Monet}}]{Batalha:2010fk}
{Batalha}, N.~M., {Borucki}, W.~J., {Koch}, D.~G., {et~al.} 2010{\natexlab{b}},
  \href{http://dx.doi.org/10.1088/2041-8205/713/2/L109}{\JournalTitle{\apjl},
  713, L109}

\bibitem[{{Beatty} {et~al.}(2016){Beatty}, {Stevens}, {Collins}, {Colon},
  {James}, {Kreidberg}, {Pepper}, {Rodriguez}, {Siverd}, {Stassun}, \&
  {Kielkopf}}]{2016arXiv161204379B}
{Beatty}, T.~G., {Stevens}, D.~J., {Collins}, K.~A., {et~al.} 2016,
  \JournalTitle{ArXiv e-prints},
  \href{http://arxiv.org/abs/1612.04379}{{\sffamily arXiv:1612.04379
  [astro-ph.EP]}}

\bibitem[{{Benedict} {et~al.}(2016){Benedict}, {Henry}, {Franz}, {McArthur},
  {Wasserman}, {Jao}, {Cargile}, {Dieterich}, {Bradley}, {Nelan}, \&
  {Whipple}}]{Benedict:2016aa}
{Benedict}, G.~F., {Henry}, T.~J., {Franz}, O.~G., {et~al.} 2016,
  \href{http://dx.doi.org/10.3847/0004-6256/152/5/141}{\JournalTitle{\aj}, 152,
  141}

\bibitem[{{Berger} {et~al.}(2006){Berger}, {Gies}, {McAlister}, {ten
  Brummelaar}, {Henry}, {Sturmann}, {Sturmann}, {Turner}, {Ridgway},
  {Aufdenberg}, \& {M{\'e}rand}}]{Berger2006}
{Berger}, D.~H., {Gies}, D.~R., {McAlister}, H.~A., {et~al.} 2006,
  \href{http://dx.doi.org/10.1086/503318}{\JournalTitle{\apj}, 644, 475}

\bibitem[{{Biddle} {et~al.}(2014){Biddle}, {Pearson}, {Crossfield}, {Fulton},
  {Ciceri}, {Eastman}, {Barman}, {Mann}, {Henry}, {Howard}, {Williamson},
  {Sinukoff}, {Dragomir}, {Vican}, {Mancini}, {Southworth}, {Greenberg},
  {Turner}, {Thompson}, {Taylor}, {Levine}, \& {Webber}}]{2014MNRAS.443.1810B}
{Biddle}, L.~I., {Pearson}, K.~A., {Crossfield}, I.~J.~M., {et~al.} 2014,
  \href{http://dx.doi.org/10.1093/mnras/stu1199}{\JournalTitle{\mnras}, 443,
  1810}

\bibitem[{{Borucki} {et~al.}(2010){Borucki}, {Koch}, {Basri}, {Batalha},
  {Brown}, {Caldwell}, {Caldwell}, {Christensen-Dalsgaard}, {Cochran},
  {DeVore}, {Dunham}, {Dupree}, {Gautier}, {Geary}, {Gilliland}, {Gould},
  {Howell}, {Jenkins}, {Kondo}, {Latham}, {Marcy}, {Meibom}, {Kjeldsen},
  {Lissauer}, {Monet}, {Morrison}, {Sasselov}, {Tarter}, {Boss}, {Brownlee},
  {Owen}, {Buzasi}, {Charbonneau}, {Doyle}, {Fortney}, {Ford}, {Holman},
  {Seager}, {Steffen}, {Welsh}, {Rowe}, {Anderson}, {Buchhave}, {Ciardi},
  {Walkowicz}, {Sherry}, {Horch}, {Isaacson}, {Everett}, {Fischer}, {Torres},
  {Johnson}, {Endl}, {MacQueen}, {Bryson}, {Dotson}, {Haas}, {Kolodziejczak},
  {Van Cleve}, {Chandrasekaran}, {Twicken}, {Quintana}, {Clarke}, {Allen},
  {Li}, {Wu}, {Tenenbaum}, {Verner}, {Bruhweiler}, {Barnes}, \&
  {Prsa}}]{Borucki2010}
{Borucki}, W.~J., {Koch}, D., {Basri}, G., {et~al.} 2010,
  \href{http://dx.doi.org/10.1126/science.1185402}{\JournalTitle{Science}, 327,
  977}

\bibitem[{{Boyajian} {et~al.}(2012){Boyajian}, {von Braun}, {van Belle},
  {McAlister}, {ten Brummelaar}, {Kane}, {Muirhead}, {Jones}, {White},
  {Schaefer}, {Ciardi}, {Henry}, {L{\'o}pez-Morales}, {Ridgway}, {Gies}, {Jao},
  {Rojas-Ayala}, {Parks}, {Sturmann}, {Sturmann}, {Turner}, {Farrington},
  {Goldfinger}, \& {Berger}}]{Boyajian2012}
{Boyajian}, T.~S., {von Braun}, K., {van Belle}, G., {et~al.} 2012,
  \href{http://dx.doi.org/10.1088/0004-637X/757/2/112}{\JournalTitle{\apj},
  757, 112}

\bibitem[{{Brewer} {et~al.}(2015){Brewer}, {Fischer}, {Basu}, {Valenti}, \&
  {Piskunov}}]{2015ApJ...805..126B}
{Brewer}, J.~M., {Fischer}, D.~A., {Basu}, S., {Valenti}, J.~A., \& {Piskunov},
  N. 2015,
  \href{http://dx.doi.org/10.1088/0004-637X/805/2/126}{\JournalTitle{\apj},
  805, 126}

\bibitem[{{Brown} {et~al.}(2011){Brown}, {Latham}, {Everett}, \&
  {Esquerdo}}]{Brown2011}
{Brown}, T.~M., {Latham}, D.~W., {Everett}, M.~E., \& {Esquerdo}, G.~A. 2011,
  \href{http://dx.doi.org/10.1088/0004-6256/142/4/112}{\JournalTitle{\aj}, 142,
  112}

\bibitem[{{Bryson} {et~al.}(2013){Bryson}, {Jenkins}, {Gilliland}, {Twicken},
  {Clarke}, {Rowe}, {Caldwell}, {Batalha}, {Mullally}, {Haas}, \&
  {Tenenbaum}}]{Bryson2013}
{Bryson}, S.~T., {Jenkins}, J.~M., {Gilliland}, R.~L., {et~al.} 2013,
  \href{http://dx.doi.org/10.1086/671767}{\JournalTitle{\pasp}, 125, 889}

\bibitem[{{Casagrande} {et~al.}(2011){Casagrande}, {Sch{\"o}nrich}, {Asplund},
  {Cassisi}, {Ram{\'{\i}}rez}, {Mel{\'e}ndez}, {Bensby}, \&
  {Feltzing}}]{2011A&A...530A.138C}
{Casagrande}, L., {Sch{\"o}nrich}, R., {Asplund}, M., {et~al.} 2011,
  \href{http://dx.doi.org/10.1051/0004-6361/201016276}{\JournalTitle{\aap},
  530, A138}

\bibitem[{{Chabrier} \& {Baraffe}(1997)}]{Chabrier:1997}
{Chabrier}, G., \& {Baraffe}, I. 1997, \JournalTitle{\aap}, 327, 1039

\bibitem[{{Chambers}(1999)}]{1999MNRAS.304..793C}
{Chambers}, J.~E. 1999,
  \href{http://dx.doi.org/10.1046/j.1365-8711.1999.02379.x}{\JournalTitle{\mnras},
  304, 793}

\bibitem[{{Charbonneau} {et~al.}(2009){Charbonneau}, {Berta}, {Irwin}, {Burke},
  {Nutzman}, {Buchhave}, {Lovis}, {Bonfils}, {Latham}, {Udry}, {Murray-Clay},
  {Holman}, {Falco}, {Winn}, {Queloz}, {Pepe}, {Mayor}, {Delfosse}, \&
  {Forveille}}]{Charbonneau:2009rt}
{Charbonneau}, D., {Berta}, Z.~K., {Irwin}, J., {et~al.} 2009,
  \href{http://dx.doi.org/10.1038/nature08679}{\JournalTitle{\nat}, 462, 891}

\bibitem[{{Christiansen} {et~al.}(2016){Christiansen}, {Clarke}, {Burke},
  {Jenkins}, {Bryson}, {Coughlin}, {Mullally}, {Thompson}, {Twicken},
  {Batalha}, {Haas}, {Catanzarite}, {Campbell}, {Kamal Uddin}, {Zamudio},
  {Smith}, \& {Henze}}]{2016ApJ...828...99C}
{Christiansen}, J.~L., {Clarke}, B.~D., {Burke}, C.~J., {et~al.} 2016,
  \href{http://dx.doi.org/10.3847/0004-637X/828/2/99}{\JournalTitle{\apj}, 828,
  99}

\bibitem[{{Ciardi} {et~al.}(2015){Ciardi}, {Beichman}, {Horch}, \&
  {Howell}}]{2015ApJ...805...16C}
{Ciardi}, D.~R., {Beichman}, C.~A., {Horch}, E.~P., \& {Howell}, S.~B. 2015,
  \href{http://dx.doi.org/10.1088/0004-637X/805/1/16}{\JournalTitle{\apj}, 805,
  16}

\bibitem[{{Cutri} {et~al.}(2003){Cutri}, {Skrutskie}, {van Dyk}, {Beichman},
  {Carpenter}, {Chester}, {Cambresy}, {Evans}, {Fowler}, {Gizis}, {Howard},
  {Huchra}, {Jarrett}, {Kopan}, {Kirkpatrick}, {Light}, {Marsh}, {McCallon},
  {Schneider}, {Stiening}, {Sykes}, {Weinberg}, {Wheaton}, {Wheelock}, \&
  {Zacarias}}]{Cutri2003}
{Cutri}, R.~M., {Skrutskie}, M.~F., {van Dyk}, S., {et~al.} 2003,
  \JournalTitle{VizieR Online Data Catalog}, 2246, 0

\bibitem[{{Dalba} {et~al.}(2017){Dalba}, {Muirhead}, {Croll}, \&
  {Kempton}}]{2017AJ....153...59D}
{Dalba}, P.~A., {Muirhead}, P.~S., {Croll}, B., \& {Kempton}, E.~M.-R. 2017,
  \href{http://dx.doi.org/10.1088/1361-6528/aa5278}{\JournalTitle{\aj}, 153,
  59}

\bibitem[{{Dawson} \& {Johnson}(2012)}]{Dawson:2012fk}
{Dawson}, R.~I., \& {Johnson}, J.~A. 2012,
  \href{http://dx.doi.org/10.1088/0004-637X/756/2/122}{\JournalTitle{\apj},
  756, 122}

\bibitem[{{Dawson} {et~al.}(2016){Dawson}, {Lee}, \&
  {Chiang}}]{2016ApJ...822...54D}
{Dawson}, R.~I., {Lee}, E.~J., \& {Chiang}, E. 2016,
  \href{http://dx.doi.org/10.3847/0004-637X/822/1/54}{\JournalTitle{\apj}, 822,
  54}

\bibitem[{{de Bruijne}(2012)}]{2012Ap&SS.341...31D}
{de Bruijne}, J.~H.~J. 2012,
  \href{http://dx.doi.org/10.1007/s10509-012-1019-4}{\JournalTitle{\apss}, 341,
  31}

\bibitem[{{Deacon} {et~al.}(2016){Deacon}, {Kraus}, {Mann}, {Magnier},
  {Chambers}, {Wainscoat}, {Tonry}, {Kaiser}, {Waters}, {Flewelling}, {Hodapp},
  \& {Burgett}}]{2016MNRAS.455.4212D}
{Deacon}, N.~R., {Kraus}, A.~L., {Mann}, A.~W., {et~al.} 2016,
  \href{http://dx.doi.org/10.1093/mnras/stv2132}{\JournalTitle{\mnras}, 455,
  4212}

\bibitem[{{Delfosse} {et~al.}(2000){Delfosse}, {Forveille}, {S{\'e}gransan},
  {Beuzit}, {Udry}, {Perrier}, \& {Mayor}}]{Delfosse2000}
{Delfosse}, X., {Forveille}, T., {S{\'e}gransan}, D., {et~al.} 2000,
  \JournalTitle{\aap}, 364, 217

\bibitem[{{Dotter} {et~al.}(2008){Dotter}, {Chaboyer}, {Jevremovi{\'c}},
  {Kostov}, {Baron}, \& {Ferguson}}]{Dotter2008}
{Dotter}, A., {Chaboyer}, B., {Jevremovi{\'c}}, D., {et~al.} 2008,
  \href{http://dx.doi.org/10.1086/589654}{\JournalTitle{\apjs}, 178, 89}

\bibitem[{{Dressing} \& {Charbonneau}(2013)}]{Dressing2013}
{Dressing}, C.~D., \& {Charbonneau}, D. 2013,
  \href{http://dx.doi.org/10.1088/0004-637X/767/1/95}{\JournalTitle{\apj}, 767,
  95}

\bibitem[{{Dressing} \& {Charbonneau}(2015)}]{Dressing2015}
---. 2015,
  \href{http://dx.doi.org/10.1088/0004-637X/807/1/45}{\JournalTitle{\apj}, 807,
  45}

\bibitem[{{Dressing} {et~al.}(2017){Dressing}, {Newton}, {Schlieder},
  {Charbonneau}, {Knutson}, {Vanderburg}, \& {Sinukoff}}]{Dressing2017}
{Dressing}, C.~D., {Newton}, E.~R., {Schlieder}, J.~E., {et~al.} 2017,
  \href{http://dx.doi.org/10.3847/1538-4357/836/2/167}{\JournalTitle{\apj},
  836, 167}

\bibitem[{{Dupuy} \& {Liu}(2012)}]{Dupuy2012}
{Dupuy}, T.~J., \& {Liu}, M.~C. 2012,
  \href{http://dx.doi.org/10.1088/0067-0049/201/2/19}{\JournalTitle{\apjs},
  201, 19}

\bibitem[{{Dupuy} \& {Liu}(2017)}]{2017arXiv170305775D}
---. 2017, \JournalTitle{ArXiv e-prints},
  \href{http://arxiv.org/abs/1703.05775}{{\sffamily arXiv:1703.05775
  [astro-ph.SR]}}

\bibitem[{{Feiden} \& {Chaboyer}(2012)}]{Feiden2012a}
{Feiden}, G.~A., \& {Chaboyer}, B. 2012,
  \href{http://dx.doi.org/10.1088/0004-637X/757/1/42}{\JournalTitle{\apj}, 757,
  42}

\bibitem[{{Feiden} \& {Chaboyer}(2013)}]{Feiden2013}
---. 2013,
  \href{http://dx.doi.org/10.1088/0004-637X/779/2/183}{\JournalTitle{\apj},
  779, 183}

\bibitem[{{Feiden} \& {Chaboyer}(2014)}]{Feiden2014a}
---. 2014,
  \href{http://dx.doi.org/10.1088/0004-637X/789/1/53}{\JournalTitle{\apj}, 789,
  53}

\bibitem[{{Foreman-Mackey} {et~al.}(2013){Foreman-Mackey}, {Hogg}, {Lang}, \&
  {Goodman}}]{Foreman-Mackey2013}
{Foreman-Mackey}, D., {Hogg}, D.~W., {Lang}, D., \& {Goodman}, J. 2013,
  \href{http://dx.doi.org/10.1086/670067}{\JournalTitle{\pasp}, 125, 306}

\bibitem[{{Gaidos} \& {Mann}(2013)}]{Gaidos2013}
{Gaidos}, E., \& {Mann}, A.~W. 2013,
  \href{http://dx.doi.org/10.1088/0004-637X/762/1/41}{\JournalTitle{\apj}, 762,
  41}

\bibitem[{{Gaidos} {et~al.}(2016{\natexlab{a}}){Gaidos}, {Mann}, \&
  {Ansdell}}]{Gaidos2016a}
{Gaidos}, E., {Mann}, A.~W., \& {Ansdell}, M. 2016{\natexlab{a}},
  \href{http://dx.doi.org/10.3847/0004-637X/817/1/50}{\JournalTitle{\apj}, 817,
  50}

\bibitem[{{Gaidos} {et~al.}(2016{\natexlab{b}}){Gaidos}, {Mann}, {Kraus}, \&
  {Ireland}}]{Gaidos2016b}
{Gaidos}, E., {Mann}, A.~W., {Kraus}, A.~L., \& {Ireland}, M.
  2016{\natexlab{b}},
  \href{http://dx.doi.org/10.1093/mnras/stw097}{\JournalTitle{\mnras}, 457,
  2877}

\bibitem[{{Gazak} {et~al.}(2012){Gazak}, {Johnson}, {Tonry}, {Dragomir},
  {Eastman}, {Mann}, \& {Agol}}]{Gazak:2012vn}
{Gazak}, J.~Z., {Johnson}, J.~A., {Tonry}, J., {et~al.} 2012,
  \href{http://dx.doi.org/10.1155/2012/697967}{\JournalTitle{Advances in
  Astronomy}, 2012}, \href{http://arxiv.org/abs/1102.1036}{{\sffamily
  arXiv:1102.1036 [astro-ph.EP]}}

\bibitem[{{Greiss} {et~al.}(2012){Greiss}, {Steeghs}, {G{\"a}nsicke},
  {Mart{\'{\i}}n}, {Groot}, {Irwin}, {Gonz{\'a}lez-Solares}, {Greimel},
  {Knigge}, {{\O}stensen}, {Verbeek}, {Drew}, {Drake}, {Jonker}, {Ripepi},
  {Scaringi}, {Southworth}, {Still}, {Wright}, {Farnhill}, {van Haaften}, \&
  {Shah}}]{Greiss2012}
{Greiss}, S., {Steeghs}, D., {G{\"a}nsicke}, B.~T., {et~al.} 2012,
  \href{http://dx.doi.org/10.1088/0004-6256/144/1/24}{\JournalTitle{\aj}, 144,
  24}

\bibitem[{{Gully-Santiago} {et~al.}(2017){Gully-Santiago}, {Herczeg},
  {Czekala}, {Somers}, {Grankin}, {Covey}, {Donati}, {Alencar}, {Hussain},
  {Shappee}, {Mace}, {Lee}, {Holoien}, {Jose}, \& {Liu}}]{Gully-Santiago2017}
{Gully-Santiago}, M.~A., {Herczeg}, G.~J., {Czekala}, I., {et~al.} 2017,
  \href{http://dx.doi.org/10.3847/1538-4357/836/2/200}{\JournalTitle{\apj},
  836, 200}

\bibitem[{{Hadden} \& {Lithwick}(2016)}]{2016arXiv161103516H}
{Hadden}, S., \& {Lithwick}, Y. 2016, \JournalTitle{ArXiv e-prints},
  \href{http://arxiv.org/abs/1611.03516}{{\sffamily arXiv:1611.03516
  [astro-ph.EP]}}

\bibitem[{{Haghighipour}(2006)}]{2006ApJ...644..543H}
{Haghighipour}, N. 2006,
  \href{http://dx.doi.org/10.1086/503351}{\JournalTitle{\apj}, 644, 543}

\bibitem[{{Haghighipour} \& {Raymond}(2007)}]{2007ApJ...666..436H}
{Haghighipour}, N., \& {Raymond}, S.~N. 2007,
  \href{http://dx.doi.org/10.1086/520501}{\JournalTitle{\apj}, 666, 436}

\bibitem[{{Henden} {et~al.}(2012){Henden}, {Levine}, {Terrell}, {Smith}, \&
  {Welch}}]{Henden:2012fk}
{Henden}, A.~A., {Levine}, S.~E., {Terrell}, D., {Smith}, T.~C., \& {Welch}, D.
  2012, \JournalTitle{Journal of the American Association of Variable Star
  Observers (JAAVSO)}, 40, 430

\bibitem[{{Henry} \& {McCarthy}(1993)}]{Henry:1993fk}
{Henry}, T.~J., \& {McCarthy}, Jr., D.~W. 1993,
  \href{http://dx.doi.org/10.1086/116685}{\JournalTitle{\aj}, 106, 773}

\bibitem[{{Huber} {et~al.}(2013){Huber}, {Chaplin}, {Christensen-Dalsgaard},
  {Gilliland}, {Kjeldsen}, {Buchhave}, {Fischer}, {Lissauer}, {Rowe},
  {Sanchis-Ojeda}, {Basu}, {Handberg}, {Hekker}, {Howard}, {Isaacson},
  {Karoff}, {Latham}, {Lund}, {Lundkvist}, {Marcy}, {Miglio}, {Silva Aguirre},
  {Stello}, {Arentoft}, {Barclay}, {Bedding}, {Burke}, {Christiansen},
  {Elsworth}, {Haas}, {Kawaler}, {Metcalfe}, {Mullally}, \&
  {Thompson}}]{2013ApJ...767..127H}
{Huber}, D., {Chaplin}, W.~J., {Christensen-Dalsgaard}, J., {et~al.} 2013,
  \href{http://dx.doi.org/10.1088/0004-637X/767/2/127}{\JournalTitle{\apj},
  767, 127}

\bibitem[{{Huber} {et~al.}(2014){Huber}, {Silva Aguirre}, {Matthews},
  {Pinsonneault}, {Gaidos}, {Garc{\'{\i}}a}, {Hekker}, {Mathur}, {Mosser},
  {Torres}, {Bastien}, {Basu}, {Bedding}, {Chaplin}, {Demory}, {Fleming},
  {Guo}, {Mann}, {Rowe}, {Serenelli}, {Smith}, \&
  {Stello}}]{2014ApJS..211....2H}
{Huber}, D., {Silva Aguirre}, V., {Matthews}, J.~M., {et~al.} 2014,
  \href{http://dx.doi.org/10.1088/0067-0049/211/1/2}{\JournalTitle{\apjs}, 211,
  2}

\bibitem[{{Huber} {et~al.}(2016){Huber}, {Bryson}, {Haas}, {Barclay},
  {Barentsen}, {Howell}, {Sharma}, {Stello}, \& {Thompson}}]{Huber2016}
{Huber}, D., {Bryson}, S.~T., {Haas}, M.~R., {et~al.} 2016,
  \href{http://dx.doi.org/10.3847/0067-0049/224/1/2}{\JournalTitle{\apjs}, 224,
  2}

\bibitem[{{Jenkins} {et~al.}(2010){Jenkins}, {Caldwell}, {Chandrasekaran},
  {Twicken}, {Bryson}, {Quintana}, {Clarke}, {Li}, {Allen}, {Tenenbaum}, {Wu},
  {Klaus}, {Van Cleve}, {Dotson}, {Haas}, {Gilliland}, {Koch}, \&
  {Borucki}}]{Jenkins:2010qy}
{Jenkins}, J.~M., {Caldwell}, D.~A., {Chandrasekaran}, H., {et~al.} 2010,
  \href{http://dx.doi.org/10.1088/2041-8205/713/2/L120}{\JournalTitle{\apjl},
  713, L120}

\bibitem[{{Johnson} {et~al.}(2011){Johnson}, {Apps}, {Gazak}, {Crepp},
  {Crossfield}, {Howard}, {Marcy}, {Morton}, {Chubak}, \&
  {Isaacson}}]{2011ApJ...730...79J}
{Johnson}, J.~A., {Apps}, K., {Gazak}, J.~Z., {et~al.} 2011,
  \href{http://dx.doi.org/10.1088/0004-637X/730/2/79}{\JournalTitle{\apj}, 730,
  79}

\bibitem[{{Johnson} {et~al.}(2012){Johnson}, {Gazak}, {Apps}, {Muirhead},
  {Crepp}, {Crossfield}, {Boyajian}, {von Braun}, {Rojas-Ayala}, {Howard},
  {Covey}, {Schlawin}, {Hamren}, {Morton}, {Marcy}, \&
  {Lloyd}}]{Johnson:2012fk}
{Johnson}, J.~A., {Gazak}, J.~Z., {Apps}, K., {et~al.} 2012,
  \href{http://dx.doi.org/10.1088/0004-6256/143/5/111}{\JournalTitle{\aj}, 143,
  111}

\bibitem[{{Jontof-Hutter} {et~al.}(2015){Jontof-Hutter}, {Rowe}, {Lissauer},
  {Fabrycky}, \& {Ford}}]{2015Natur.522..321J}
{Jontof-Hutter}, D., {Rowe}, J.~F., {Lissauer}, J.~J., {Fabrycky}, D.~C., \&
  {Ford}, E.~B. 2015,
  \href{http://dx.doi.org/10.1038/nature14494}{\JournalTitle{\nat}, 522, 321}

\bibitem[{{Kaib} {et~al.}(2013){Kaib}, {Raymond}, \& {Duncan}}]{Kaib:2013}
{Kaib}, N.~A., {Raymond}, S.~N., \& {Duncan}, M. 2013,
  \href{http://dx.doi.org/10.1038/nature11780}{\JournalTitle{\nat}, 493, 381}

\bibitem[{{Kane} {et~al.}(2012){Kane}, {Ciardi}, {Gelino}, \& {von
  Braun}}]{2012MNRAS.425..757K}
{Kane}, S.~R., {Ciardi}, D.~R., {Gelino}, D.~M., \& {von Braun}, K. 2012,
  \href{http://dx.doi.org/10.1111/j.1365-2966.2012.21627.x}{\JournalTitle{\mnras},
  425, 757}

\bibitem[{{Kane} {et~al.}(2017){Kane}, {von Braun}, {Henry}, {Waters},
  {Boyajian}, \& {Mann}}]{Kane2017}
{Kane}, S.~R., {von Braun}, K., {Henry}, G.~W., {et~al.} 2017,
  \href{http://dx.doi.org/10.3847/1538-4357/835/2/200}{\JournalTitle{\apj},
  835, 200}

\bibitem[{{Kinemuchi} {et~al.}(2012){Kinemuchi}, {Barclay}, {Fanelli},
  {Pepper}, {Still}, \& {Howell}}]{2012PASP..124..963K}
{Kinemuchi}, K., {Barclay}, T., {Fanelli}, M., {et~al.} 2012,
  \href{http://dx.doi.org/10.1086/667603}{\JournalTitle{\pasp}, 124, 963}

\bibitem[{{Kipping}(2010)}]{Kipping:2010lr}
{Kipping}, D.~M. 2010,
  \href{http://dx.doi.org/10.1111/j.1365-2966.2010.17242.x}{\JournalTitle{\mnras},
  408, 1758}

\bibitem[{{Kipping}(2013)}]{Kipping2013}
---. 2013,
  \href{http://dx.doi.org/10.1093/mnras/stt1435}{\JournalTitle{\mnras}, 435,
  2152}

\bibitem[{{Kipping}(2014{\natexlab{a}})}]{2014MNRAS.444.2263K}
---. 2014{\natexlab{a}},
  \href{http://dx.doi.org/10.1093/mnras/stu1561}{\JournalTitle{\mnras}, 444,
  2263}

\bibitem[{{Kipping}(2014{\natexlab{b}})}]{2014MNRAS.440.2164K}
---. 2014{\natexlab{b}},
  \href{http://dx.doi.org/10.1093/mnras/stu318}{\JournalTitle{\mnras}, 440,
  2164}

\bibitem[{{Kipping} {et~al.}(2014){Kipping}, {Nesvorn{\'y}}, {Buchhave},
  {Hartman}, {Bakos}, \& {Schmitt}}]{2014ApJ...784...28K}
{Kipping}, D.~M., {Nesvorn{\'y}}, D., {Buchhave}, L.~A., {et~al.} 2014,
  \href{http://dx.doi.org/10.1088/0004-637X/784/1/28}{\JournalTitle{\apj}, 784,
  28}

\bibitem[{{Kratter} \& {Perets}(2012)}]{2012ApJ...753...91K}
{Kratter}, K.~M., \& {Perets}, H.~B. 2012,
  \href{http://dx.doi.org/10.1088/0004-637X/753/1/91}{\JournalTitle{\apj}, 753,
  91}

\bibitem[{{Kraus} {et~al.}(2016){Kraus}, {Ireland}, {Huber}, {Mann}, \&
  {Dupuy}}]{Kraus2016a}
{Kraus}, A.~L., {Ireland}, M.~J., {Huber}, D., {Mann}, A.~W., \& {Dupuy}, T.~J.
  2016, \href{http://dx.doi.org/10.3847/0004-6256/152/1/8}{\JournalTitle{\aj},
  152, 8}

\bibitem[{{Kraus} {et~al.}(2011){Kraus}, {Tucker}, {Thompson}, {Craine}, \&
  {Hillenbrand}}]{Kraus2011}
{Kraus}, A.~L., {Tucker}, R.~A., {Thompson}, M.~I., {Craine}, E.~R., \&
  {Hillenbrand}, L.~A. 2011,
  \href{http://dx.doi.org/10.1088/0004-637X/728/1/48}{\JournalTitle{\apj}, 728,
  48}

\bibitem[{{Kreidberg}(2015)}]{Kreidberg2015}
{Kreidberg}, L. 2015,
  \href{http://dx.doi.org/10.1086/683602}{\JournalTitle{\pasp}, 127, 1161}

\bibitem[{{Laughlin} {et~al.}(1997){Laughlin}, {Bodenheimer}, \&
  {Adams}}]{Laughlin1997}
{Laughlin}, G., {Bodenheimer}, P., \& {Adams}, F.~C. 1997, \JournalTitle{\apj},
  482, 420

\bibitem[{{L{\'e}pine} \& {Bongiorno}(2007)}]{Lepine:2007qy}
{L{\'e}pine}, S., \& {Bongiorno}, B. 2007,
  \href{http://dx.doi.org/10.1086/510333}{\JournalTitle{\aj}, 133, 889}

\bibitem[{{L{\'e}pine} \& {Gaidos}(2011)}]{Lepine:2011vn}
{L{\'e}pine}, S., \& {Gaidos}, E. 2011,
  \href{http://dx.doi.org/10.1088/0004-6256/142/4/138}{\JournalTitle{\aj}, 142,
  138}

\bibitem[{{Limbach} \& {Turner}(2015)}]{2015PNAS..112...20L}
{Limbach}, M.~A., \& {Turner}, E.~L. 2015,
  \href{http://dx.doi.org/10.1073/pnas.1406545111}{\JournalTitle{Proceedings of
  the National Academy of Science}, 112, 20}

\bibitem[{{Lloyd}(2011)}]{2011ApJ...739L..49L}
{Lloyd}, J.~P. 2011,
  \href{http://dx.doi.org/10.1088/2041-8205/739/2/L49}{\JournalTitle{\apjl},
  739, L49}

\bibitem[{{Mandel} \& {Agol}(2002)}]{MandelAgol2002}
{Mandel}, K., \& {Agol}, E. 2002,
  \href{http://dx.doi.org/10.1086/345520}{\JournalTitle{\apjl}, 580, L171}

\bibitem[{{Mann} {et~al.}(2013{\natexlab{a}}){Mann}, {Brewer}, {Gaidos},
  {L{\'e}pine}, \& {Hilton}}]{Mann2013a}
{Mann}, A.~W., {Brewer}, J.~M., {Gaidos}, E., {L{\'e}pine}, S., \& {Hilton},
  E.~J. 2013{\natexlab{a}},
  \href{http://dx.doi.org/10.1088/0004-6256/145/2/52}{\JournalTitle{\aj}, 145,
  52}

\bibitem[{{Mann} {et~al.}(2015){Mann}, {Feiden}, {Gaidos}, {Boyajian}, \& {von
  Braun}}]{Mann2015b}
{Mann}, A.~W., {Feiden}, G.~A., {Gaidos}, E., {Boyajian}, T., \& {von Braun},
  K. 2015,
  \href{http://dx.doi.org/10.1088/0004-637X/804/1/64}{\JournalTitle{\apj}, 804,
  64}

\bibitem[{{Mann} {et~al.}(2013{\natexlab{b}}){Mann}, {Gaidos}, \&
  {Ansdell}}]{Mann2013c}
{Mann}, A.~W., {Gaidos}, E., \& {Ansdell}, M. 2013{\natexlab{b}},
  \href{http://dx.doi.org/10.1088/0004-637X/779/2/188}{\JournalTitle{\apj},
  779, 188}

\bibitem[{{Mann} {et~al.}(2010){Mann}, {Gaidos}, \& {Gaudi}}]{Mann:2010fj}
{Mann}, A.~W., {Gaidos}, E., \& {Gaudi}, B.~S. 2010,
  \href{http://dx.doi.org/10.1088/0004-637X/719/2/1454}{\JournalTitle{\apj},
  719, 1454}

\bibitem[{{Mann} {et~al.}(2013{\natexlab{c}}){Mann}, {Gaidos}, {Kraus}, \&
  {Hilton}}]{Mann2013b}
{Mann}, A.~W., {Gaidos}, E., {Kraus}, A., \& {Hilton}, E.~J.
  2013{\natexlab{c}},
  \href{http://dx.doi.org/10.1088/0004-637X/770/1/43}{\JournalTitle{\apj}, 770,
  43}

\bibitem[{{Mann} {et~al.}(2016){Mann}, {Gaidos}, {Mace}, {Johnson}, {Bowler},
  {LaCourse}, {Jacobs}, {Vanderburg}, {Kraus}, {Kaplan}, \&
  {Jaffe}}]{Mann2016a}
{Mann}, A.~W., {Gaidos}, E., {Mace}, G.~N., {et~al.} 2016,
  \href{http://dx.doi.org/10.3847/0004-637X/818/1/46}{\JournalTitle{\apj}, 818,
  46}

\bibitem[{{Mann} {et~al.}(2017){Mann}, {Gaidos}, {Vanderburg}, {Rizzuto},
  {Ansdell}, {Medina}, {Mace}, {Kraus}, \& {Sokal}}]{Mann:2017aa}
{Mann}, A.~W., {Gaidos}, E., {Vanderburg}, A., {et~al.} 2017,
  \href{http://dx.doi.org/10.1088/1361-6528/aa5276}{\JournalTitle{\aj}, 153,
  64}

\bibitem[{{Martinez} {et~al.}(2017){Martinez}, {Crossfield}, {Schlieder},
  {Dressing}, {Obermeier}, {Livingston}, {Ciceri}, {Peacock}, {Beichman},
  {L{\'e}pine}, {Aller}, {Chance}, {Petigura}, {Howard}, \&
  {Werner}}]{Martinez:2017aa}
{Martinez}, A.~O., {Crossfield}, I.~J.~M., {Schlieder}, J.~E., {et~al.} 2017,
  \href{http://dx.doi.org/10.3847/1538-4357/aa56c7}{\JournalTitle{\apj}, 837,
  72}

\bibitem[{{Mazeh} {et~al.}(2013){Mazeh}, {Nachmani}, {Holczer}, {Fabrycky},
  {Ford}, {Sanchis-Ojeda}, {Sokol}, {Rowe}, {Zucker}, {Agol}, {Carter},
  {Lissauer}, {Quintana}, {Ragozzine}, {Steffen}, \&
  {Welsh}}]{2013ApJS..208...16M}
{Mazeh}, T., {Nachmani}, G., {Holczer}, T., {et~al.} 2013,
  \href{http://dx.doi.org/10.1088/0067-0049/208/2/16}{\JournalTitle{\apjs},
  208, 16}

\bibitem[{{Moorhead} {et~al.}(2011){Moorhead}, {Ford}, {Morehead}, {Rowe},
  {Borucki}, {Batalha}, {Bryson}, {Caldwell}, {Fabrycky}, {Gautier}, {Koch},
  {Holman}, {Jenkins}, {Li}, {Lissauer}, {Lucas}, {Marcy}, {Quinn}, {Quintana},
  {Ragozzine}, {Shporer}, {Still}, \& {Torres}}]{Moorhead:2011lr}
{Moorhead}, A.~V., {Ford}, E.~B., {Morehead}, R.~C., {et~al.} 2011,
  \href{http://dx.doi.org/10.1088/0067-0049/197/1/1}{\JournalTitle{\apjs}, 197,
  1}

\bibitem[{{Moriarty} \& {Ballard}(2016)}]{2016ApJ...832...34M}
{Moriarty}, J., \& {Ballard}, S. 2016,
  \href{http://dx.doi.org/10.3847/0004-637X/832/1/34}{\JournalTitle{\apj}, 832,
  34}

\bibitem[{{Morton}(2012)}]{2012ApJ...761....6M}
{Morton}, T.~D. 2012,
  \href{http://dx.doi.org/10.1088/0004-637X/761/1/6}{\JournalTitle{\apj}, 761,
  6}

\bibitem[{{Morton}(2015)}]{2015ascl.soft03011M}
---. 2015, {VESPA: False positive probabilities calculator}, Astrophysics
  Source Code Library, \href{http://arxiv.org/abs/1503.011}{{\sffamily
  ascl:1503.011}}

\bibitem[{{Morton} {et~al.}(2016){Morton}, {Bryson}, {Coughlin}, {Rowe},
  {Ravichandran}, {Petigura}, {Haas}, \& {Batalha}}]{2016ApJ...822...86M}
{Morton}, T.~D., {Bryson}, S.~T., {Coughlin}, J.~L., {et~al.} 2016,
  \href{http://dx.doi.org/10.3847/0004-637X/822/2/86}{\JournalTitle{\apj}, 822,
  86}

\bibitem[{{Mould}(1976)}]{1976A&A....48..443M}
{Mould}, J.~R. 1976, \JournalTitle{\aap}, 48, 443

\bibitem[{{Moutou} {et~al.}(2017){Moutou}, {Vigan}, {Mesa}, {Desidera},
  {Thebault}, {Zurlo}, \& {Salter}}]{2017arXiv170105664M}
{Moutou}, C., {Vigan}, A., {Mesa}, D., {et~al.} 2017, \JournalTitle{ArXiv
  e-prints}, \href{http://arxiv.org/abs/1701.05664}{{\sffamily arXiv:1701.05664
  [astro-ph.SR]}}

\bibitem[{{Mui{\~n}os} \& {Evans}(2014)}]{2014AN....335..367M}
{Mui{\~n}os}, J.~L., \& {Evans}, D.~W. 2014,
  \href{http://dx.doi.org/10.1002/asna.201312045}{\JournalTitle{Astronomische
  Nachrichten}, 335, 367}

\bibitem[{{Muirhead} {et~al.}(2012{\natexlab{a}}){Muirhead}, {Hamren},
  {Schlawin}, {Rojas-Ayala}, {Covey}, \& {Lloyd}}]{Muirhead2012a}
{Muirhead}, P.~S., {Hamren}, K., {Schlawin}, E., {et~al.} 2012{\natexlab{a}},
  \href{http://dx.doi.org/10.1088/2041-8205/750/2/L37}{\JournalTitle{\apjl},
  750, L37}

\bibitem[{{Muirhead} {et~al.}(2012{\natexlab{b}}){Muirhead}, {Johnson}, {Apps},
  {Carter}, {Morton}, {Fabrycky}, {Pineda}, {Bottom}, {Rojas-Ayala},
  {Schlawin}, {Hamren}, {Covey}, {Crepp}, {Stassun}, {Pepper}, {Hebb}, {Kirby},
  {Howard}, {Isaacson}, {Marcy}, {Levitan}, {Diaz-Santos}, {Armus}, \&
  {Lloyd}}]{Muirhead2012}
{Muirhead}, P.~S., {Johnson}, J.~A., {Apps}, K., {et~al.} 2012{\natexlab{b}},
  \href{http://dx.doi.org/10.1088/0004-637X/747/2/144}{\JournalTitle{\apj},
  747, 144}

\bibitem[{{Muirhead} {et~al.}(2014){Muirhead}, {Becker}, {Feiden},
  {Rojas-Ayala}, {Vanderburg}, {Price}, {Thorp}, {Law}, {Riddle}, {Baranec},
  {Hamren}, {Schlawin}, {Covey}, {Johnson}, \& {Lloyd}}]{Muirhead2014}
{Muirhead}, P.~S., {Becker}, J., {Feiden}, G.~A., {et~al.} 2014,
  \href{http://dx.doi.org/10.1088/0067-0049/213/1/5}{\JournalTitle{\apjs}, 213,
  5}

\bibitem[{{Muirhead} {et~al.}(2015){Muirhead}, {Mann}, {Vanderburg}, {Morton},
  {Kraus}, {Ireland}, {Swift}, {Feiden}, {Gaidos}, \& {Gazak}}]{Muirhead2015}
{Muirhead}, P.~S., {Mann}, A.~W., {Vanderburg}, A., {et~al.} 2015,
  \href{http://dx.doi.org/10.1088/0004-637X/801/1/18}{\JournalTitle{\apj}, 801,
  18}

\bibitem[{{Mullan} \& {MacDonald}(2001)}]{2001ApJ...559..353M}
{Mullan}, D.~J., \& {MacDonald}, J. 2001,
  \href{http://dx.doi.org/10.1086/322336}{\JournalTitle{\apj}, 559, 353}

\bibitem[{{Neves} {et~al.}(2012){Neves}, {Bonfils}, {Santos}, {Delfosse},
  {Forveille}, {Allard}, {Nat{\'a}rio}, {Fernandes}, \&
  {Udry}}]{2012A&A...538A..25N}
{Neves}, V., {Bonfils}, X., {Santos}, N.~C., {et~al.} 2012,
  \href{http://dx.doi.org/10.1051/0004-6361/201118115}{\JournalTitle{\aap},
  538, A25}

\bibitem[{{Newton} {et~al.}(2014){Newton}, {Charbonneau}, {Irwin},
  {Berta-Thompson}, {Rojas-Ayala}, {Covey}, \& {Lloyd}}]{Newton:2014}
{Newton}, E.~R., {Charbonneau}, D., {Irwin}, J., {et~al.} 2014,
  \href{http://dx.doi.org/10.1088/0004-6256/147/1/20}{\JournalTitle{\aj}, 147,
  20}

\bibitem[{{Newton} {et~al.}(2015){Newton}, {Charbonneau}, {Irwin}, \&
  {Mann}}]{Newton2015A}
{Newton}, E.~R., {Charbonneau}, D., {Irwin}, J., \& {Mann}, A.~W. 2015,
  \href{http://dx.doi.org/10.1088/0004-637X/800/2/85}{\JournalTitle{\apj}, 800,
  85}

\bibitem[{{Parviainen} \& {Aigrain}(2015)}]{2015MNRAS.453.3821P}
{Parviainen}, H., \& {Aigrain}, S. 2015,
  \href{http://dx.doi.org/10.1093/mnras/stv1857}{\JournalTitle{\mnras}, 453,
  3821}

\bibitem[{{Perryman} {et~al.}(2001){Perryman}, {de Boer}, {Gilmore}, {H{\o}g},
  {Lattanzi}, {Lindegren}, {Luri}, {Mignard}, {Pace}, \& {de
  Zeeuw}}]{2001A&A...369..339P}
{Perryman}, M.~A.~C., {de Boer}, K.~S., {Gilmore}, G., {et~al.} 2001,
  \href{http://dx.doi.org/10.1051/0004-6361:20010085}{\JournalTitle{\aap}, 369,
  339}

\bibitem[{{Pineda} {et~al.}(2013){Pineda}, {Bottom}, \&
  {Johnson}}]{Pineda:2013fk}
{Pineda}, J.~S., {Bottom}, M., \& {Johnson}, J.~A. 2013,
  \href{http://dx.doi.org/10.1088/0004-637X/767/1/28}{\JournalTitle{\apj}, 767,
  28}

\bibitem[{{Plavchan} {et~al.}(2014){Plavchan}, {Bilinski}, \&
  {Currie}}]{2014PASP..126...34P}
{Plavchan}, P., {Bilinski}, C., \& {Currie}, T. 2014,
  \href{http://dx.doi.org/10.1086/674819}{\JournalTitle{\pasp}, 126, 34}

\bibitem[{{Puget} {et~al.}(2004){Puget}, {Stadler}, {Doyon}, {Gigan},
  {Thibault}, {Luppino}, {Barrick}, {Benedict}, {Forveille}, {Rambold},
  {Thomas}, {Vermeulen}, {Ward}, {Beuzit}, {Feautrier}, {Magnard}, {Mella},
  {Preis}, {Vallee}, {Wang}, {Lin}, {Hall}, \& {Hodapp}}]{2004SPIE.5492..978P}
{Puget}, P., {Stadler}, E., {Doyon}, R., {et~al.} 2004,
  \href{http://dx.doi.org/10.1117/12.551097}{in \procspie, Vol. 5492,
  Ground-based Instrumentation for Astronomy, ed. A.~F.~M. {Moorwood} \&
  M.~{Iye}}, 978

\bibitem[{{Quintana} {et~al.}(2007){Quintana}, {Adams}, {Lissauer}, \&
  {Chambers}}]{2007ApJ...660..807Q}
{Quintana}, E.~V., {Adams}, F.~C., {Lissauer}, J.~J., \& {Chambers}, J.~E.
  2007, \href{http://dx.doi.org/10.1086/512542}{\JournalTitle{\apj}, 660, 807}

\bibitem[{{Quintana} {et~al.}(2014){Quintana}, {Barclay}, {Raymond}, {Rowe},
  {Bolmont}, {Caldwell}, {Howell}, {Kane}, {Huber}, {Crepp}, {Lissauer},
  {Ciardi}, {Coughlin}, {Everett}, {Henze}, {Horch}, {Isaacson}, {Ford},
  {Adams}, {Still}, {Hunter}, {Quarles}, \& {Selsis}}]{Quintana2014}
{Quintana}, E.~V., {Barclay}, T., {Raymond}, S.~N., {et~al.} 2014,
  \href{http://dx.doi.org/10.1126/science.1249403}{\JournalTitle{Science}, 344,
  277}

\bibitem[{{Ricker}(2014)}]{Ricker2014}
{Ricker}, G.~R. 2014, \JournalTitle{Journal of the American Association of
  Variable Star Observers (JAAVSO)}, 42, 234

\bibitem[{{Robin} {et~al.}(2003){Robin}, {Reyl{\'e}}, {Derri{\`e}re}, \&
  {Picaud}}]{2003A&A...409..523R}
{Robin}, A.~C., {Reyl{\'e}}, C., {Derri{\`e}re}, S., \& {Picaud}, S. 2003,
  \href{http://dx.doi.org/10.1051/0004-6361:20031117}{\JournalTitle{\aap}, 409,
  523}

\bibitem[{{Rojas-Ayala} {et~al.}(2010){Rojas-Ayala}, {Covey}, {Muirhead}, \&
  {Lloyd}}]{2010ApJ...720L.113R}
{Rojas-Ayala}, B., {Covey}, K.~R., {Muirhead}, P.~S., \& {Lloyd}, J.~P. 2010,
  \href{http://dx.doi.org/10.1088/2041-8205/720/1/L113}{\JournalTitle{\apjl},
  720, L113}

\bibitem[{{Rojas-Ayala} {et~al.}(2012){Rojas-Ayala}, {Covey}, {Muirhead}, \&
  {Lloyd}}]{Rojas-Ayala:2012uq}
---. 2012,
  \href{http://dx.doi.org/10.1088/0004-637X/748/2/93}{\JournalTitle{\apj}, 748,
  93}

\bibitem[{{Rowe} \& {Thompson}(2015)}]{2015arXiv150400707R}
{Rowe}, J.~F., \& {Thompson}, S.~E. 2015, \JournalTitle{ArXiv e-prints},
  \href{http://arxiv.org/abs/1504.00707}{{\sffamily arXiv:1504.00707
  [astro-ph.EP]}}

\bibitem[{{Rowe} {et~al.}(2014){Rowe}, {Bryson}, {Marcy}, {Lissauer},
  {Jontof-Hutter}, {Mullally}, {Gilliland}, {Issacson}, {Ford}, {Howell},
  {Borucki}, {Haas}, {Huber}, {Steffen}, {Thompson}, {Quintana}, {Barclay},
  {Still}, {Fortney}, {Gautier}, {Hunter}, {Caldwell}, {Ciardi}, {Devore},
  {Cochran}, {Jenkins}, {Agol}, {Carter}, \& {Geary}}]{Rowe2014}
{Rowe}, J.~F., {Bryson}, S.~T., {Marcy}, G.~W., {et~al.} 2014,
  \href{http://dx.doi.org/10.1088/0004-637X/784/1/45}{\JournalTitle{\apj}, 784,
  45}

\bibitem[{{Seager} \& {Mall{\'e}n-Ornelas}(2003)}]{Seager:2003lr}
{Seager}, S., \& {Mall{\'e}n-Ornelas}, G. 2003,
  \href{http://dx.doi.org/10.1086/346105}{\JournalTitle{\apj}, 585, 1038}

\bibitem[{{Silva Aguirre} {et~al.}(2015){Silva Aguirre}, {Davies}, {Basu},
  {Christensen-Dalsgaard}, {Creevey}, {Metcalfe}, {Bedding}, {Casagrande},
  {Handberg}, {Lund}, {Nissen}, {Chaplin}, {Huber}, {Serenelli}, {Stello}, {Van
  Eylen}, {Campante}, {Elsworth}, {Gilliland}, {Hekker}, {Karoff}, {Kawaler},
  {Kjeldsen}, \& {Lundkvist}}]{2015MNRAS.452.2127S}
{Silva Aguirre}, V., {Davies}, G.~R., {Basu}, S., {et~al.} 2015,
  \href{http://dx.doi.org/10.1093/mnras/stv1388}{\JournalTitle{\mnras}, 452,
  2127}

\bibitem[{{Skrutskie} {et~al.}(2006){Skrutskie}, {Cutri}, {Stiening},
  {Weinberg}, {Schneider}, {Carpenter}, {Beichman}, {Capps}, {Chester},
  {Elias}, {Huchra}, {Liebert}, {Lonsdale}, {Monet}, {Price}, {Seitzer},
  {Jarrett}, {Kirkpatrick}, {Gizis}, {Howard}, {Evans}, {Fowler}, {Fullmer},
  {Hurt}, {Light}, {Kopan}, {Marsh}, {McCallon}, {Tam}, {Van Dyk}, \&
  {Wheelock}}]{Skrutskie2006}
{Skrutskie}, M.~F., {Cutri}, R.~M., {Stiening}, R., {et~al.} 2006,
  \href{http://dx.doi.org/10.1086/498708}{\JournalTitle{\aj}, 131, 1163}

\bibitem[{{Spada} {et~al.}(2013){Spada}, {Demarque}, {Kim}, \&
  {Sills}}]{Spada2013}
{Spada}, F., {Demarque}, P., {Kim}, Y.-C., \& {Sills}, A. 2013,
  \href{http://dx.doi.org/10.1088/0004-637X/776/2/87}{\JournalTitle{\apj}, 776,
  87}

\bibitem[{{Still} \& {Barclay}(2012)}]{Still2012}
{Still}, M., \& {Barclay}, T. 2012, {PyKE: Reduction and analysis of Kepler
  Simple Aperture Photometry data}, Astrophysics Source Code Library,
  \href{http://arxiv.org/abs/1208.004}{{\sffamily ascl:1208.004}}

\bibitem[{{Stumpe} {et~al.}(2012){Stumpe}, {Smith}, {Van Cleve}, {Twicken},
  {Barclay}, {Fanelli}, {Girouard}, {Jenkins}, {Kolodziejczak}, {McCauliff}, \&
  {Morris}}]{Stumpe2012}
{Stumpe}, M.~C., {Smith}, J.~C., {Van Cleve}, J.~E., {et~al.} 2012,
  \href{http://dx.doi.org/10.1086/667698}{\JournalTitle{\pasp}, 124, 985}

\bibitem[{{Swift} {et~al.}(2015){Swift}, {Montet}, {Vanderburg}, {Morton},
  {Muirhead}, \& {Johnson}}]{Swift2015}
{Swift}, J.~J., {Montet}, B.~T., {Vanderburg}, A., {et~al.} 2015,
  \href{http://dx.doi.org/10.1088/0067-0049/218/2/26}{\JournalTitle{\apjs},
  218, 26}

\bibitem[{{Terrien} {et~al.}(2012){Terrien}, {Mahadevan}, {Bender},
  {Deshpande}, {Ramsey}, \& {Bochanski}}]{Terrien:2012lr}
{Terrien}, R.~C., {Mahadevan}, S., {Bender}, C.~F., {et~al.} 2012,
  \href{http://dx.doi.org/10.1088/2041-8205/747/2/L38}{\JournalTitle{\apjl},
  747, L38}

\bibitem[{{Torres} {et~al.}(2015){Torres}, {Kipping}, {Fressin}, {Caldwell},
  {Twicken}, {Ballard}, {Batalha}, {Bryson}, {Ciardi}, {Henze}, {Howell},
  {Isaacson}, {Jenkins}, {Muirhead}, {Newton}, {Petigura}, {Barclay},
  {Borucki}, {Crepp}, {Everett}, {Horch}, {Howard}, {Kolbl}, {Marcy},
  {McCauliff}, \& {Quintana}}]{Torres2015}
{Torres}, G., {Kipping}, D.~M., {Fressin}, F., {et~al.} 2015,
  \href{http://dx.doi.org/10.1088/0004-637X/800/2/99}{\JournalTitle{\apj}, 800,
  99}

\bibitem[{{Van Eylen} \& {Albrecht}(2015)}]{Van-Eylen2015}
{Van Eylen}, V., \& {Albrecht}, S. 2015,
  \href{http://dx.doi.org/10.1088/0004-637X/808/2/126}{\JournalTitle{\apj},
  808, 126}

\bibitem[{{von Braun} {et~al.}(2014){von Braun}, {Boyajian}, {van Belle},
  {Kane}, {Jones}, {Farrington}, {Schaefer}, {Vargas}, {Scott}, {ten
  Brummelaar}, {Kephart}, {Gies}, {Ciardi}, {L{\'o}pez-Morales}, {Mazingue},
  {McAlister}, {Ridgway}, {Goldfinger}, {Turner}, \&
  {Sturmann}}]{2014MNRAS.438.2413V}
{von Braun}, K., {Boyajian}, T.~S., {van Belle}, G.~T., {et~al.} 2014,
  \href{http://dx.doi.org/10.1093/mnras/stt2360}{\JournalTitle{\mnras}, 438,
  2413}

\bibitem[{{Wang} {et~al.}(2015){Wang}, {Fischer}, {Horch}, \&
  {Xie}}]{2015ApJ...806..248W}
{Wang}, J., {Fischer}, D.~A., {Horch}, E.~P., \& {Xie}, J.-W. 2015,
  \href{http://dx.doi.org/10.1088/0004-637X/806/2/248}{\JournalTitle{\apj},
  806, 248}

\bibitem[{{Wang} {et~al.}(2014){Wang}, {Fischer}, {Xie}, \&
  {Ciardi}}]{2014ApJ...791..111W}
{Wang}, J., {Fischer}, D.~A., {Xie}, J.-W., \& {Ciardi}, D.~R. 2014,
  \href{http://dx.doi.org/10.1088/0004-637X/791/2/111}{\JournalTitle{\apj},
  791, 111}

\bibitem[{{Weiss} \& {Marcy}(2014)}]{2014ApJ...783L...6W}
{Weiss}, L.~M., \& {Marcy}, G.~W. 2014,
  \href{http://dx.doi.org/10.1088/2041-8205/783/1/L6}{\JournalTitle{\apjl},
  783, L6}

\bibitem[{{Wu} \& {Lithwick}(2013)}]{2013ApJ...772...74W}
{Wu}, Y., \& {Lithwick}, Y. 2013,
  \href{http://dx.doi.org/10.1088/0004-637X/772/1/74}{\JournalTitle{\apj}, 772,
  74}

\bibitem[{{Xie} {et~al.}(2016){Xie}, {Dong}, {Zhu}, {Huber}, {Zheng}, {De Cat},
  {Fu}, {Liu}, {Luo}, {Wu}, {Zhang}, {Zhang}, {Zhou}, {Cao}, {Hou}, {Wang}, \&
  {Zhang}}]{2016PNAS..11311431X}
{Xie}, J.-W., {Dong}, S., {Zhu}, Z., {et~al.} 2016,
  \href{http://dx.doi.org/10.1073/pnas.1604692113}{\JournalTitle{Proceedings of
  the National Academy of Science}, 113, 11431}

\bibitem[{{Zhou} {et~al.}(2014){Zhou}, {Bayliss}, {Hartman}, {Bakos}, {Penev},
  {Csubry}, {Tan}, {Jord{\'a}n}, {Mancini}, {Rabus}, {Brahm}, {Espinoza},
  {Mohler-Fischer}, {Ciceri}, {Suc}, {Cs{\'a}k}, {Henning}, \&
  {Schmidt}}]{2014MNRAS.437.2831Z}
{Zhou}, G., {Bayliss}, D., {Hartman}, J.~D., {et~al.} 2014,
  \href{http://dx.doi.org/10.1093/mnras/stt2100}{\JournalTitle{\mnras}, 437,
  2831}

\end{thebibliography}

\clearpage
\rotate
\begin{deluxetable*}{lcccclllclllcc}
\setlength{\tabcolsep}{0.050in}
\tabletypesize{\tiny}
\tabletypesize{\scriptsize}
\tablecaption{CFHT/WIRCam Parallax and Proper Motion Results \label{tab:plx}}
\tablewidth{0pt}
\tablehead{                            
\colhead{} &
\colhead{} &
\colhead{} &
\colhead{} &
\colhead{} &
\multicolumn{3}{c}{Relative} & 
\colhead{} &
\multicolumn{3}{c}{Absolute} & 
\colhead{} &
\colhead{} \\
\cline{6-8}
\cline{10-12}
\colhead{KOI \#} &
\colhead{$\alpha_{J2000}$} &
\colhead{$\delta_{J2000}$} &
\colhead{Epoch} &
\colhead{} &
\colhead{$\pi_{\rm rel}$} &
\colhead{$\mu_{\alpha, {\rm rel}}\cos{\delta}$} &
\colhead{$\mu_{\delta, {\rm rel}}$} &
\colhead{} &
\colhead{$\pi_{\rm abs}$} &
\colhead{$\mu_{\alpha, {\rm abs}}\cos{\delta}$} &
\colhead{$\mu_{\delta, {\rm abs}}$} &
\colhead{} &
\colhead{$\chi^2$/dof} \\
\colhead{} &
\colhead{(deg)} &
\colhead{(deg)} &
\colhead{(MJD)} &
\colhead{} &
\colhead{(\arcsec)} &
\colhead{(\arcsec\ yr$^{-1}$)} &
\colhead{(\arcsec\ yr$^{-1}$)} &
\colhead{} &
\colhead{(\arcsec)} &
\colhead{(\arcsec\ yr$^{-1}$)} &
\colhead{(\arcsec\ yr$^{-1}$)} &
\colhead{} &
\colhead{}}
\startdata
1725A     & 283.628483 & $+$48.390960              & 56170.29 && 0.0130(12)    &   $-$0.0109(11)    &   $-$0.0447(13)    &&  0.0145(12)    &   $-$0.0187(15)    &   $-$0.0460(16)    &&     10.8/21     \\
1725B\tablenotemark{a}& 283.630162 & $+$48.390793 & 56170.29 && 0.0129(11)    &   $-$0.0131(11)    &   $-$0.0433(12)    &&  0.0143(11)    &   $-$0.0207(15)    &   $-$0.0448(15)    &&     17.8/21     \\
314      & 290.381624 & $+$43.293224              & 56019.65 && 0.0176(14)    &   $-$0.0165(14)    &  \phs0.0246(15)    &&  0.0191(14)    &   $-$0.0236(20)    &  \phs0.0229(19)    &&     14.4/15     \\
2705A     & 290.956619 & $+$49.366951              & 56409.60 && 0.0072(13)    &  \phs0.0556(17)    &  \phs0.1276(21)    &&  0.0086(13)    &  \phs0.0483(21)    &  \phs0.1261(23)    &&     16.6/13     \\
2705B\tablenotemark{b}& 290.955954 & $+$49.367244 & 56409.60 && 0.0069(14)    &  \phs0.0575(17)    &  \phs0.1277(19)    &&  0.0083(15)    &  \phs0.0505(20)    &  \phs0.1261(21)    &&     22.0/13     \\
961      & 292.219673 & $+$44.617832              & 56023.64 && 0.0232(10)\phn &  \phs0.0934(10)    &   $-$0.4124(14)    && 0.0247(10)\phn &  \phs0.0863(20)    &   $-$0.4143(17)    &&     21.6/19     \\
2453B\tablenotemark{c}& 294.466973 & $+$44.751947 & 56409.63 && 0.0066(13)    &   $-$0.0024(15)    &  \phs0.1751(20)    &&  0.0082(13)    &   $-$0.0096(19)    &  \phs0.1738(22)    && \phn 8.5/13     \\
2453A     & 294.468887 & $+$44.754908              & 56409.63 && 0.0061(13)    &   $-$0.0011(18)    &  \phs0.1752(19)    &&  0.0077(13)    &   $-$0.0083(22)    &  \phs0.1739(20)    && \phn 9.5/13     \\
1702     & 297.729464 & $+$42.866881              & 56151.39 && 0.0070(11)    &   $-$0.0000(11)    &   $-$0.0454(13)    &&  0.0083(11)    &   $-$0.0072(12)    &   $-$0.0471(14)    &&     24.8/19     \\
2704     & 298.736673 & $+$46.499255              & 56410.62 && 0.0050(14)    &  \phs0.0438(18)    &  \phs0.1387(19)    &&  0.0066(14)    &  \phs0.0361(21)    &  \phs0.1366(21)    &&     17.2/15     \\
463B      & 300.206877 & $+$45.018466              & 56175.35 && 0.0070(15)    &  \phs0.0889(14)    &  \phs0.0239(18)    &&  0.0085(15)    &  \phs0.0827(16)    &  \phs0.0217(19)    &&     15.6/15     \\
463A\tablenotemark{d} & 300.222476 & $+$45.016320 & 56175.35 && 0.0094(12)    &  \phs0.0871(10)\phn &  \phs0.0244(12)    && 0.0109(12)    &  \phs0.0809(13)    &  \phs0.0222(14)    &&      16.5/15     \\
\enddata

\tablecomments{This table gives all the astrometric parameters derived
  from our MCMC analysis for each target. For parameters in units of arcseconds, errors are given in parentheses in units of $10^{-4}$~arcsec. ($\alpha$, $\delta$, MJD):~Coordinates that correspond to the epoch listed, which is the
  first epoch of our observations for that target.  ($\pi$, $\mu_{\alpha}\cos{\delta}$, $\mu_{\delta}$):~Parallax and proper motion parameters are listed as both as relative and absolute values.  Relative values are our directly fitted results.  Absolute
  values include a correction for the mean parallax and proper motion
  of our reference stars, which includes an additional uncertainty
  added in quadrature to our direct fitting results (see
  Section~\ref{sec:obs}).  $\chi^2$/dof: The lowest $\chi^2$ in each set
  of MCMC chains along with the degrees of freedom (dof).}

\tablenotetext{a}{KIC~10905748}
\tablenotetext{b}{KIC~11453591}
\tablenotetext{c}{KIC~8631743}
\tablenotetext{d}{KIC~8845251; KOI-463B is the planet host. }
\end{deluxetable*}

\rotate
\renewcommand{\arraystretch}{1.3}
\begin{deluxetable*}{l l c c c c c c c c c c c c c c c}
\tablecaption{Planet and Planet-candidate Parameters from New Transit Fits}
\tablehead{
\colhead{KOI} & \colhead{Planet } & \colhead{$T_0$} & \colhead{$P$} & \colhead{$R_P / R_\star$}  & \colhead{$|b|$}  & \colhead{$\sqrt{e}\cos(\omega)$} & \colhead{$\sqrt{e}\sin(\omega)$} & \colhead{$\rho_\star$} & \colhead{$e$\tablenotemark{a} (mode)} & \colhead{$e$ (68\%)}& \colhead{$R_P$\tablenotemark{b}} \\
\colhead{} & \colhead{} & \colhead{(BJD-2454833)} & \colhead{(days)} & \colhead{(\%)} & \colhead{} & \colhead{} & \colhead{} & \colhead{($\rho_\odot$)} & \colhead{} & &  \colhead{($R_\earth$)} }
\startdata
314.01 & Kepler-138 c & $122.72666^{+0.00049}_{-0.00049}$ & $13.78110882^{+0.00000750}_{-0.00000745}$ & $2.367^{+0.048}_{-0.044}$ & $0.50^{+0.10}_{-0.15}$ & $0.05^{+0.41}_{-0.43}$ & $-0.01^{+0.13}_{-0.12}$ & $4.89^{+0.40}_{-0.41}$ & 0.01 & [0.00, 0.19] & $1.17^{+0.07}_{-0.07}$ \\
314.02 & Kepler-138 d & $124.83521^{+0.00086}_{-0.00084}$ & $23.0889203^{+0.0000232}_{-0.0000247}$ & $2.42^{+0.21}_{-0.13}$ & $0.85^{+0.08}_{-0.08}$ & $0.00^{+0.17}_{-0.17}$ & $0.01^{+0.31}_{-0.35}$ & $4.89^{+0.40}_{-0.41}$ & 0.01 & [0.00, 0.15] & $1.20^{+0.11}_{-0.10}$ \\
314.03 & Kepler-138 b & $123.20467^{+0.00386}_{-0.00349}$ & $10.3132068^{+0.00003440}_{-0.0000451}$ & $0.933^{+0.044}_{-0.042}$ & $0.61^{+0.10}_{-0.21}$ & $-0.06^{+0.37}_{-0.34}$ & $-0.01^{+0.27}_{-0.23}$ & $4.89^{+0.40}_{-0.41}$ & 0.00 & [0.00, 0.22] & $0.46^{+0.03}_{-0.03}$ \\
463.01 & Kepler-560 b & $148.31404^{+0.00068}_{-0.00068}$ & $18.4776196^{+0.0000130}_{-0.0000132}$ & $4.646^{+0.202}_{-0.108}$ & $0.46^{+0.22}_{-0.30}$ & $0.03^{+0.49}_{-0.54}$ & $0.02^{+0.25}_{-0.27}$ & $7.11^{+0.88}_{-0.89}$ & 0.17 & [0.13, 0.24] & $1.93^{+0.15}_{-0.14}$ \\
961.01 & Kepler-42 b & $131.64297^{+0.00021}_{-0.00020}$ & $1.21377060^{+0.00000023}_{-0.00000025}$ & $3.993^{+0.069}_{-0.069}$ & $0.53^{+0.08}_{-0.12}$ & $-0.02^{+0.14}_{-0.12}$ & $-0.22^{+0.64}_{-0.28}$ & $27.06^{+1.47}_{-1.41}$ & 0.04 & [0.00, 0.24] & $0.76^{+0.03}_{-0.03}$ \\
961.02 & Kepler-42 c & $131.60346^{+0.00013}_{-0.00013}$ & $0.45328731^{+0.00000005}_{-0.00000005}$ & $3.837^{+0.058}_{-0.058}$ & $0.43^{+0.11}_{-0.14}$ & $-0.02^{+0.32}_{-0.26}$ & $-0.03^{+0.12}_{-0.13}$ & $27.06^{+1.47}_{-1.41}$ & 0.00 & [0.00, 0.10] & $0.73^{+0.03}_{-0.03}$ \\
961.03 & Kepler-42 d & $131.92800^{+0.00039}_{-0.00041}$ & $1.86511236^{+0.00000075}_{-0.00000071}$ & $3.493^{+0.141}_{-0.133}$ & $0.71^{+0.09}_{-0.15}$ & $0.02^{+0.49}_{-0.34}$ & $-0.14^{+0.20}_{-0.19}$ & $27.06^{+1.47}_{-1.41}$ & 0.02 & [0.00, 0.24] & $0.67^{+0.04}_{-0.03}$ \\
1702.01 &  & $168.73959^{+0.00088}_{-0.00088}$ & $1.53818001^{+0.00000155}_{-0.00000154}$ & $2.625^{+0.094}_{-0.053}$ & $0.31^{+0.28}_{-0.21}$ & $-0.01^{+0.33}_{-0.32}$ & $0.00^{+0.20}_{-0.19}$ & $8.56^{+1.48}_{-1.46}$ & 0.01 & [0.00, 0.16] & $0.96^{+0.11}_{-0.10}$ \\
1725.01 &  & $128.53395^{+0.00126}_{-0.00134}$ & $9.87863917^{+0.00001067}_{-0.00001010}$ & $3.355^{+0.098}_{-0.072}$ & $0.45^{+0.17}_{-0.26}$ & $0.12^{+0.41}_{-0.53}$ & $0.04^{+0.22}_{-0.25}$ & $4.12^{+0.44}_{-0.44}$ & 0.13 & [0.10, 0.19] & $1.84^{+0.11}_{-0.11}$ \\
2453.01 &  & $131.86771^{+0.00121}_{-0.00112}$ & $1.53051509^{+0.00000202}_{-0.00000215}$ & $2.565^{+0.267}_{-0.179}$ & $0.59^{+0.26}_{-0.39}$ & $0.00^{+0.63}_{-0.62}$ & $0.01^{+0.39}_{-0.39}$ & $10.42^{+2.57}_{-2.53}$ & 0.33 & [0.26, 0.43] & $0.82^{+0.14}_{-0.12}$ \\
2704.01 & Kepler-445 c & $537.95379^{+0.00138}_{-0.00137}$ & $4.87122714^{+0.00000636}_{-0.00000638}$ & $7.17^{+0.21}_{-0.14}$ & $0.33^{+0.22}_{-0.22}$ & $-0.01^{+0.34}_{-0.34}$ & $-0.00^{+0.16}_{-0.16}$ & $8.46^{+1.87}_{-1.72}$ & 0.01 & [0.00, 0.17] & $2.72^{+0.44}_{-0.43}$ \\
2704.02 & Kepler-445 b & $538.96375^{+0.00292}_{-0.00290}$ & $2.98416640^{+0.00000891}_{-0.00000936}$ & $4.56^{+0.23}_{-0.16}$ & $0.40^{+0.24}_{-0.26}$ & $0.01^{+0.22}_{-0.21}$ & $0.02^{+0.34}_{-0.36}$ & $8.46^{+1.87}_{-1.72}$ & 0.02 & [0.00, 0.18] & $1.74^{+0.29}_{-0.28}$ \\
2704.03 & Kepler-445 d & $533.67969^{+0.00928}_{-0.00940}$ & $8.15272856^{+0.00006453}_{-0.00007041}$ & $3.50^{+0.30}_{-0.25}$ & $0.43^{+0.26}_{-0.28}$ & $-0.00^{+0.35}_{-0.34}$ & $0.01^{+0.19}_{-0.19}$ & $8.46^{+1.87}_{-1.72}$ & 0.01 & [0.00, 0.17] & $1.33^{+0.25}_{-0.23}$ \\
2705.01A & Kepler-1319A b & $538.69953^{+0.00136}_{-0.00128}$ & $2.88676239^{+0.00000377}_{-0.00000377}$ & $2.439^{+0.280}_{-0.126}$ & $0.65^{+0.22}_{-0.41}$ & $-0.09^{+0.59}_{-0.52}$ & $-0.05^{+0.38}_{-0.36}$ & $3.60^{+0.71}_{-0.73}$ & 0.33 & [ 0.25, 0.40] & $1.41^{+0.20}_{-0.17}$ \\
2705.01B & Kepler-1319B b & $538.69938^{+0.00128}_{-0.00126}$ & $2.88676250^{+0.00000367}_{-0.00000372}$ & $13.390^{+0.889}_{-0.565}$ & $0.40^{+0.28}_{-0.27}$ & $-0.01^{+0.45}_{-0.46}$ & $0.01^{+0.23}_{-0.22}$ & $20.80^{+4.43}_{-4.46}$ & 0.03 & [ 0.00, 0.26] & $2.95^{+0.47}_{-0.44}$ \\
\enddata
\tablenotetext{a}{$e$ is calculated from $\sqrt{e}\sin(\omega)$ and $\sqrt{e}\cos(\omega)$ after the MCMC is complete.}
\tablenotetext{b}{$R_P$ is calculated by combining the posteriors from the transit fit and the stellar radius (Section~\ref{sec:params} and Table~\ref{tab:stellar}).}
\tablenotetext{c}{Planet parameters are listed assuming they orbit the primary and secondary separately.}
\label{tab:planets}
\end{deluxetable*}

\clearpage

\appendix

\setcounter{figure}{0}  
\renewcommand\thefigure{A.\arabic{figure}}    

\begin{figure*}[h]
  \centering
\includegraphics[width=0.45\textwidth]{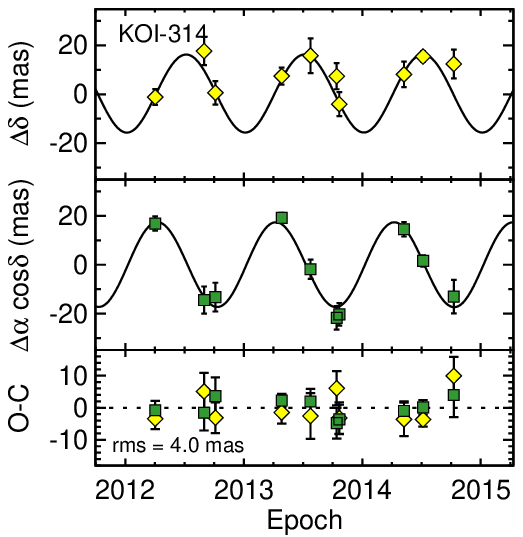}
\includegraphics[width=0.45\textwidth]{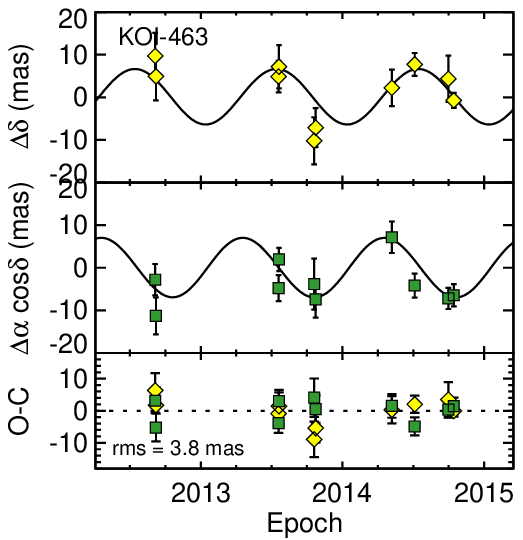}
\includegraphics[width=0.45\textwidth]{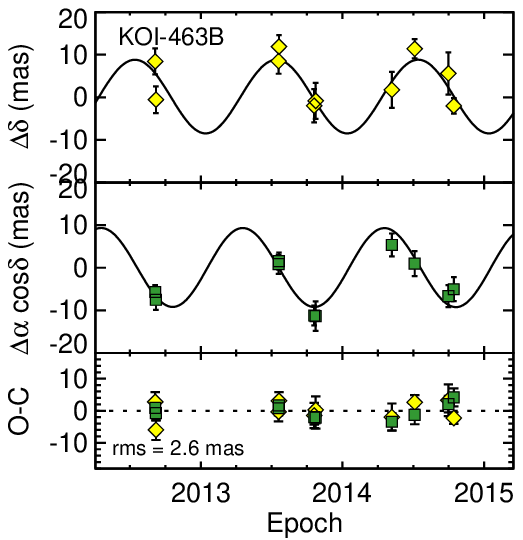}
\includegraphics[width=0.45\textwidth]{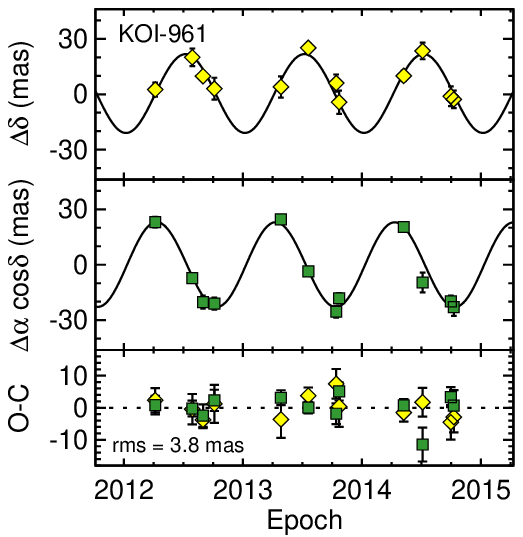}
\caption{Relative astrometry for each object as a function of their Julian year. Within each figure, the top panel shows $\delta$ (yellow diamonds), the middle $\alpha$ (green squares), and the bottom the residuals for both, with the RMS of the fit noted. We fit proper motion and parallax simultaneously; however, the proper motion has been removed from these plots for simplicity. Each figure represents a different object, including 'B' components for confirmed binaries.}
 \label{fig:plx}
\end{figure*}

\setcounter{figure}{0}  

\begin{figure*}
  \centering
\includegraphics[width=0.45\textwidth]{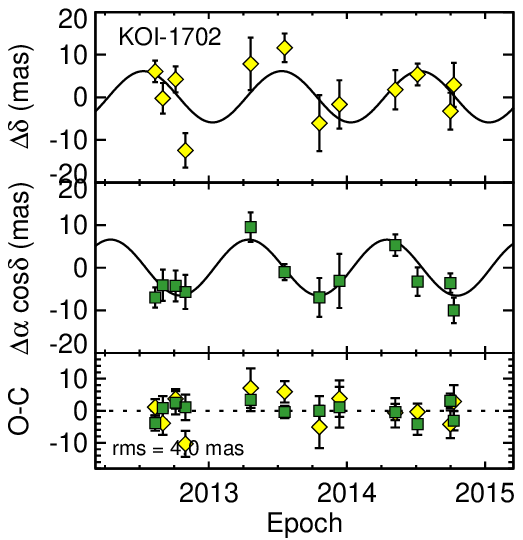}
\includegraphics[width=0.45\textwidth]{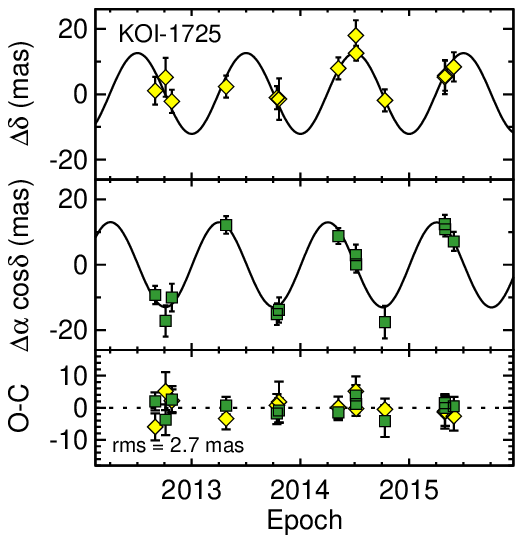}
\includegraphics[width=0.45\textwidth]{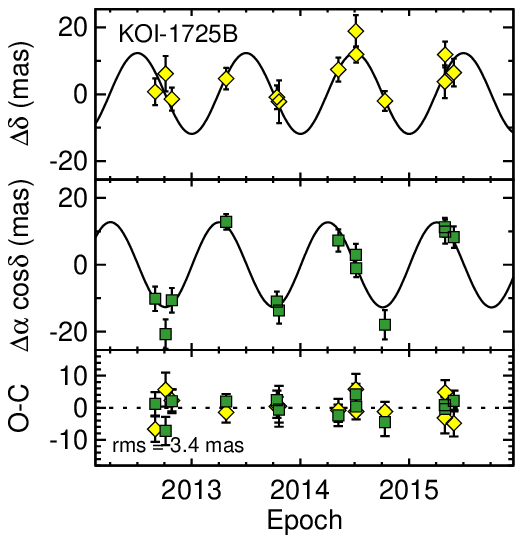}
\includegraphics[width=0.45\textwidth]{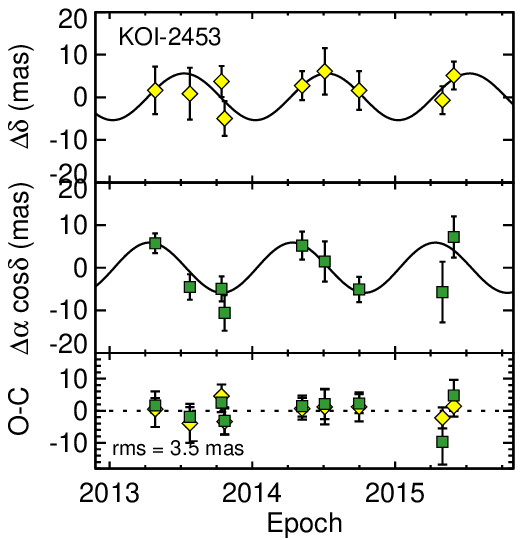}
\caption{Continued}
 \label{fig:plx}
\end{figure*}

\setcounter{figure}{0}  

\begin{figure*}
  \centering
  \includegraphics[width=0.45\textwidth]{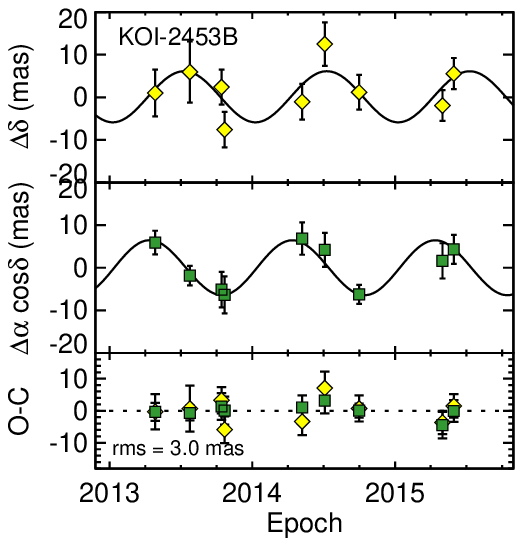}
\includegraphics[width=0.45\textwidth]{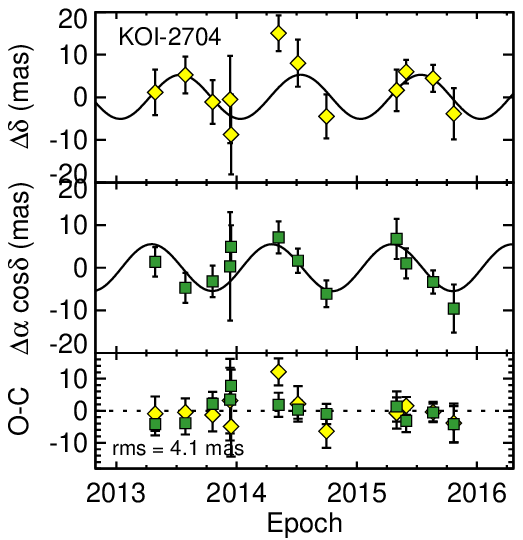}
\includegraphics[width=0.45\textwidth]{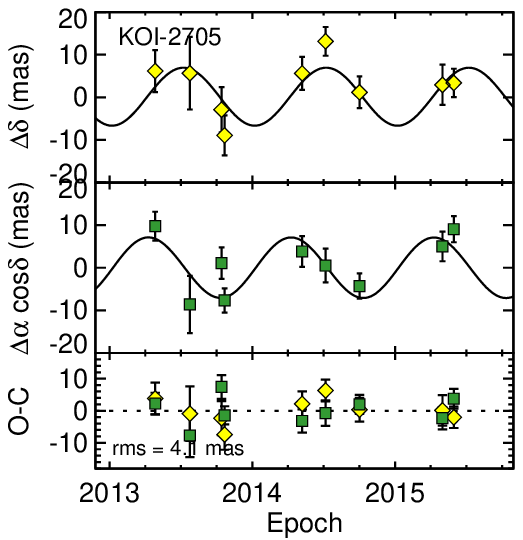}
\includegraphics[width=0.45\textwidth]{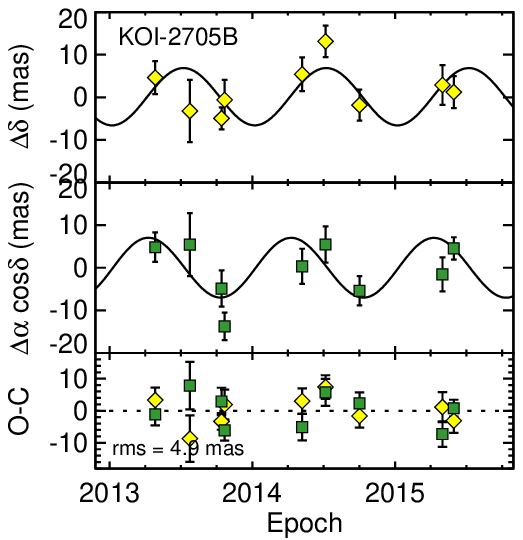}
  \caption{}
 \label{fig:plx}
\end{figure*}

\begin{figure*}
\includegraphics[width=0.5\textwidth]{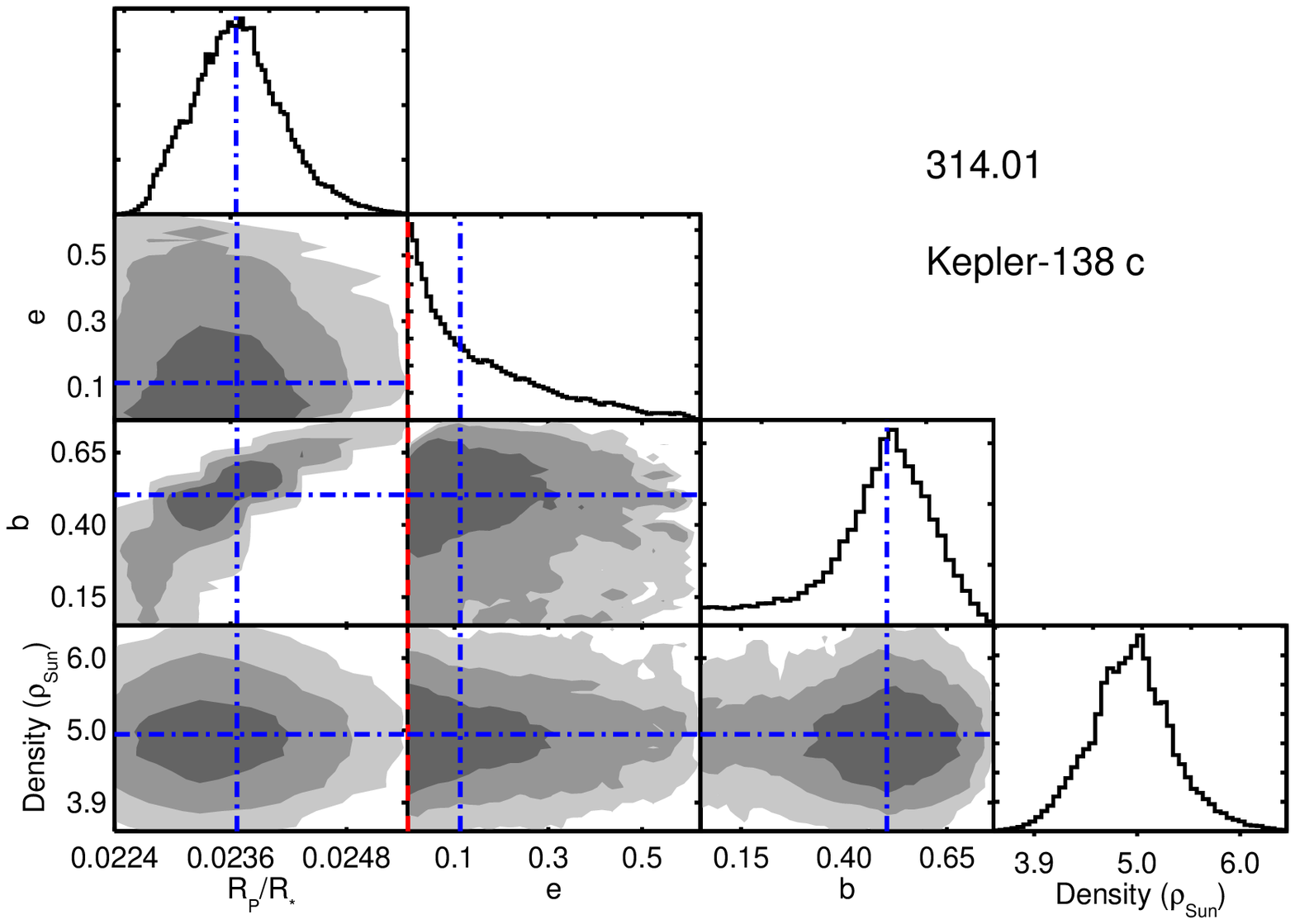}
\includegraphics[width=0.5\textwidth]{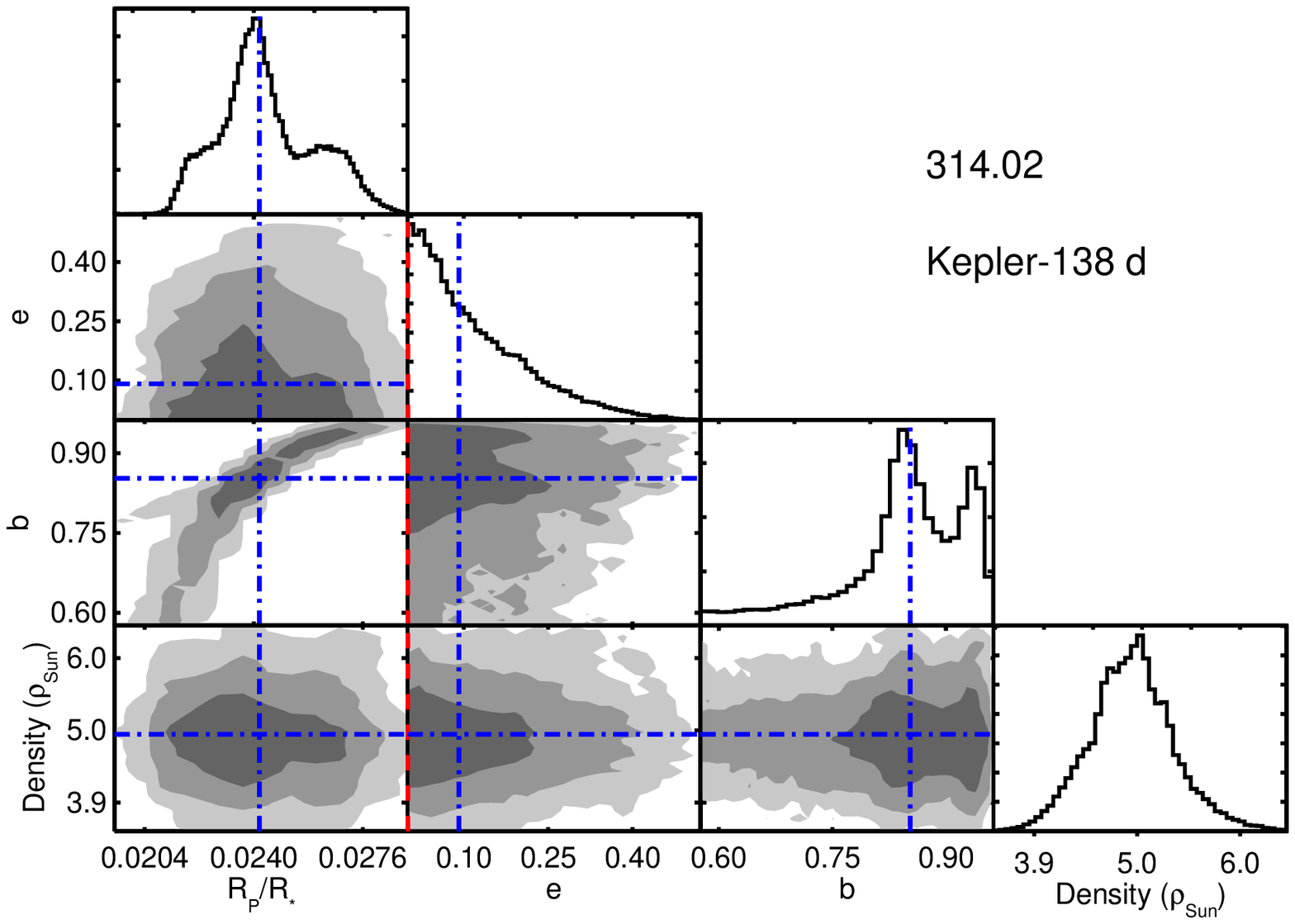}
\includegraphics[width=0.5\textwidth]{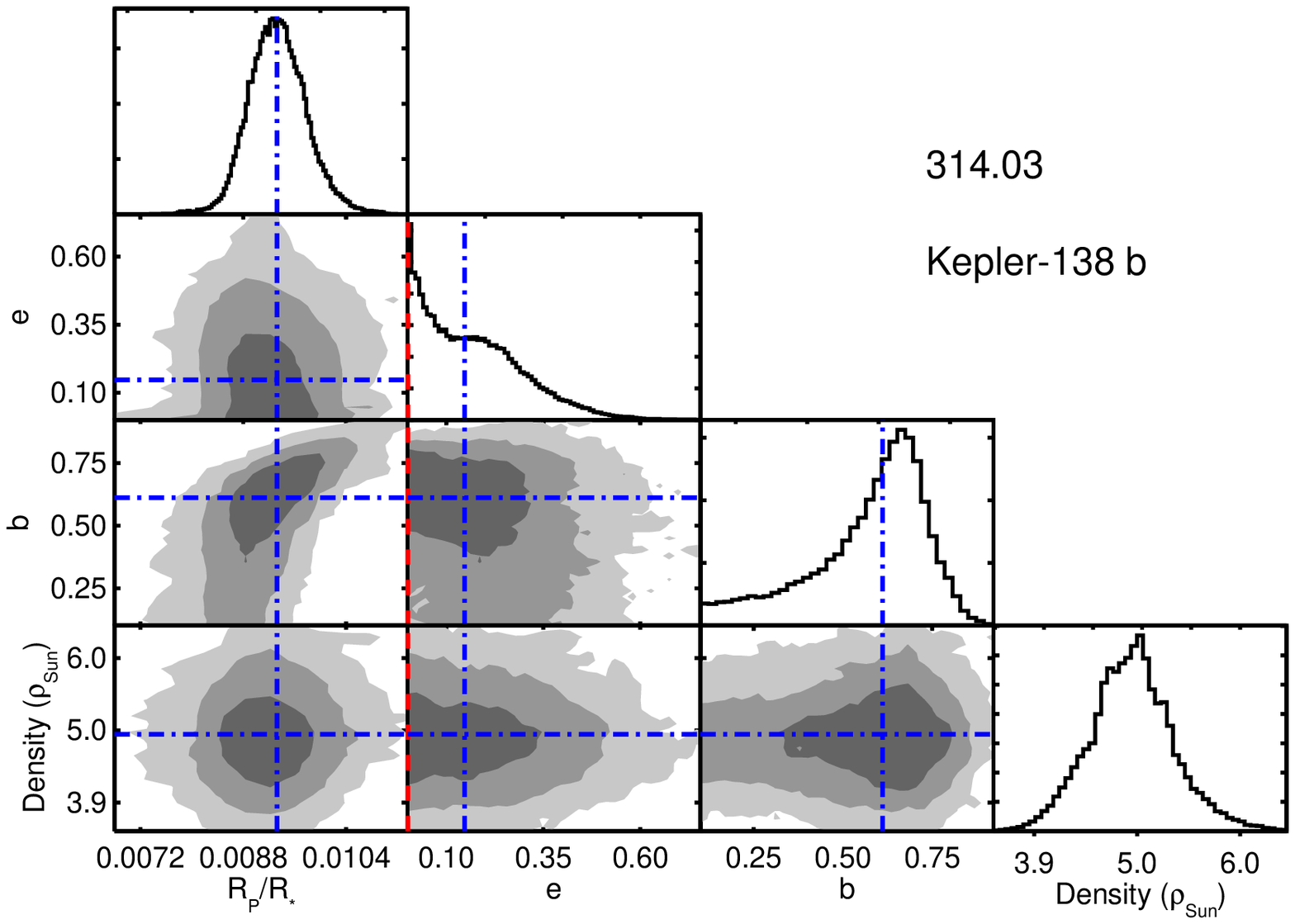}
\includegraphics[width=0.5\textwidth]{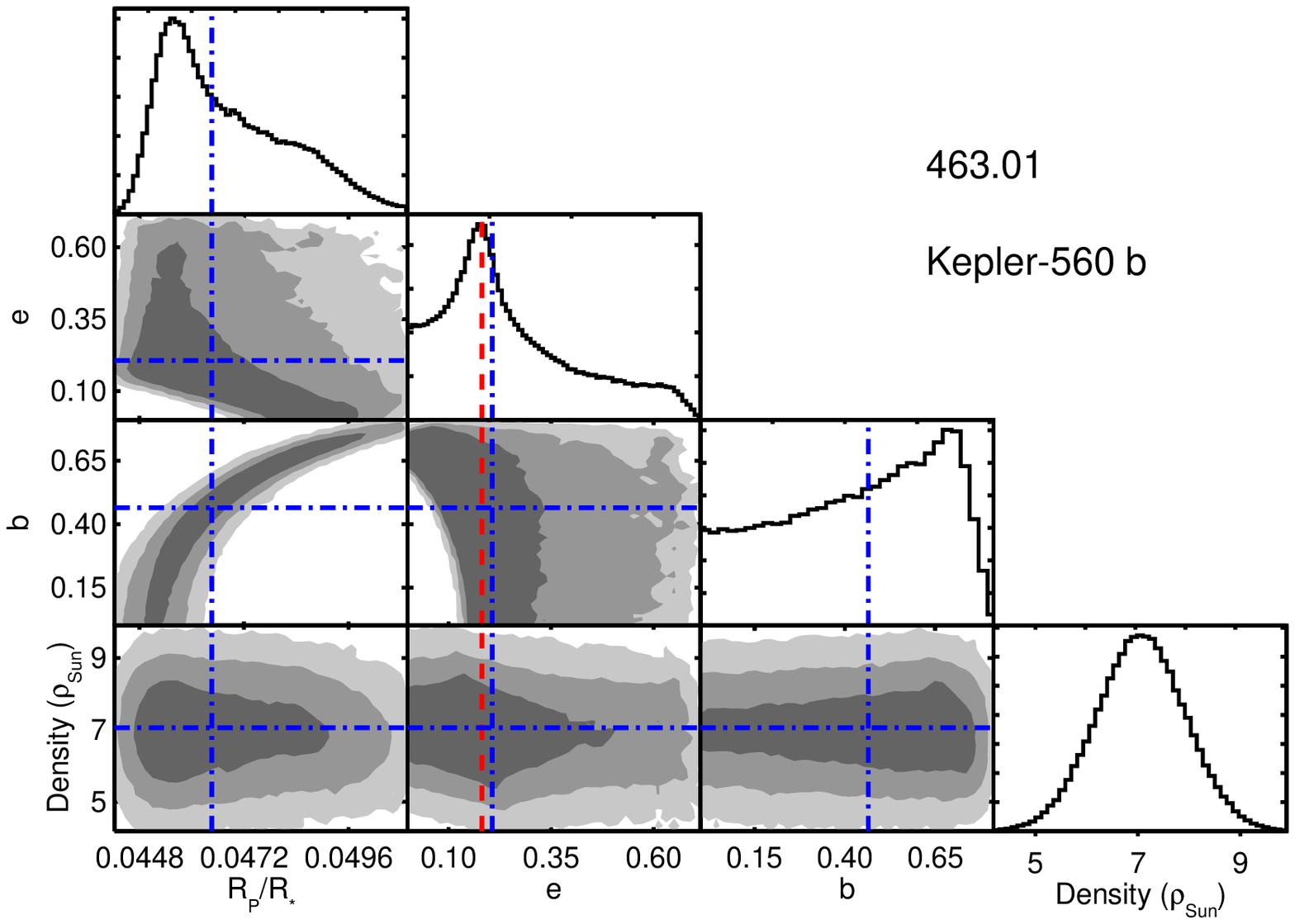}
\includegraphics[width=0.5\textwidth]{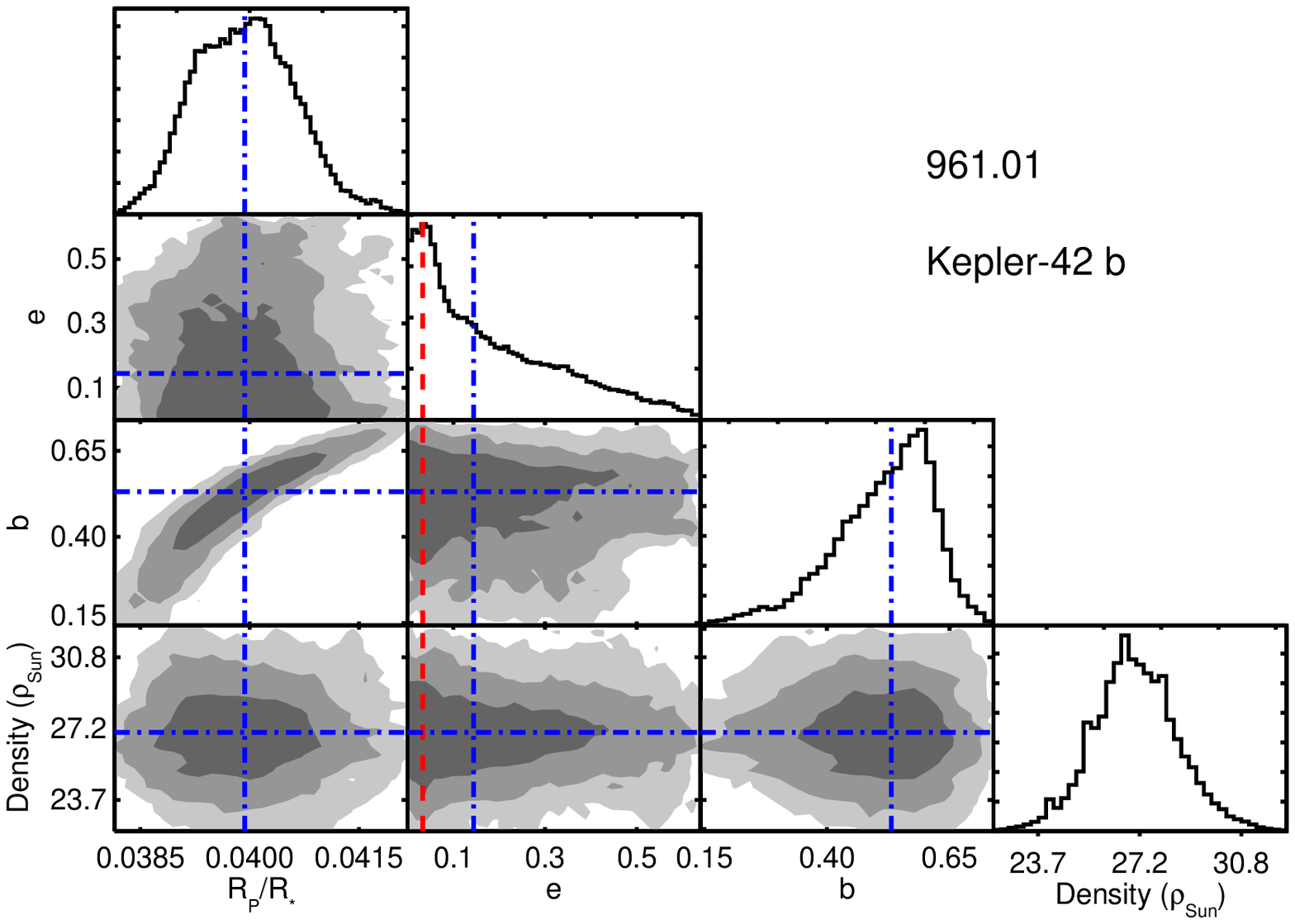}
\includegraphics[width=0.5\textwidth]{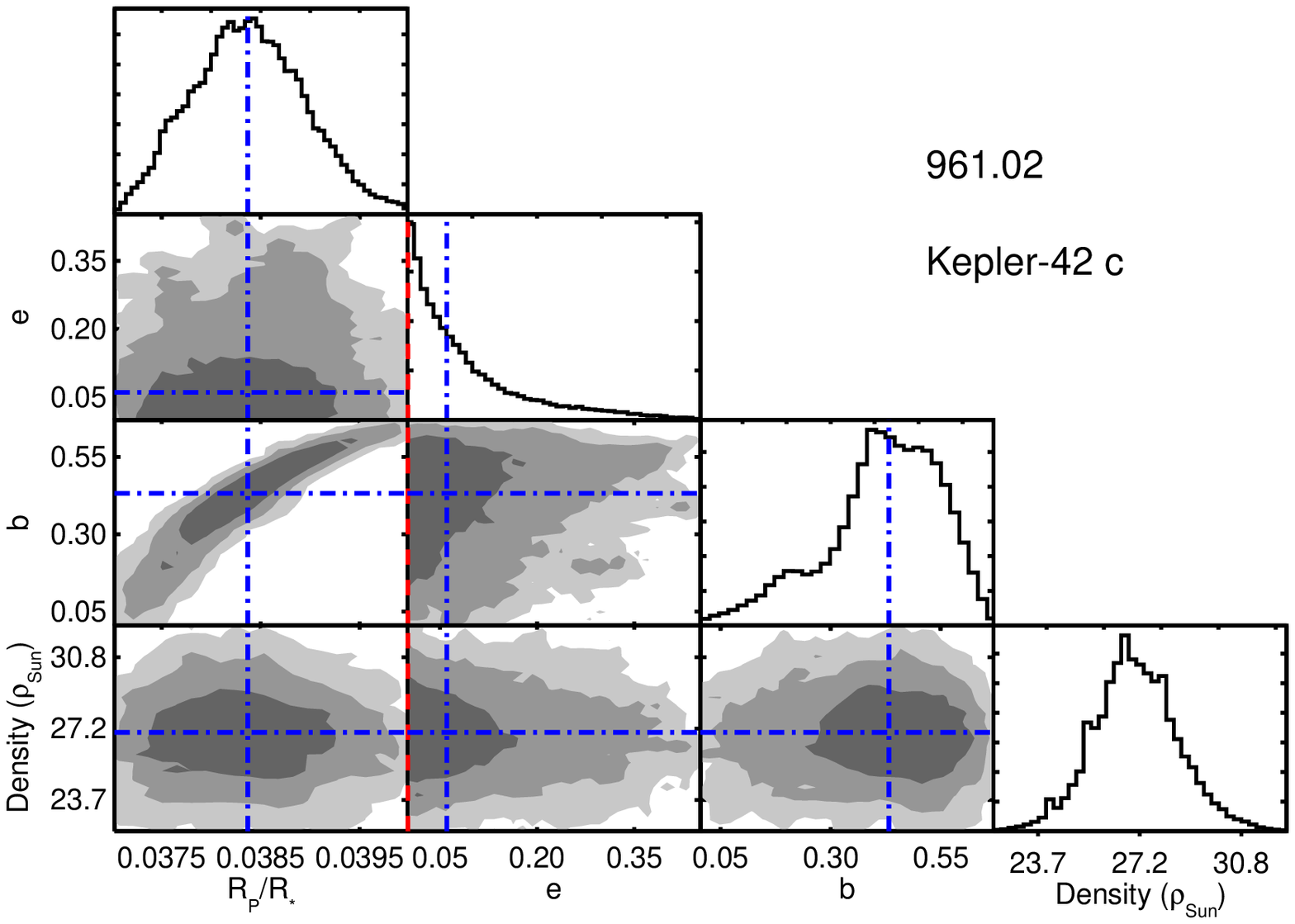}
\caption{Posteriors and correlations for $R_P/R_\star$, $b$, $e$ and $\rho_\star$ for all systems. Gray regions surround 68\%, 95\%, and 99.7\% of the data points (1, 2, and 3$\sigma$ for Gaussian distributions) from darkest to lightest. Values of $b$ tend to be symmetric around zero, so the absolute value is shown. The red dashed line indicates the statistical mode, and the dotted-dashed blue line the median. Planet names are shown in the upper right corner. Axis ranges are selected to exclude the most extreme 0.1\% of points to make the remaining points more clear, so some contours are cut on the edges. Note that $e$ is not fit in the MCMC, but calculated from the $\sqrt{e}\sin(\omega)$ and $\sqrt{e}\cos(\omega)$ distributions.}
 \label{fig:transitfit}
\end{figure*}

\setcounter{figure}{1}  

\begin{figure*}
\includegraphics[width=0.5\textwidth]{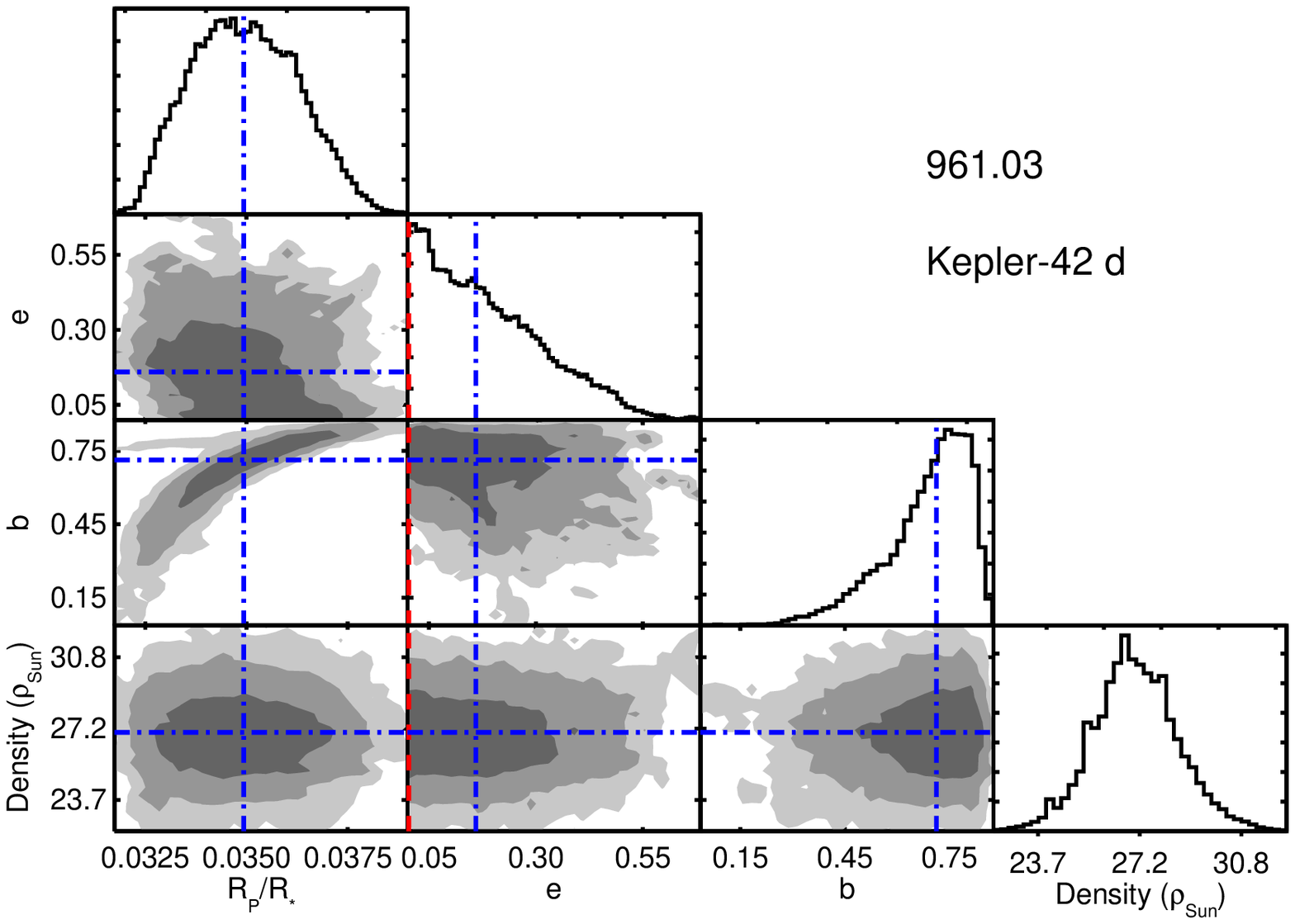}
\includegraphics[width=0.5\textwidth]{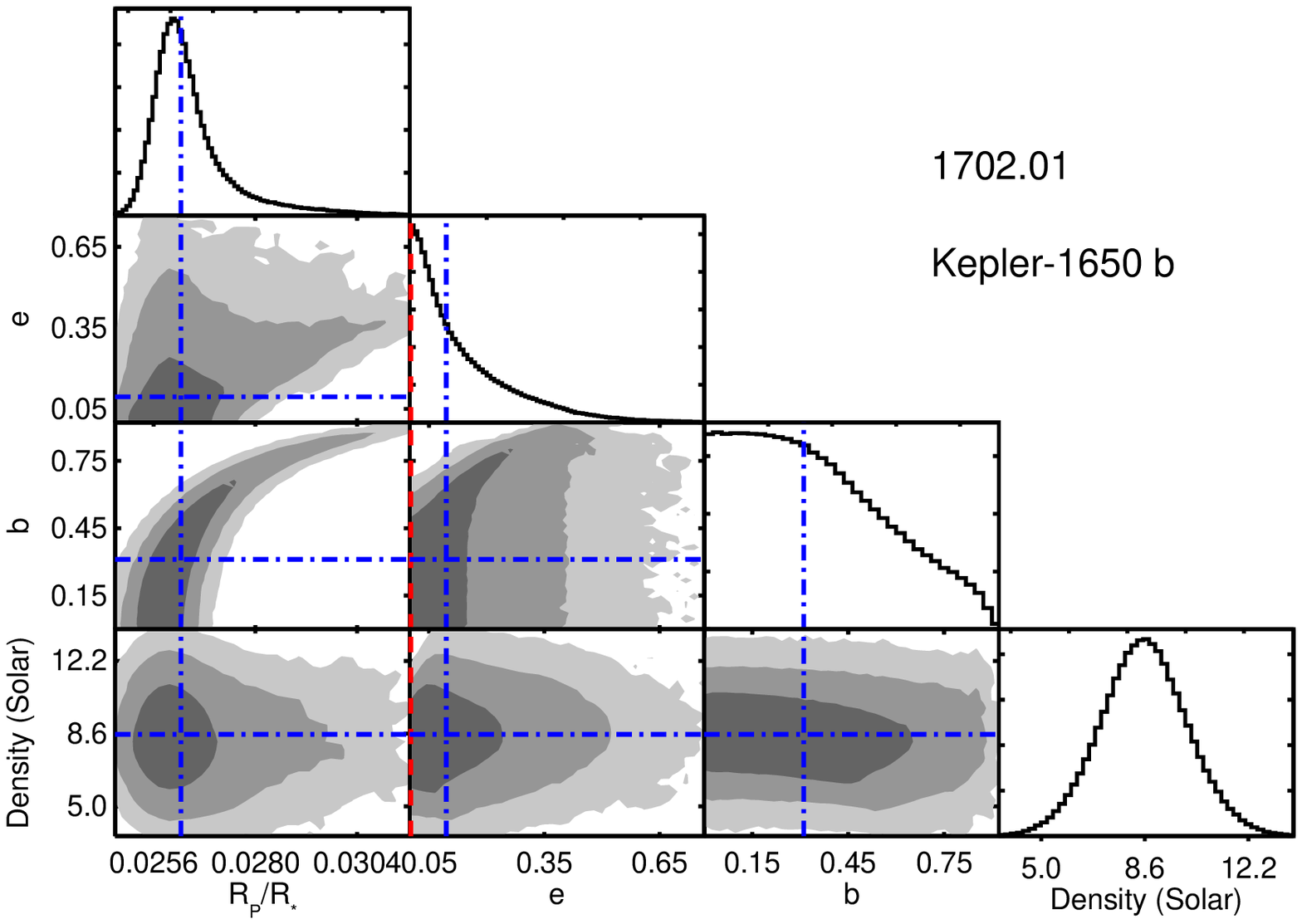}
\includegraphics[width=0.5\textwidth]{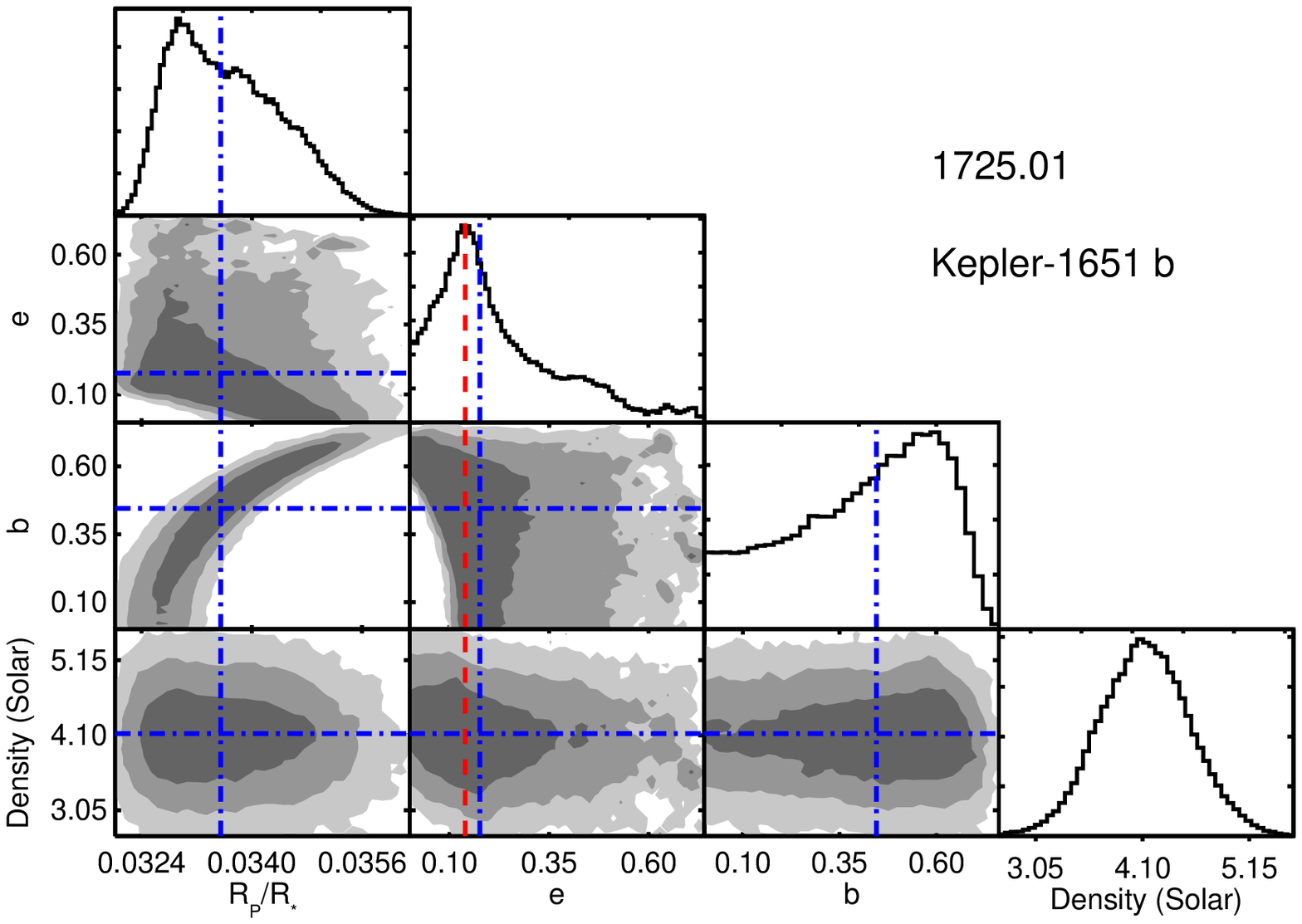}
\includegraphics[width=0.5\textwidth]{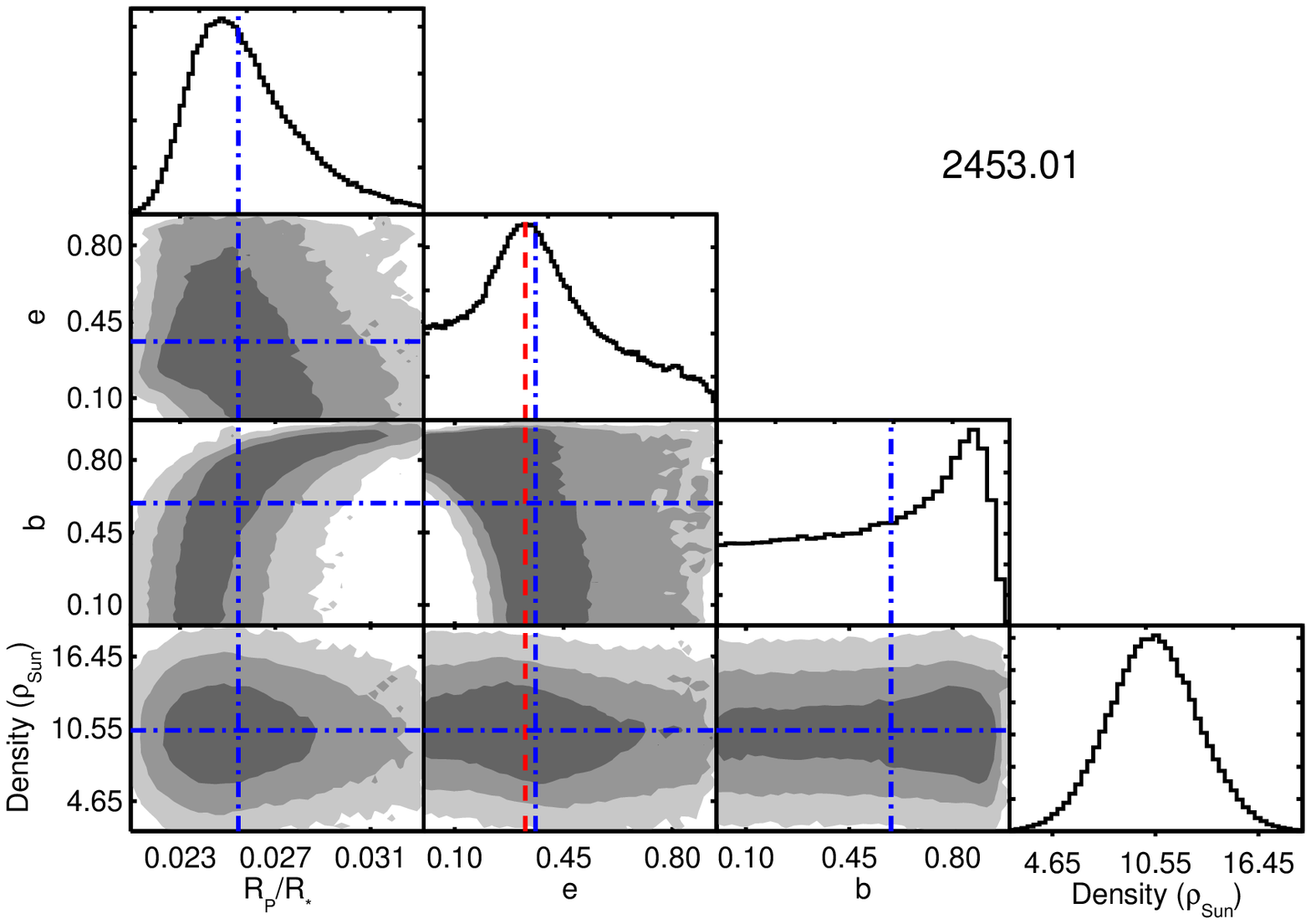}
\includegraphics[width=0.5\textwidth]{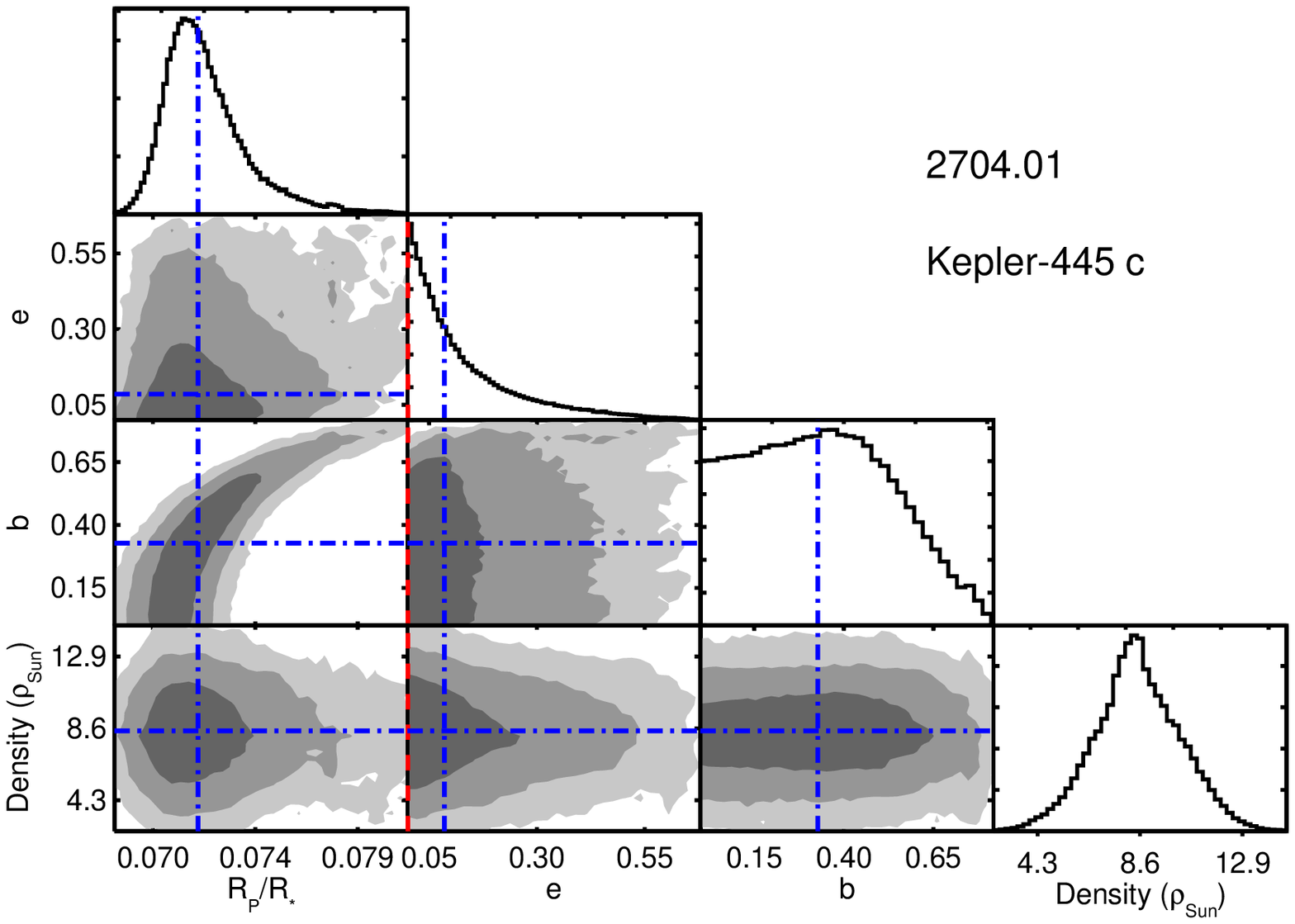}
\includegraphics[width=0.5\textwidth]{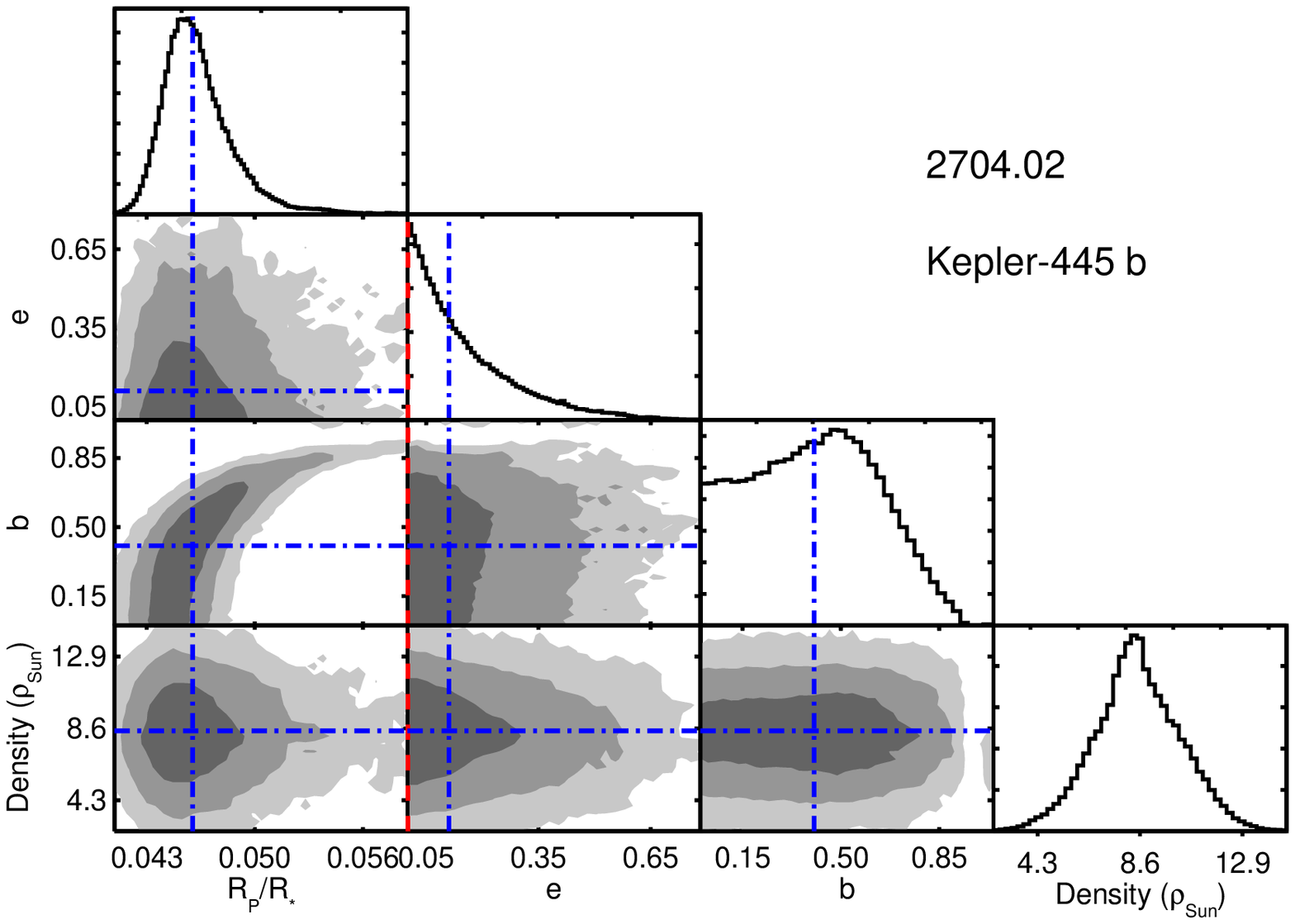}
\caption{Continued}
 \label{fig:transitfit}
\end{figure*}

\setcounter{figure}{1}  

\begin{figure*}
\includegraphics[width=0.5\textwidth]{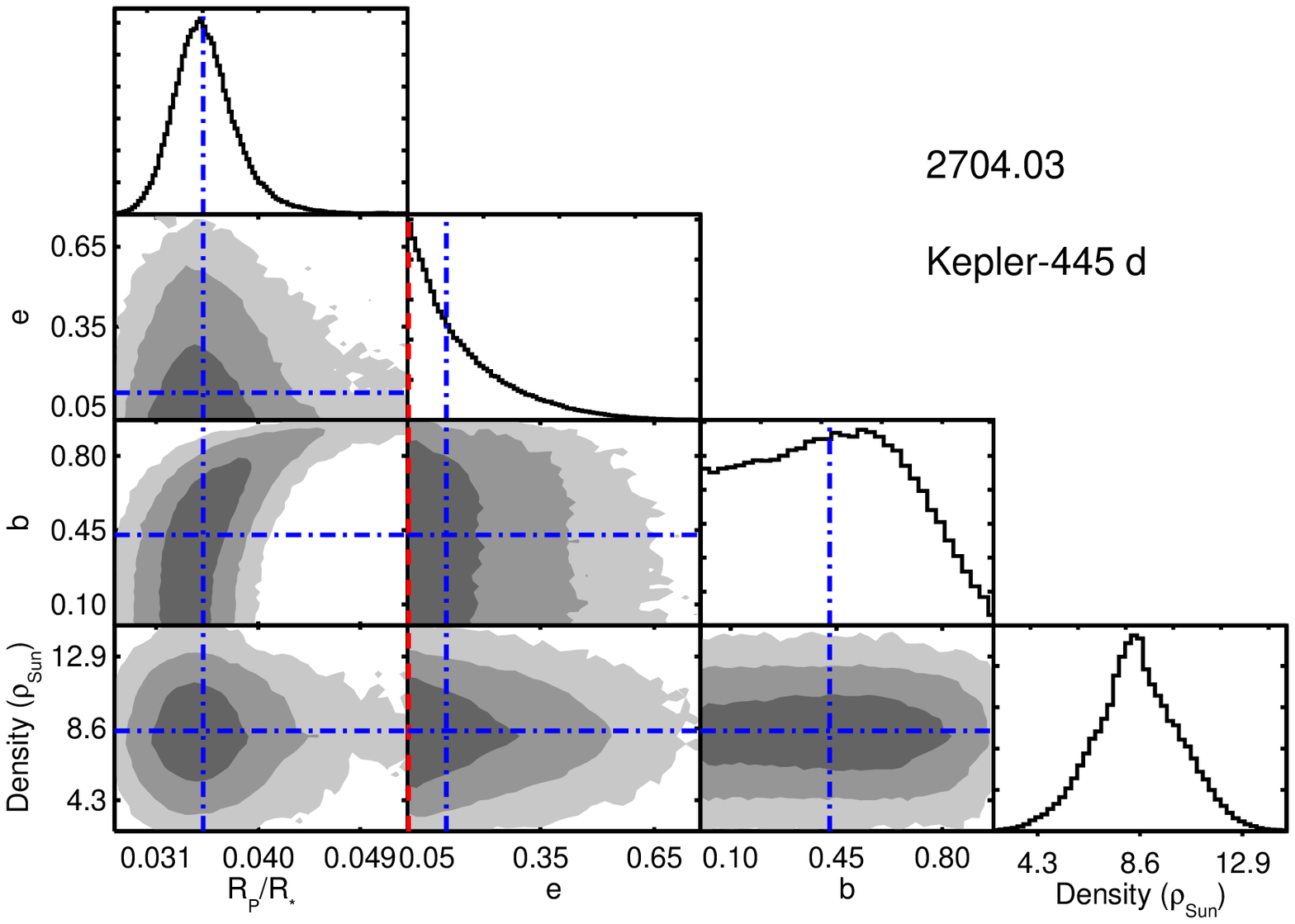}
\includegraphics[width=0.5\textwidth]{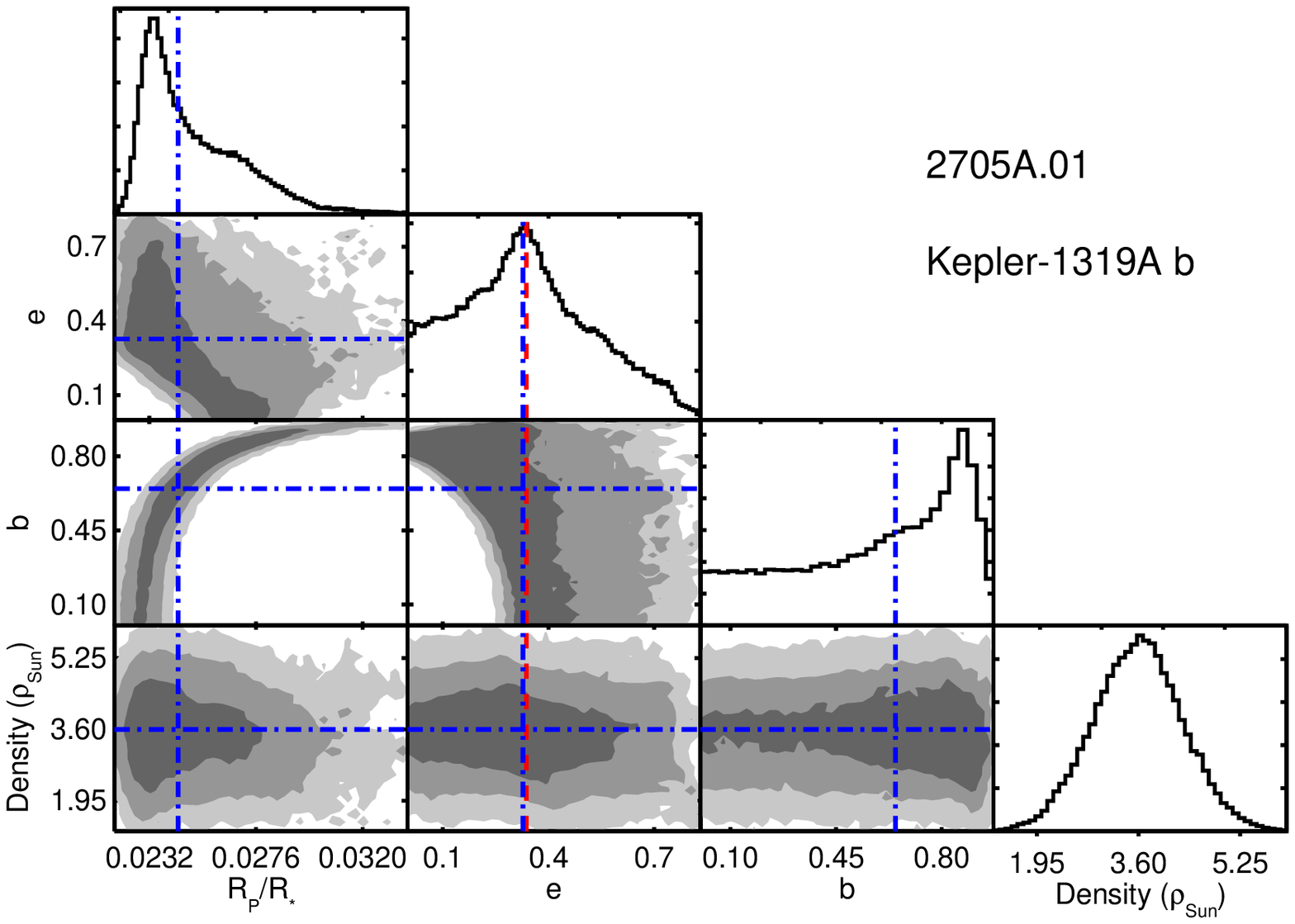}
\includegraphics[width=0.5\textwidth]{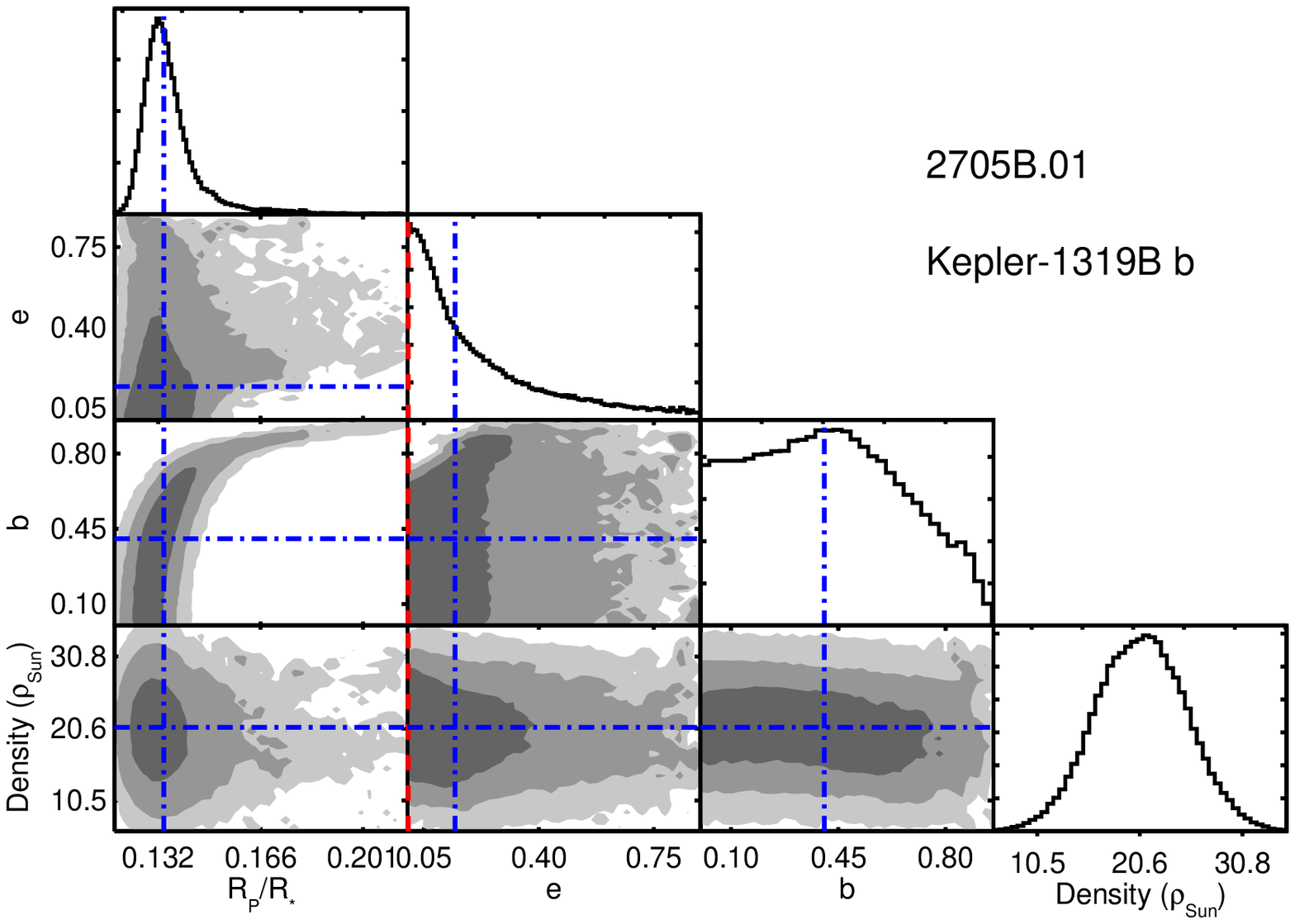}
\caption{Continued}
 \label{fig:transitfit}
\end{figure*}

\end{document}